\documentclass[aps, twocolumn, superscriptaddress, showpacs]{revtex4-1}

\usepackage{amsmath}
\usepackage{amssymb}
\usepackage{xfrac}
\usepackage{xcolor}
\usepackage{physics}
\usepackage{braket}
\usepackage{breqn}
\usepackage{dcolumn}
\usepackage{ragged2e}
\usepackage{tabularx}
\usepackage{multirow}
\usepackage{cancel}
\usepackage{soul}
\usepackage{comment}
\renewcommand{\v}{\mathbf} 

\begin{document}

\title{Effect of Hilbert space truncation on Anderson localization}
\author{Akshay Krishna and R.\ N.\ Bhatt}
\affiliation{Department of Electrical Engineering$,$
Princeton University$,$ Princeton$,$ NJ 08544}
\date{\today}

\begin{abstract}
The 1-D Anderson model possesses a completely localized spectrum of eigenstates for all values of the disorder. We consider the effect of projecting the Hamiltonian to a truncated Hilbert space, destroying time reversal symmetry. We analyze the ensuing eigenstates using different measures such as inverse participation ratio and sample-averaged moments of the position operator. In addition, we examine amplitude fluctuations in detail to detect the possibility of multifractal behavior (characteristic of mobility edges) that may arise as a result of the truncation procedure.
\end{abstract}
\maketitle

\section{Introduction}
It is well established that in single particle models with short range hopping and uncorrelated on-site energies \cite{Anderson58}, all eigenstates are localized for arbitrary small disorder in one dimension \cite{MottTwose61}. Further, extensive work following the advent of the scaling theory of localization \cite{GangofFour79}, showed clearly the importance of time-reversed paths or the symmetry between time-reversed states (see e.g. references \cite{LeeRamakrishnan85} and \cite{AltshulerAronov}), in the theory of weak localization. This helped establish the absence of extended states in the Anderson model \cite{Anderson58} in two dimensions with pure potential scattering. Extensive theoretical and numerical work \cite{MacKinnonKramer1981, SlevinOhtsuki1997, SlevinOhtsuki1999, SlevinOhtsuki2001} has determined quantitatively universal quantities associated with the Anderson problem in $d = 1, 2 \text{ and } 3$ dimensions, which have been summarized in several reviews, e.g.\ \cite{MacKinnonKramer1993, EversMirlin08}. 

	In contrast, in a two-dimensional system subject to a strong perpendicular magnetic field (in the single Landau level or quantum Hall limit), states remain extended at the (one-dimensional) boundaries, even in the strong disorder limit \cite{Halperin1982, HuckesteinKramer1990, HuoBhatt1992, ChalkerCoddington, SlevinOhtsukiQH}. This again is attributable to the absence of time-reversal symmetry in the presence of a magnetic field. Even for weak magnetic fields, one finds enhancement of conductance in experimental systems, in two-dimensions \cite{Bishop83, Bergman83} as well as in bulk, three-dimensional materials \cite{Rosenbaum81, Rosenbaum83}. This negative magnetoresistance in the weak disorder regime is quantitatively attributable to the suppression of weak localization due to the time-reversal symmetry breaking magnetic field \cite{Gorkov79, Hikami80, Kawabata80, LeeRamakrishnan85, AltshulerAronov}.

	Given the crucial effect of time-reversal symmetry (TRS) on Anderson Localization, a study of models which can examine the effect of symmetry breaking, and investigate the interpolation between the extreme cases of TRS (the standard Anderson model) and complete breaking of TRS (as in the Landau level limit) would be very desirable. In the case of magnetic field, the breaking up of the band into Landau levels in any finite field complicates in the interpretation of the results \cite{LiuDas1994, XieLiu1996, LiuXieNiu1996, YangBhatt1996, WanBhatt2001}. Consequently, in this study, we break TRS in a different manner -- namely, by projecting out parts of the Hilbert space, in a controlled manner. This could potentially provide such a platform which is more accessible than the low field, multi Landau level problem.

	The model we consider is essentially the original Anderson model on a one-dimensional lattice, characterized by a nearest neighbor hopping and an onsite energy drawn independently for each site from a uniform distribution from $-W$ to $W$. We then project out a certain fraction of eigenstates of the non-disordered lattice (which can be characterized by the wavevector $\v{k}$), and look at the eigenstates of the resulting truncated Hamiltonian. 

	The technique of projecting out parts of the Hilbert space is extensively used in condensed matter problems. For example, in correlated electron systems, the full many-body Hilbert space is often truncated to that arising from one or two electronic bands near the Fermi level, e.g.\ in the Hubbard model \cite{Hubbard63, Gutzwiller65} or the periodic Anderson model \cite{Anderson61, RiceUeda1985, Jarrell1995}. Such a procedure can be justified on the basis of Renormalization Group (RG) arguments \cite{Wilson75}. The RG method is found to be crucial in analyzing many problems such as the Kondo effect \cite{Kondo64, Anderson70, Krishnamurthy80}, and affords a more complete understanding of Landau Fermi Liquid Theory \cite{Landau57, BaymPethick, Shankar94}. 

However, our purpose in this paper is somewhat different, and what we do here is more radical. We simply project out states that are neither separated from others by a gap, nor are taken into account by any RG procedure. Thus in effect, we add to the conventional Anderson Hamiltonian $H_0$ a non-perturbative term of the form $H' = V \sum\nolimits' \ket{k} \bra{k}$, (where the primed sum is over a \emph{subset} of plane wave eigenstates of the tight-binding Hamiltonian without disorder), and take the limit   $V \to \infty$. In essence, we study a new generalized Hamiltonian of which the standard Anderson Hamiltonian is a special case.

	If we project out all states with negative $\v{k}$, we have a Hilbert space with only right-moving states, much like a two-dimensional disordered system in high magnetic field, which has states with only one chirality (clockwise or counterclockwise, depending on the sign of the magnetic field). As a result, there is complete breaking of TRS, and we find all states remain extended for all values of the disorder parameter $W$. 

	We then consider the case of projecting out only a fraction $F$ of the left-moving states which we call ``partial TRS breaking'', and examine the evolution of properties of the eigenstates as the fraction F  is changed. In this case, the Hilbert space retains some $\v{k}$-states along with their time-reversed partners ($\v{-k}$), while other $\v{k}$-states do not have their time-reversed partner. Using the analogy with weak localization, one may expect the former to create localized states, and the latter to remain extended. With this intuition, one may anticipate that the model interpolates between a totally extended and a totally localized band of states, as the fraction $F$ is reduced from 1 to 0. Thus this model could be expected to interpolate between itinerant (metallic) and localized (insulating) behavior already in one-dimension and possibly show signs of a metal-insulator transition (single particle localization). Earlier attempts to create transitions between localized and delocalized phases in 1-D include correlated potentials \cite{Phillips90, MouraLyra1998, Izrailev99, Moura99}, incommensurate potentials \cite{AubryAndre80} and long range hopping \cite{ZhouBhatt, Mourearxiv}.  

	Our results show that this intuitive reasoning is not entirely correct. While some properties of the eigenstates are roughly in accord with this expectation, other aspects of eigenstates and their distribution and correlation belie this logic. We find, in particular, that different measures of the eigenstates give contradictory results for the ``nature'' of eigenstates, and eigenvalue correlations (e.g. distribution of eigenvalues splittings) turn out to be quite different. Nevertheless, as we argue in the concluding section, the model offers interesting new insights into the physics of Anderson localization.

	A study of the effect of Hilbert space truncation on \emph{many-body} systems, similar in its formulation to the current work, was undertaken recently \cite{Geraedts17} to study its effect on the phenomenon of many-body localization. While the goals of that study were somewhat different, it showed that it is possible to study many-body localization in incomplete Hilbert spaces, provided one is willing to sacrifice some fidelity. Since most experimental results have imperfect fidelity, e.g. due to noise from extraneous sources, this does not in itself prove to be the limiting issue in many situations. 

	The plan of this paper is as follows. In Section II, we define the model and describe the quantities we calculate for different values of $F$. In Section III, we provide results of our investigation of the nature (localized or extended) of our eigenstates using different measures used in past literature. In section IV, we compare our exact numerical results with a na\"{i}ve expectation of the effect of Hilbert space truncation using perturbative analysis. In Section V, we investigate in more detail the statistics of the magnitude of the wavefunction over many orders of magnitude, and thereby motivate and evaluate the multifractal distribution function $f(\alpha)$, used to study eigenstates in the vicinity of a metal-insulator transition, for states arising in our model. In Section VI, we calculate the ensemble averaged current carried by the eigenstates as a function of energy. We also study the distribution of eigenvalue spacing, and compare with the well-known universal Wigner-Dyson results for unprojected Hamiltonians with different symmetry classes (GOE/GUE). In Section VII, we discuss the case of large disorder. Finally, in Section VIII, we summarize our results and conclusions.
	
\section{The model}

	The model we study is the standard 1-D tight-binding Hamiltonian with a constant nearest-neighbour hoppping term $t$ and variable on-site disorder $\epsilon_i$. The Hamiltonian can be written, in second quantized form, as
	\begin{align}
 	H_0 &= \sum_{n=1}^N \bigg[ \epsilon_m c_m^\dagger c_m -t \Big( c_m^\dagger c_{m+1} + c_{m+1}^\dagger c_m \Big) \bigg]
 	\end{align}
	or equivalently, in the position space basis $\{\Ket{x_m}\}$, as
	\begin{align}
	H_0 &= \sum_{m=1}^N \bigg[ \epsilon_m \Ket{x_m}\Bra{x_m} \\ &- t \Big( \Ket{x_m}\Bra{x_{m+1}} + \Ket{x_{m+1}}\Bra{x_m} \Big) \bigg] \label{eqAnd}.
 	\end{align}
 	
	In the standard Anderson model, the on-site disorder is drawn randomly from a uniform distribution of width $W$.
	\begin{align}
	P(\epsilon_m) &= \frac{1}{W}, \qquad -\frac{W}{2} < \epsilon_m < \frac{W}{2}.
	\end{align}
	Henceforth, we set the hopping parameter $t$ to 1. The parameter $W$ then denotes the effective disorder scale in the problem. The zero-disorder bandwidth $B = 4$ is also another important energy scale, and in later sections we use the quantity $W/B$ to distinguish the small and large disorder regimes.
	
\begin{figure}[ht!]
\centering
 \includegraphics[width=1.0\columnwidth]{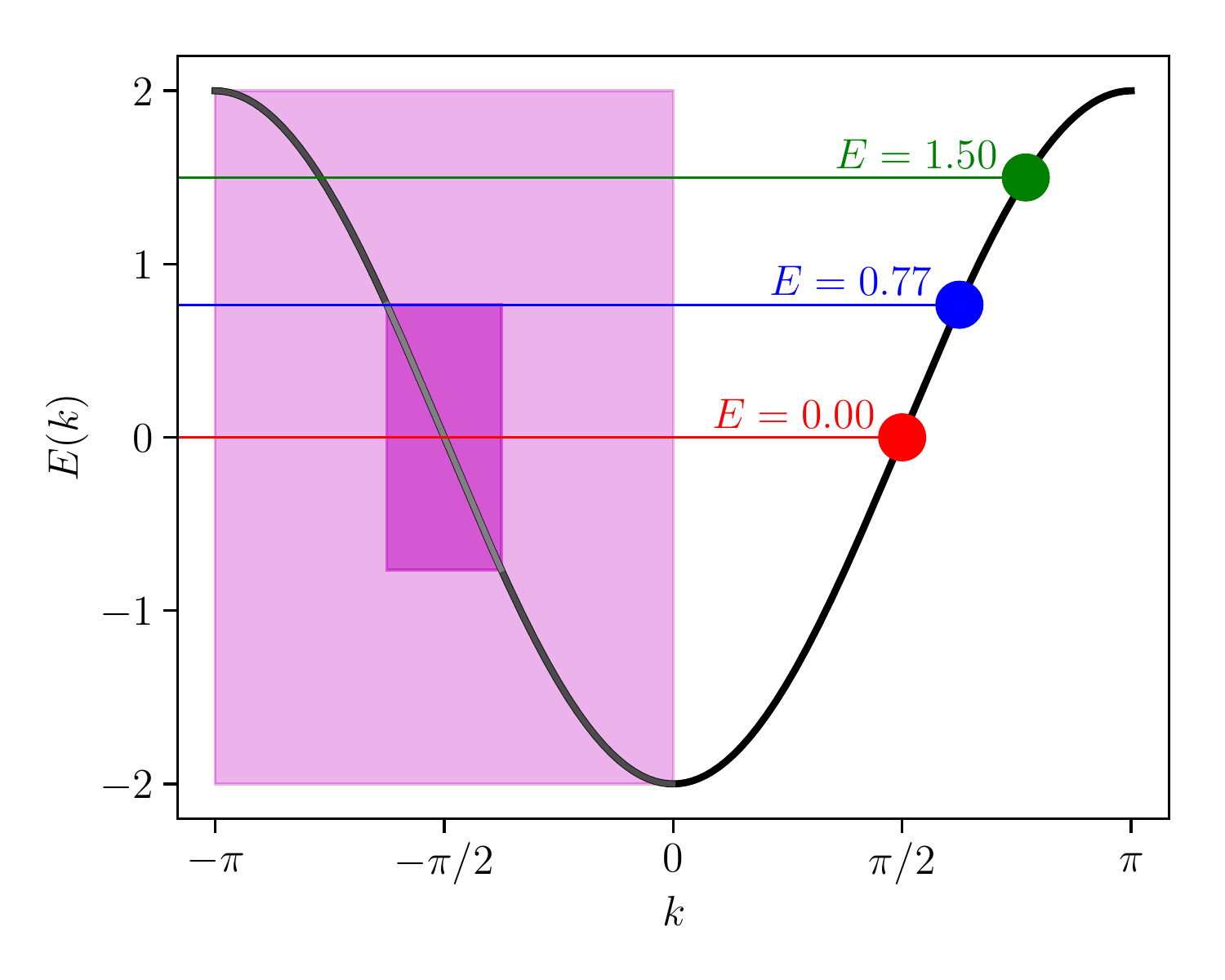} 
\caption{Schematic of the energy-momentum dispersion relation of the zero-disorder 1-D nearest neighbour hopping problem. The solid black line indicates the tight-binding band $E(k) = -2 \cos(k)$. The light pink shaded region represents the part of the Hamiltonian projected out in the case of complete TRS breaking ($F = 1$). Likewise, the dark pink region represents the portion of the Hamiltonian projected out in the partial TRS breaking case ($F = 1/4$). We also depict the colour code used to represent the three characteristic energies ($E = 0$, $E = E_c = 2 \cos \left( \frac{3 \pi}{8} \right) \approx 0.77$ and $E = 1.5$) for which we calculate various ensemble-averaged quantities in the paper. } \label{figschem}
\end{figure}

	We impose periodic boundary conditions, and identify the ($N+1$)\textsuperscript{th} site with the $1$\textsuperscript{st} site. The zero-disorder ($W = 0$) case of the Hamiltonian gives a tight-binding band (see Fig.\ \ref{figschem}) \begin{align*}
	E(k_r) &= -2 \cos(k_r)\\ 
	\intertext{where}
	 k_r &= \frac{2 \pi r}{N},  \quad r \in \left\lbrace -\frac{N}{2}, -\frac{N}{2} + 1, \cdots, \frac{N}{2} -1 \right\rbrace.
	\end{align*} All eigenstates $\Ket{k_r}$ are Bloch states with a position space representation $\Braket{x_m | k_r} = \frac{1}{\sqrt{N}} \exp(\frac{i 2 \pi m r}{N})$. When disorder is non-zero ($W \neq 0$), translational invariance breaks down for every disorder realization. As a result, all eigenstates become localized for all realizations of disorder but for a set of measure zero.
	
	By a unitary transform to momentum space, one can write down the Hamiltonian in terms of the basis $\{\Ket{k_r}\}$ as 
\begin{dmath}
	\hat{H} = \sum_{r, s} \frac{\sum_m \epsilon_m}{N} e^{-\frac{i 2 \pi m(r-s)}{N}} \Ket{k_r} \Bra{k_s} - 2 \sum_{r} \cos(\frac{2 \pi r}{N}) \Ket{k_r} \Bra{k_r}. \label{eqHamK}
\end{dmath}

If we let the indices $r$ and $s$ in Equation (\ref{eqHamK}) run over the full range $-\frac{N}{2}$ to $\frac{N}{2}-1$, we have the Anderson model. To this, we add the second term $H' = V \sum\nolimits' \ket{k_r} \bra{k_r}$, where the primed sum over $r$ is over a subset of its allowed values. The limit $V \to \infty$ implies a projection onto the remaining $k$ states. Within this paradigm, we look at a couple of different cases. By restricting this sum to be only over non-negative indices, i.e., $0 \leq r,s < \frac{N}{2}$, one can project out all states with negative $k$. This corresponds to the case $F=1$ of complete TRS breaking. For partial TRS breaking, we restrict the indices to exclude a fraction $F$ of negative $k$ states near the centre of the band, i.e., $r,s \in \left\lbrace -\frac{N}{2}, \cdots, -\frac{N(1+F)}{4} - 1 \right\rbrace \cup \left\lbrace -\frac{N(1-F)}{4}, \cdots, \frac{N}{2} -1 \right\rbrace$. For this choice, particle-hole symmetry is retained even in the truncated Hamiltonian. This truncation procedure is equivalent to eliminating some Fourier components $k$ from the Hamiltonian. 

	When $F = 0$ (Anderson model) or $F = 1$ (complete TRS breaking), the disorder strength $W$ is the only energy scale in the problem. For intermediate values of $F$, another energy scale $E_c$ is introduced by the fourier-space cutoff. This scale separates states with and without their time-reversed partners, in the no-disorder case, and is given by $E_c = 2 \cos \left( \frac{\pi}{2} (1-F) \right)$. We focus primarily on two illustrative cases, namely, $F=1$ (complete TRS breaking) and $F = 1/4$ (partial TRS breaking). Results for typical $F$ ($0 < F < 1$) are qualitatively similar to $F = 1/4$ except for a change in the characteristic energy $E_c$, which controls the various disorder regimes. As shown in Fig.\ \ref{figschem}, $E_c = 2 \cos \left( \frac{3 \pi}{8} \right) = 0.7654 \cdots$ for the case $F = 1/4$. \emph{In figures in the rest of the paper, we represent this value to two decimal places as $E = 0.77$ for brevity}.
	
	To study the localization behaviour of eigenstates, two metrics that can be used to quantify the localization length ($\xi$) are the inverse participation ratio and the second moment of the probability density of the wavefunction.  In later sections, we also use other measures such as eigenvalue statistics and currents to study localization properties of wavefunctions. Alternative classification schemes based on the shape of the wavefunction itself have also been used \cite{Varga1992}.
	
	For a normalized wavefunction, the inverse participation ratio ($P_2$) is defined as \cite{MacKinnonKramer1993, EversMirlin08} \begin{align}
	P_2 &\equiv \sum\limits_m |\psi_m|^4 \label{eqIPR2}, \\ \intertext{where} \psi_m &= \Braket{x_m | \psi} 
	\end{align} measures the amplitude of the wavefunction $\Ket{\psi}$ at the $m$\textsuperscript{th}  site. The IPR localization length can be defined as \begin{align}
	\xi_{IPR} &\equiv \frac{1}{2 P_2} \label{xiIPR}.
	\end{align}
	
	The second moment $M_2$, or variance, of a variable $x$ with probability density $p(x)$ measures its spread from the mean. It is defined as \begin{align} M_2 &\equiv  \int\limits_{-\infty}^{\infty} \mathrm{d}x \ p(x) (x-a)^2 - \left[ \int\limits_{-\infty}^{\infty} \mathrm{d}x \ p(x) (x-a) \right]^2 ,
	\end{align} where $a$ is the coordinate of some arbitrary `origin'. Note that $M_2$ is independent of the choice of $a$, and hence $a$ is usually set to zero \footnote{This is just the customary definition $M_2 = \bar{x^2} - \bar{x}^2$ where $\bar{x} = \int\limits_{-\infty}^{\infty} (x-a) p(x) \mathrm{d}x$ measured with respect to arbitrary origin $a$. While $\bar{x}$ depends on the choice of origin, $M_2$ does not.}.
	
	One can use a similar quantity to measure the spread of a wavefunction. However for a finite sized system with periodic boundary conditions, the distance $x$ is defined only modulo system size $L$. The second moment $M_2$ is no longer independent of the choice of origin $a$ and $a$ is chosen to minimize the value of $M_2$. The formula for second moment is then modified as \cite{Heyer81, Levy39} \begin{dmath}
	M_2 = \min\limits_{a: 0 \leq a < L} \left\lbrace \int_0^L \mathrm{d}x \ p(x) ( (x-a) \text{ mod } L)^2 - \left[ \int_0^L \mathrm{d}x \ p(x) ((x-a) \text{ mod } L) \right]^2 \right\rbrace.
	\end{dmath}  In dealing with wavefunctions on a discrete lattice of $N$ sites one may write the analogous formula as \footnote{ As an example, consider a wavefunction with support only on two sites $|\psi_1|^2 = |\psi_3|^2 = 1/2$. Its second moment $M_2 = 1$. The same wavefunction, with a shifted origin, is $|\psi_2|^2 = |\psi_N|^2 = 1/2$. The second moment of this wavefunction is $N/2-1$. This example illustrates the need to choose an appropriate origin to minimize $M_2$. Physically this choice implies that the boundary sites $n=1$ and $n=N$ are as far as possible from the bulk of the wavefunction.} \begin{dmath}
	M_2 = \min\limits_a \sum\limits_{m=1}^N ((m-a) \text{ mod }N)^2 |\psi_m|^2 - \left[ \sum\limits_{m=1}^N ((m-a) \text{ mod } N) |\psi_m|^2 \right]^2 .\end{dmath} The corresponding localization length derived from this quantity is \begin{align}
	\xi_{M_2} \equiv \sqrt{2 M_2} \label{xiM2}.
\end{align}
	A similar quantity, based on the second moment, was proposed as a measure of the localization length in \cite{Raffaele1999}.

	With the numerical factors of $1/2$ in Eq.\ \ref{xiIPR} and $\sqrt{2}$ in Eq.\ \ref{xiM2}, the definitions of $\xi_{IPR}$ and $\xi_{M_2}$ coincide with the exponential decay length for a purely exponential wavefunction. The two localization lengths defined above measure slightly different quantities. $\xi_{IPR}$ measures the  number of sites on which the amplitude of the wavefunction is significant. On the other hand, $\xi_{M_2}$ measures how far the wavefunction extends from a central site before its amplitude decays significantly. This distinction between $\xi_{IPR}$ and $\xi_{M_2}$ will be essential to understand the discussion in later sections of this paper. Table \ref{tabxi} summarizes their behaviour in the extended and localized wavefunctions. As stated earlier, for the exponentially localized wavefunction, $\xi_{IPR} = \xi_{M_2} = \xi$ with the numerical factors in Eq.\ \ref{xiIPR} and Eq. \ref{xiM2}.
	
 \begin{table}[ht!]
\begin{ruledtabular}
\begin{tabular}{l | c c c}
 & $|\psi_m|$ & $\xi_{IPR}$ & $\xi_{M_2}$\\ \hline
 Exponentially localized  & \multirow{2}{*}{$\mathcal{C} \exp(-\frac{m}{\xi} )$} & \multirow{2}{*}{$\xi$} & \multirow{2}{*}{$\xi$}\\
 state ($1 \ll \xi \ll N$)  &  &  & \\ \hline
  Gaussian localized & \multirow{2}{*}{$\mathcal{C} \exp(-\frac{m^2}{2 \xi^2} )$} & \multirow{2}{*}{$\sqrt{\frac{\pi}{2}} \xi$} & \multirow{2}{*}{$\xi$} \\ 
 ($1 \ll \xi \ll N$)  &  &  & \\ \hline
 Power-law localized & \multirow{2}{*}{$\mathcal{C} m^{-\alpha}$} & \multirow{2}{*}{$\frac{2 (1 + 2 \zeta(2 \alpha))^2}{1 + 2 \zeta(4 \alpha)}$} & \multirow{2}{*}{$\sqrt{\frac{4 \zeta(2 \alpha-2)}{1 + 2 \zeta(2 \alpha)}}$} \\ 
 ($N \to \infty, \alpha > 3/2$)  &  &  & \\ \hline
 Periodic (Bloch) & \multirow{2}{*}{$\frac{1}{\sqrt{N}}$}\ & \multirow{2}{*}{$\frac{N}{2}$} & \multirow{2}{*}{$\frac{N}{\sqrt{6}}$} \\ 
 extended state  &  &  & \\
\end{tabular}
\caption{Analytic expressions for localization lengths for various kinds of localized and extended wavefunctions on $N$ sites in 1-D. The symbol $\mathcal{C}$ above denotes a normalization constant  and $\zeta(s)$ denotes the Riemann zeta function.}\label{tabxi}
\end{ruledtabular}
\end{table}

	In the next section, we use these metrics to make the case that projecting out a fraction of the Hilbert space has a dramatic effect on the nature of eigenstates, including inducing a transition from localized to extended states in some domain of the spectrum. 

\section{Nature of eigenstates of the truncated Hamiltonian}

	First, to establish that our numerical calculations give sensible results and to establish a basis for comparison with the truncated Hamiltonian, we plot the localization lengths of eigenstates in the standard Anderson model. For our numerical calculations, we first focus on the small disorder case $W = 1$ in system sizes up to $N = 8192$ sites. For this value of disorder, the spectrum broadens from $[-2, 2]$ in the zero-disorder case to $[-2.5, 2.5]$. Due to particle-hole symmetry in ensemble-averaged quantities, we plot energy-resolved quantities as a function of the absolute value of the energy only. At each value of system size considered, we compute 2,048,000 eigenenergies and wavefunctions by exact diagonalization, as summarized in Table \ref{tab_num}, All quantities plotted are then obtained by ensemble averaging.
	 
\begin{table}[ht!]
\begin{ruledtabular}
\begin{tabular}{c c}
System size ($N$) & Number of realizations \\ \hline
64 & 32000\\
128 & 16000\\
256 & 8000\\
512 & 4000\\
1024 & 2000\\
2048 & 1000\\
4096 & 500\\
8192 & 250
\end{tabular}
\caption{A summary of the number of numerical realizations of random disorder performed in this work. The numbers above are applicable to all disorder strengths and truncation windows discussed in this text.}\label{tab_num}
\end{ruledtabular}
\end{table}	 
	 
	 \begin{figure}[ht!]
\centering
  \includegraphics[width=1.0\columnwidth]{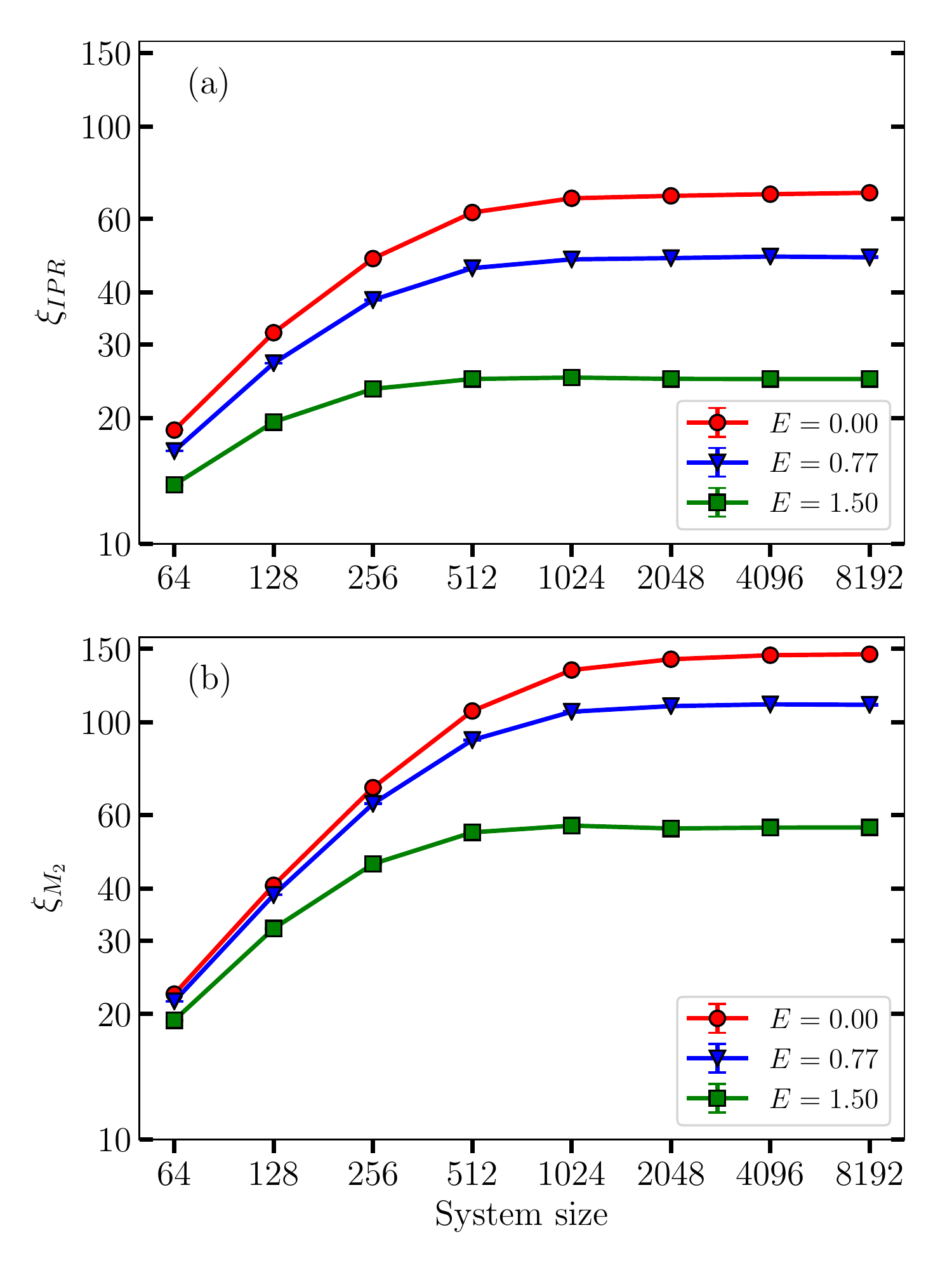} 
\caption{ Log-log plot of the IPR localization length $\xi_{IPR}$ (a) and the second moment based localization length $\xi_{M_2}$ (b) as a function of system size $N$ in the Anderson model in the small disorder regime ($W = 1$). We have chosen to plot the localization lengths for three representative energy values in the middle of the spectrum, namely $E = 0, 0.77 \text{ and } 1.5$. The error bars are not visible on this scale.}\label{figxiAnd}
\end{figure}

	 In this regime, the energy-resolved mean localization lengths are much larger than the lattice spacing (of the order of few tens to a hundred) but well within the largest few system sizes that we analyze. For a fixed system size, the localization lengths are the largest for states closest to the band centre ($E = 0$), and decrease as one moves away to the band edges. In Fig.\ \ref{figxiAnd}, we see that the localization lengths increase as a function of system size at first for the smallest system sizes considered. This effect is due to the finite size of the system. But at the largest few system sizes (for $N \gtrsim 1024$), both measures of localization length saturate to constant values as summarized in Table \ref{tabxiAnd}. Error bars are calculated from one standard deviation in the usual manner by dividing by the square root of the number of independent samples.
	 
	 As can be seen, the two measures of localization roughly scale with each other, with $\xi_{M_2}$ somewhat larger than twice $\xi_{IPR}$ \footnote{They do not equal each other, as may be expected for a perfectly exponentially localized wavefunction, as in table \ref{tabxi}. This is due to the fact that Anderson localized wavefunctions are exponentials modulated by a sinusoidal component (an example of an Anderson localized wavefunction is given in Fig.\ \ref{figAndW1}).  In such a case, the wavefunction is not purely monotonic and typically $\xi_{M_2} > \xi_{IPR}$.}. Since these localization lengths $\xi_{IPR}$ and $\xi_{M_2}$ are much smaller than the largest system size in our study, $N_{max} = 8192$, we believe that finite-size effects do not affect our results.
	 
	 \begin{table}[ht!]
\begin{ruledtabular}
\begin{tabular}{c c c}
Energy ($E$) & $\xi_{IPR}$ & $\xi_{M_2}$\\ \hline
  $0.00$ & $69.3 \pm 0.3$ & $145.6 \pm 0.7$ \\
  $0.77$ & $48.6 \pm 0.2$ & $110.2 \pm 0.5$ \\
  $1.50$ & $24.8 \pm 0.1$ & $56.0 \pm 0.2$
\end{tabular}
\caption{Ensemble averaged localization lengths of the Anderson Hamiltonian at $W = 1$ for $N = 8192$ sites at three representative values of the energy.}\label{tabxiAnd}
\end{ruledtabular}
\end{table}	 
	
\begin{figure}[ht!]
\centering
 \includegraphics[width=1.0\columnwidth]{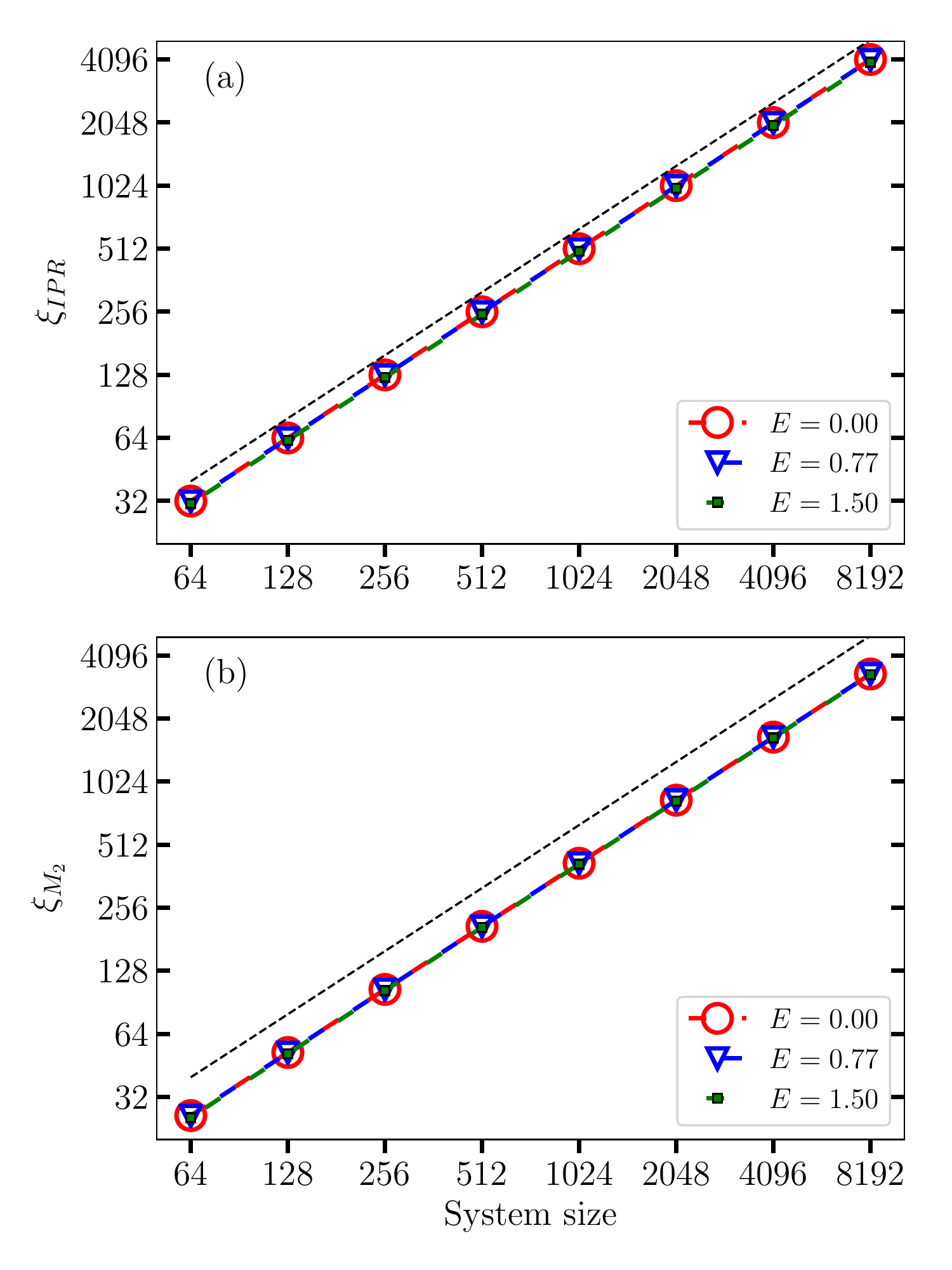} 
\caption{ Log-log plot of the IPR localization length $\xi_{IPR}$ (a) and the second moment based localization length $\xi_{M_2}$ (b) as a function of system size  $N$ in the Anderson model with complete TRS breaking ($F = 1$). All $k$ states in $[-\pi, 0]$ are projected out. The disorder strength and energy windows are the same as in Fig.\ \ref{figxiAnd}. It is clear that the localization lengths are independent of energy. The dotted black line with slope 1 is drawn as a guide to the eye, and is the same in both panels. The error bars are not visible on this scale.} \label{figxiTRS}
\end{figure}
	 
\begin{figure}[ht!]
\centering
 \includegraphics[width=1.0\columnwidth]{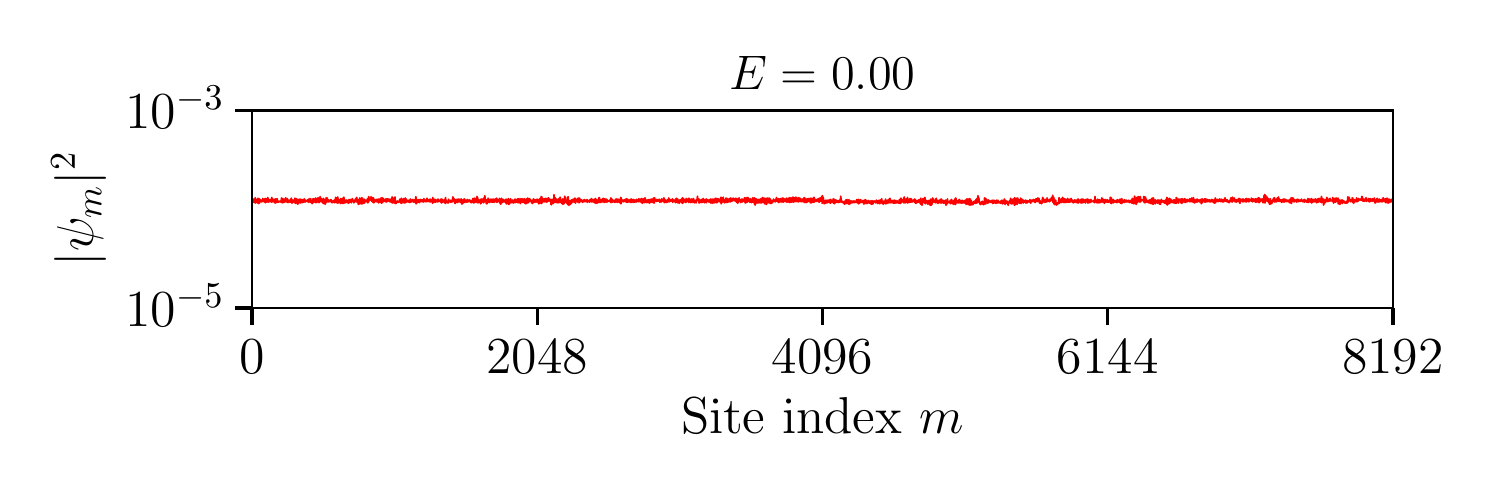}
\caption{A typical wavefunction at the centre of the band ($E=0$) of the Anderson model with complete TRS breaking ($F=1$) at small disorder ($W = 1$) for a system with $N = 8192$ sites. Wavefunctions for other values of $E$ (except in the tail of the density of states) are very similar.}\label{figwfsTRS1}
\end{figure}	 
	 
	 Next, we calculate the same quantities for the case of $F=1$ (Fig.\ \ref{figxiTRS}), where time reversal symmetry is completely broken. As Fig.\ \ref{figxiTRS} suggests, the entire bulk of the spectrum is affected in an identical manner by the truncation procedure, without any apparent energy-resolved differences in behaviour. Both measures of localization length $\xi$ scale as a linear function of system size $N$ within our error bars. We conclude that  the eigenstates of these Hamiltonians are extended (see Fig.\ \ref{figwfsTRS1} for an example of a wavefunction at $E = 0$). They are extended because these Hamiltonians have no $-k$ states.

\begin{figure}[ht!]
\centering
 \includegraphics[width=1.0\columnwidth]{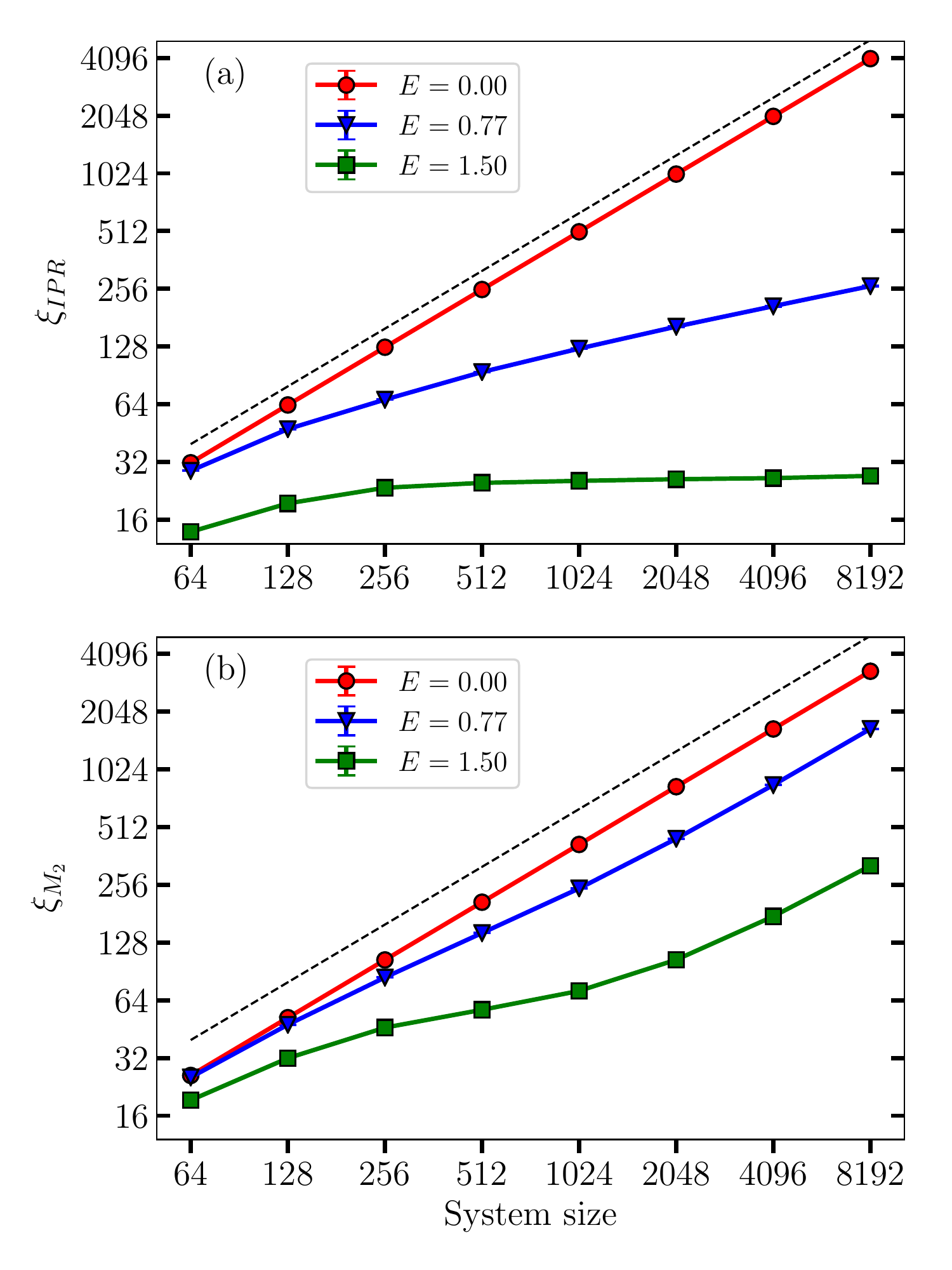}
\caption{ Log-log plot of the IPR localization length $\xi_{IPR}$ (a) and the second moment based localization length $\xi_{M_2}$ (b) as a function of system size $N$ for the case of partially broken TRS ($F = 1/4$, with $k \in [-\frac{5 \pi}{8}, -\frac{3 \pi}{8}]$ removed) at small disorder ($W=1$). The three chosen energies plotted correspond to different regimes of behaviour, as described in the text. The dotted black line with slope 1 is drawn as a guide to the eye, and is the same in both panels.}\label{figxiTRS025}
\end{figure}

Finally, we calculate the localization lengths in the case $F = 1/4$, i.e., when $k$-states in $[-\frac{5\pi}{8}, \frac{3\pi}{8}]$ are projected out. In this case, the energy cut-off $E_c = 2 \cos \left( \frac{3 \pi}{8} \right) \approx 0.77$. The localization  lengths plotted in Fig.\ \ref{figxiTRS025} suggest that the eigenstates respond in an energy-resolved manner to the $F = 1/4$ truncation procedure in different ways.  States near the centre of the band (at $E = 0$) seem extended-like, with both $\xi_{IPR}$ and $\xi_{M_2}$ scaling linearly with system size. This is just like in Fig.\ 2. 

	Looking at the IPR alone, we are tempted to conclude, as our qualitative argument in the introduction would suggest, that there is a transition between extended states at $E = 0$, and localized states above the cutoff energy, at $E = 1.50$, which have a saturating $\xi_{IPR}$ (solid green line), similar to that seen in Fig.\ \ref{figxiTRS}. In between, at the cut-off $E_c = 0.77$ (solid blue line), the IPR localization length grows as some power-law with a non-trivial exponent ($\xi_{IPR} \sim N^{\gamma }$, with $\gamma = 0.37 \pm 0.01$), suggestive of a critical state. 

	However, the second moment in Fig.\ \ref{figxiTRS025} suggests a different story. In this case, all energy windows show an increase in localization length with system size. Given this dichotomy, it is apparent, therefore, that the truncation procedure leads to eigenstates that do not fall into the standard paradigm of localized and extended states. At the cut-off energy $E_c = 0.77$, the second moment localization length scales almost linearly ($\xi_{M_2} \sim N^{(0.92 \pm 0.03)}$). At larger energies ($E = 1.5$), $\xi_{M_2}$ shows negative curvature for small sizes on the double logarithmic pot, and shows a tendency towards saturation like $\xi_{IPR}$. However, it returns to a linear-like scaling at large sizes. We surmise that in the thermodynamic limit $N \to \infty$, all energy windows would show linear scaling of $\xi_{M_2}$ with system size $N$.

	To understand the effect of truncation, we consider approaching from the low-disorder limit. Start from the zero-disorder $W = 0$ case. Here, the eigenstates are momentum states $\Ket{k}$, with $E_k = -2 \cos (k) $. When we turn on disorder, by first-order perturbation theory, the states $\Ket{k'}$ that mix in the most with $\Ket{k}$ are those that maximize the energy denominator $|E_k - E_{k'}|$. Therefore, the eigenstates are spread out in a region around $k$ and $-k$, with significant weight in a region $k - \delta k$ to $k + \delta k$, and $-k - \delta k$ to $-k + \delta k$ where $\delta k$ increases as disorder $W$ increases. 
		
	\begin{figure}[ht!]
\centering
 \includegraphics[width=1.0\columnwidth]{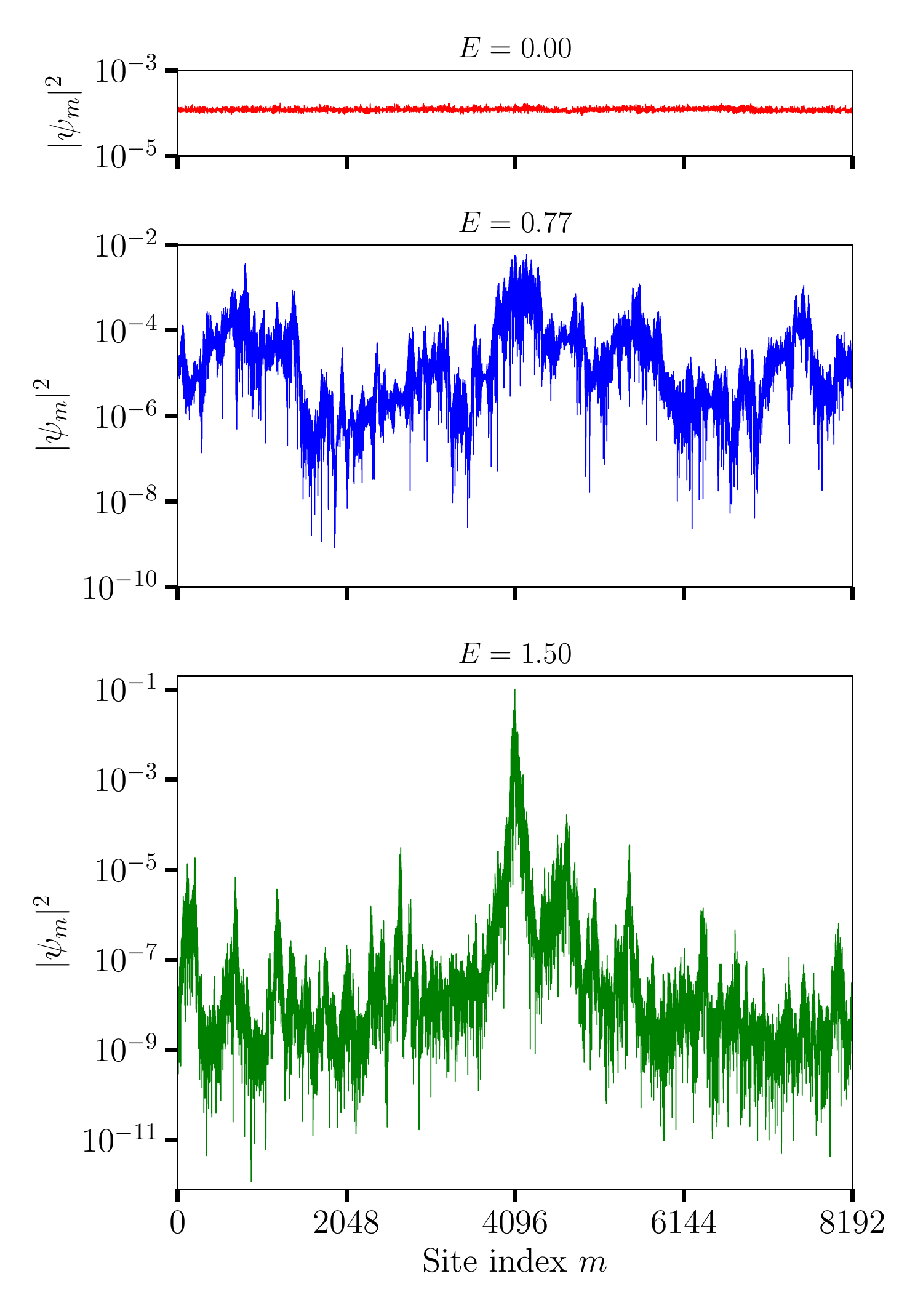}
\caption{Typical Wavefunctions at three characteristic eneries in the Anderson model with partially broken TRS ($F = 1/4$, with $k \in [-\frac{5 \pi}{8}, -\frac{3 \pi}{8}]$ removed) at small disorder ($W=1$) for a system with $N = 8192$ sites. States at the centre of the band (top) are clearly extended, with all sites having nearly equal amplitude. States at $E = 1.50$ (bottom) seem to have a single localized-like peak with a sinusoidal background (see text for discussion). States at $E_c = 0.77$ (middle) seem to have fluctuations over several orders of magnitude, suggestive of critical states.}\label{figwfsW1}
\end{figure}

	When we truncate the Hilbert space by projecting out states $\Ket{k}$ lying in an interval $k_{min} < k < k_{max}$, we may expect that Anderson-localized states with significant weight in that region of $k$ will suddenly be forced to delocalize due to the loss of their time-reversed partners. On the other hand, Anderson-localized states with significant weight outside this range will only be marginally affected by the truncation procedure. We may expect that these eigenstates essentially appear localized, with some residual `sinusoidal' background that persists due to the incompleteness of the Hilbert space. Therefore, one would expect to see extended states in the regime $E < E_c$, and `localized states' in $E > E_c$.
	
	\begin{figure}[ht!]
\centering
 \includegraphics[width=1.0\columnwidth]{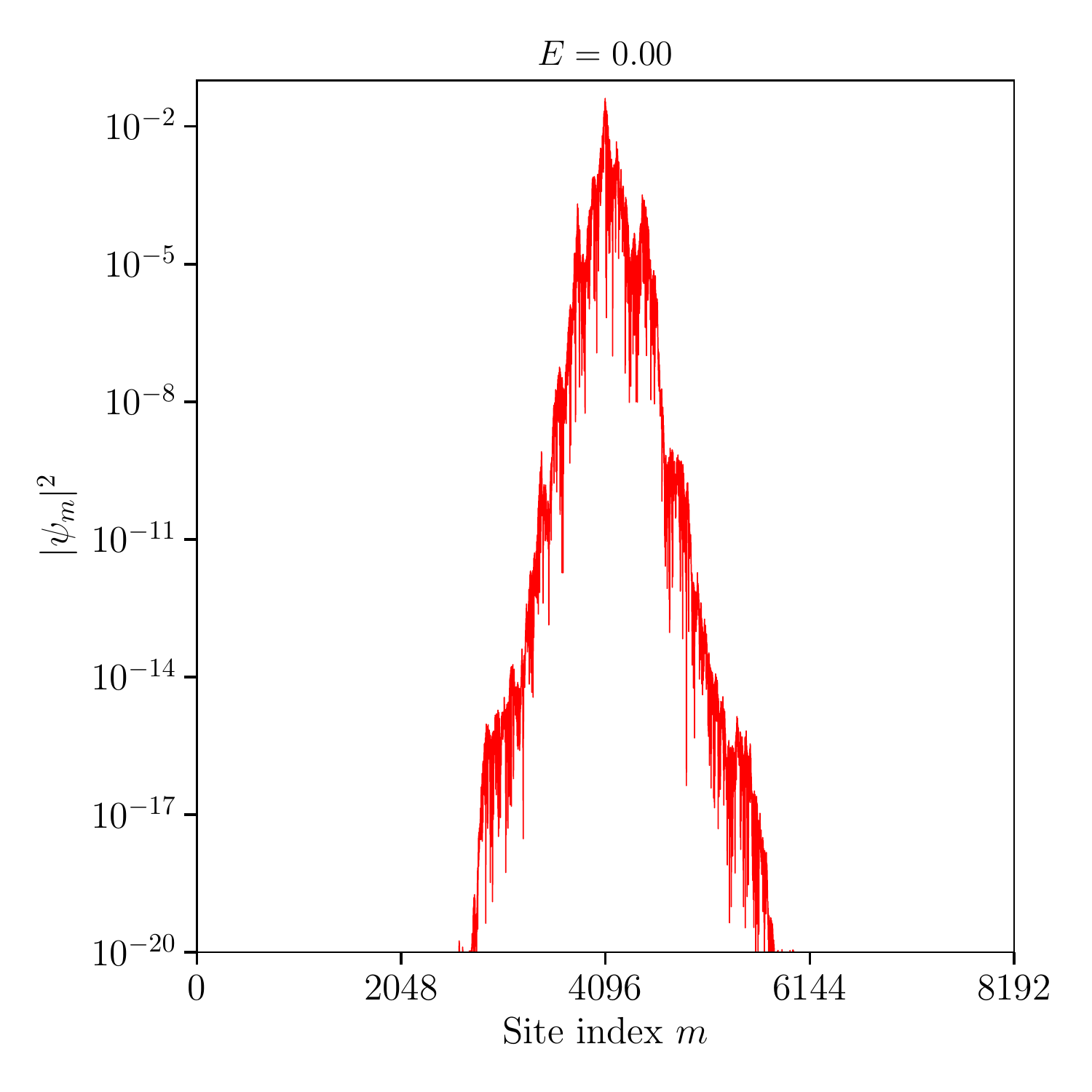}
\caption{An Anderson localized wavefunction at the same disorder strength ($W = 1$) as Fig.\ \ref{figwfsW1} at $N = 8192$ sites. The wavefunction has one central peak and is exponentially decaying away from it. The probability density $|\psi_m|^2$ drops by approximately 18 orders of magnitude over a span of nearly 1600 sites. An envelope of the functional form $e^{-2x/\xi}$, with this decay rate has $\xi \approx 75$, consistent with the values tabulated in table \ref{tabxiAnd}. }\label{figAndW1}
\end{figure}

	In Fig.\ \ref{figwfsW1}, we verify this intuition by plotting the wavefunctions themselves. The wavefunctions plotted are typical, have been randomly chosen and are not cherry-picked to illustrate our point. The perturbation theory based argument above seems to hold for this case, with states of the centre of the band (upper panel) looking essentially extended-like with constant $|\psi|^2$ and having both $\xi_{IPR}$ and $\xi_{M_2}$ scale as in the fully broken TRS case as in figure \ref{figxiTRS}. However, states above $E_c$ (lower panel of Fig.\ \ref{figwfsW1}) appear to have one large localized peak, with a background $|\psi^2| \approx 10^{-8}$ that extends over the rest of the system. This explains why the IPR localization length is small but the second moment localization length is large -- there are only a few sites with large amplitude, but the wavefunction never truly decays. There seems to be a gradual transition between the two kinds of behaviour around $E = E_c$ at which wavefunction amplitude seems to fluctuate, like in a critical state at a metal-insulator transition (middle panel). For reference, we also show a typical Anderson localized wavefunction at the centre of the band in Fig.\ \ref{figAndW1}. In a section \ref{secmfa}, we do a multifractal analysis to systematically examine the nature of the wavefunctions.  
	
	Note that this perturbative argument works only in the case of small disorder, i.e.\ in the regime where the localization length is much larger than the lattice spacing. In this regime, the wavefunctions are relatively localized in Fourier space, allowing us to argue that mixing between different $\Ket{k}$ modes is relatively well-controlled ($\delta k \ll 2 \pi$).  Therefore, we have to choose our parameters carefullly, arranging disorder $W$ to be small enough that the cut-off scale $2 E_c$ is larger than it, yet not so small that finite-size effects become important.

\section{Comparison of exact numerical results with perturbative analysis}

	To put the foregoing discussion on a more sound mathematical footing, let us denote the eigenstates of the truncated Hamiltonian by  $\Ket{\phi^{(t)}_j}, j = 1, \cdots, M$, and those of the original Anderson Hamiltonian by $\Ket{\psi^{(A)}_j}, j = 1, \cdots, N$. The number of Fourier components discarded equals $N - M$. Note that we consider eignestates corresponding to the same disorder realization for both Hamiltonians. The quantity  $\left| \Braket{\phi^{(t)}_i |\psi^{(A)}_j} \right|^2$  measures the overlap of the $i$\textsuperscript{th} eigenstate of the truncated Hamiltonian and the $j$\textsuperscript{th} eigenstate of the Anderson Hamiltonian. For a fixed $i$, $v_i \equiv \sup\limits_j \left| \Braket{\phi^{(t)}_i |\psi^{(A)}_j} \right|^2$ quantifies the extent to which the new eigenstate  $\Ket{\phi^{(t)}_i}$ is mappable to an old eigenstate. If $v_i$ is very nearly equal to unity, then one can make the case that the perturbation has little effect and $\Ket{\phi^{(t)}_i}$ is equivalent to  $\Ket{\psi^{(t)}_{j'}}$ , where $j' = \arg \sup\limits_j \left| \Braket{\phi^{(t)}_i |\psi^{(A)}_j} \right|^2$. If $v$ is much smaller than 1, it implies that discarding Fourier components has a strong effect, and completely alters the eigenstate. 
	
	\begin{figure}[ht!]
\centering
 \includegraphics[width=0.45\textwidth]{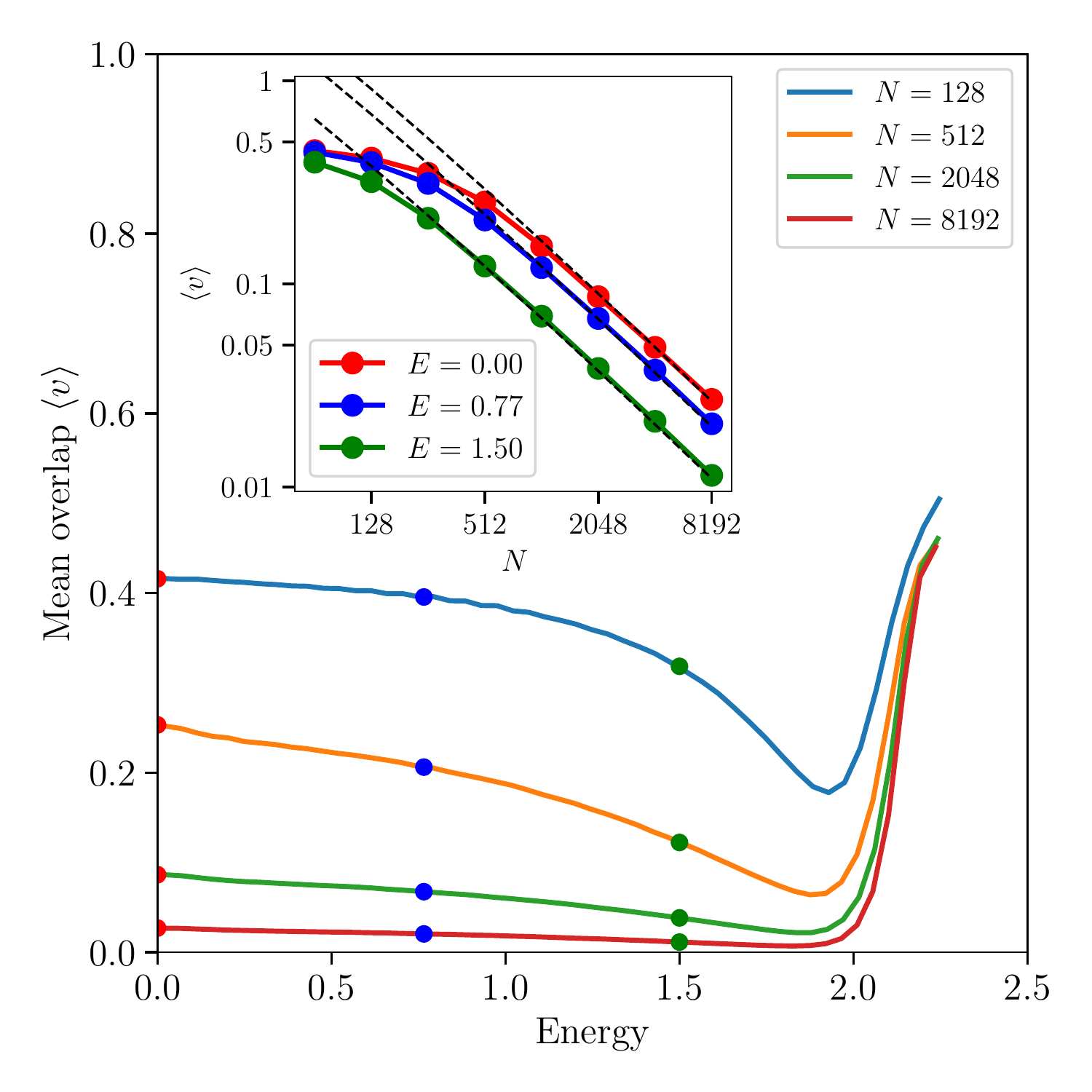} 
\caption{Energy-resolved value of the ensemble averaged overlap $\langle v \rangle$ (see text for definition) between the truncated eigenstates and the Anderson localized eigenstates for the case $F = 1$  at $W=1$ (small disorder) for four different system sizes. Here TRS is completely broken (all $k \in [-\pi, 0]$ are projected out). Red, blue and green dots identify the energies $E = 0$, $E = 0.77$ and $E = 1.5$ respectively. In the inset, we show a log-log plot of how $\langle v \rangle$ changes as a function of system size $N$. Black dashed lines indicate fits proportional to $\ln N/N$. } \label{figovTRS}
\end{figure}

	In Fig.\ \ref{figovTRS}, we plot the ensemble averaged overlap $\langle v \rangle$ for the case of full TRS breaking ($F=1$) as a function of energy for different sizes. We see that for all energies $E < 2$, the ensemble averaged overlap $\langle v \rangle$ decreases as the system size is increased, rapidly approaching zero in the thermodynamic limit. The intuition is that when TRS is completely broken, all states lose their time-reversed partners, and the eigenstates of the truncated Hamiltonian have very little overlap with those of the Anderson Hamiltonian. We limit our discussion to the main part of the band, with typical Anderson localized states \cite{EdwardsThouless1971}, or $|E| \  \leq 1.5$. The rise in $v$ at $E > 2$ may be related to the known peculiar properties \cite{JohriBhattPRL12, JohriBhattPRB12} of the tail states ($E > 2$). The  response of these states to the truncation procedure may be very different from that of typical states in the bulk.
	
	The inset of Fig.\ \ref{figovTRS} shows the scaling behaviour of $\langle v \rangle$ as a function of system size $N$ for three energies indicated. These are well fit by the functional form $\langle v \rangle \sim \ln N/N$, for all three energy windows considered. This finding is related to the problem of random projections in high-dimensional spaces as studied in applied mathematics and computer science. Consider a normalized vector $\mathbf{x} = (x_1, \cdots x_d)$ that is uniformly distributed on the surface of a $d-1$ dimensionsal hypersphere. The Johnson-Lindenstrauss lemma \cite{JohnsonLindenstrauss} provides bounds on the errors made by projecting $\mathbf{x}$ to a lower dimensional Euclidean subspace, and treats the issue from the perspective of approximations made in computer science to compress data. The issue has also been studied in the context of quantum information processing and the statistics of random quantum states. A key result \cite{Lakshminarayan2008} is that for this ensemble of random vectors, the quantity $t \equiv \max\limits_i |x_i|^2$ has an expectation value given by \begin{align}
	\mathbb{E} (t) = \frac{\ln d}{d} + \frac{\gamma}{d}+ \mathcal{O} \left(\frac{1}{d^2} \right), \end{align}
	where $\gamma = 0.5772 \cdots$ is the Euler-Mascheroni constant. Since the mean overlap $\langle v \rangle$ in our model appears to follow the same scaling behaviour, we claim that main consequence of the truncation procedure is to scramble the eigenstates completely and render them effectively random in the basis of the Anderson localized wavefunctions. We may then identify $\langle v \rangle$ as the same quantity as $\mathbb{E}(\max\limits_i |x_i|^2 )$, where the $x_i$'s are coefficients of the the eigenstates of the truncated Hamiltonian expressed in the basis of Anderson states. 
	
\begin{figure}[ht!]
\centering
 \includegraphics[width=0.45\textwidth]{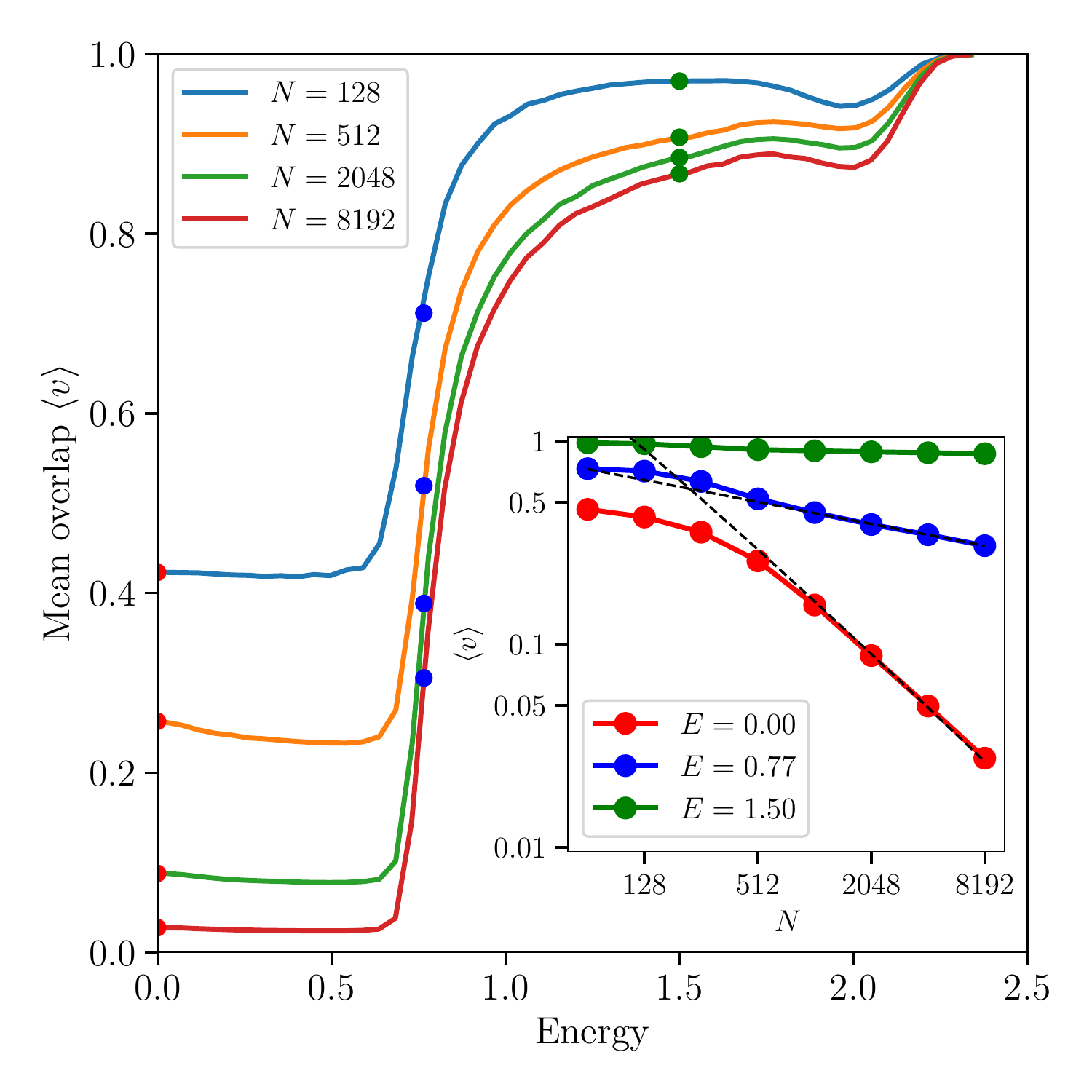} 
\caption{Energy-resolved value of the ensemble averaged overlap $\langle v \rangle$ between the truncated eigenstates and the Anderson localized eigenstates for the case $F = 1/4$ at $W=1$ (small disorder) for four different system sizes. Here of TRS is partially broken by projecting out all $k \in [-\frac{5\pi}{8}, -\frac{3 \pi}{8}]$. Red, blue and green dots identify the energies $E = 0$, $E = 0.77$ and $E = 1.5$ respectively. In the inset, we show a log-log plot of how $\langle v \rangle$ changes as a function of system size $N$. At $E = 0$ (red line), $\langle v \rangle$ is best fit by a function proportional to $\ln N / N$. At the cut-off energy $E = E_c = 0.77$ (blue line), we use a power law fit $\langle v \rangle \sim N^{-0.18}$. } \label{figov_nok_-0625_-0375}
\end{figure}	
	
	In Fig. \ref{figov_nok_-0625_-0375}, we plot the ensemble averaged overlap $\langle v \rangle$ as a function of the energy for the case $F=0.25$. We notice the effect of the cut-off energy scale $E_c$, introduced by projecting out $k \in [-\frac{5 \pi}{8}, -\frac{3 \pi}{8}]$, which seems to demarcate two different regimes of behaviour. States for which $E < E_c$ in the partially broken TRS model have a mean overlap that approaches zero as system size is increased. The scaling behaviour in this regime (see inset) appears to mirror that in the case of complete TRS breaking with  $\langle v \rangle \sim \ln N/N$.    Therefore, the extended behaviour of eigenstates in $E < E_c$ may be understood as a consequence of their complete scrambling from the Anderson-localized basis. However, states in $E > E_c$ have a much slower variation in $\langle v \rangle$ as a function of $N$, indicating that they retain a large amount of Anderson-localized character even after truncation. From our numerics, $\langle v \rangle = 0.866$ for $E = 1.50$ at $N = 8192$ sites. The fact that the truncated Hilbert space has a dimension $0.875 N$ suggests that most of the drop in overlap may be explained by the incompleteness of the basis induced by the truncation. There is a sharp transition between the two regimes around $E = E_c$, where the overlap $\langle v \rangle$ falls as a power law in the system size ($\langle v \rangle \sim N^{-\gamma}$, with $\gamma = 0.18 \pm 0.01$).
	
	These results, therefore, bolster our assertion that at \emph{low} disorder, the truncation procedure can be understood in terms of a perturbation theory starting from the non-disordered case and that the impact on eigenstates is drastically different depending on where in the spectrum they lie.
	
	In this and the preceding section, we have established that the $E = E_c$ eigenstates in the Anderson model with partial TRS breaking lie at the boundary between two kinds of behaviour -- extended-like with very little resemblance to the Anderson localized wavefunctions at the centre of the band $E < E_c$ and localized-like at energies $E>E_c$. In the next section, we examine in detail the putative critical states at $E_c$ through the lens of the multifractal spectrum and try to compare these states with known critical states in other models.
	
\section{Multifractal analysis}\label{secmfa}
	
	We first briefly recap concepts from multifractal analysis.  More comprehensive reviews may be found in \cite{Huckestein1995, EversMirlin08, Rodriguezetal11}. Critical states, such as those at a mobility edge, show structure at many length scales and the site probability densities $|\psi_m|^2$ fluctuate over several orders of magnitude in a systematic way. Such ``multifractal'' objects are characterized by two functions $f(q)$ and $\alpha(q)$ \cite{Halsey86}, commonly plotted against each other to give the ``multifractal spectrum'' $f(\alpha)$.  The value of $f$ is related to the frequency or rarity of finding a site with a given wavefunction density, i.e.\ to the probability density function of wavefunction probabilities $P(|\psi_m|^2)$. The value of $\alpha$ is related to the value of the site probability $|\psi_m|^2$ itself.

	 The multifractal spectrum is connected to the ensemble average of the generalized IPR (Eq.\ \ref{eqIPR2}) or higher moments of the probability density. The generalized IPR $P_q \equiv \sum_m |\psi_m|^{2q}$, whose ensemble average obeys the scaling relation \begin{align}
\langle P_q \rangle \sim L^{- \tau_q},  \label{eqIPRtauq}
\end{align} where $\tau_q$ is known as the mass exponent \cite{Rodriguezetal11}. The exponents $\tau_q$ are related to the multifractal spectrum \cite{EversMirlin08} by a Legendre transform \begin{align}
\alpha_q = \frac{d \tau_q}{d q} \text{\qquad and \quad}
f_q = q \alpha_q - \tau_q. \label{legendrefa}
\end{align}	Therefore the two functions $f(\alpha)$ and $\tau_q$ provide equivalent information about the critical wavefunction.
	
	A metallic (extended) wavefunction has amplitudes on all sites of the same order of magnitude. As a result, the multifractal spectrum $f(\alpha)$ reduces to a single point $(d,d)$. As one approaches the metal-insulator transition, the spectrum $f(\alpha)$ spreads out and attains a downward concave, approximately parabolic shape, with its peak at $\alpha = \alpha_0 > d$ \cite{EversMirlin08}, shifted from that of an extended state.  

\begin{figure}[ht!]
\centering
\begin{minipage}{0.04\columnwidth}
(a)
\end{minipage}
\begin{minipage}{0.93\columnwidth}
\includegraphics[trim={2cm, 0cm, 1cm, 4cm}, clip, width=0.93\columnwidth]{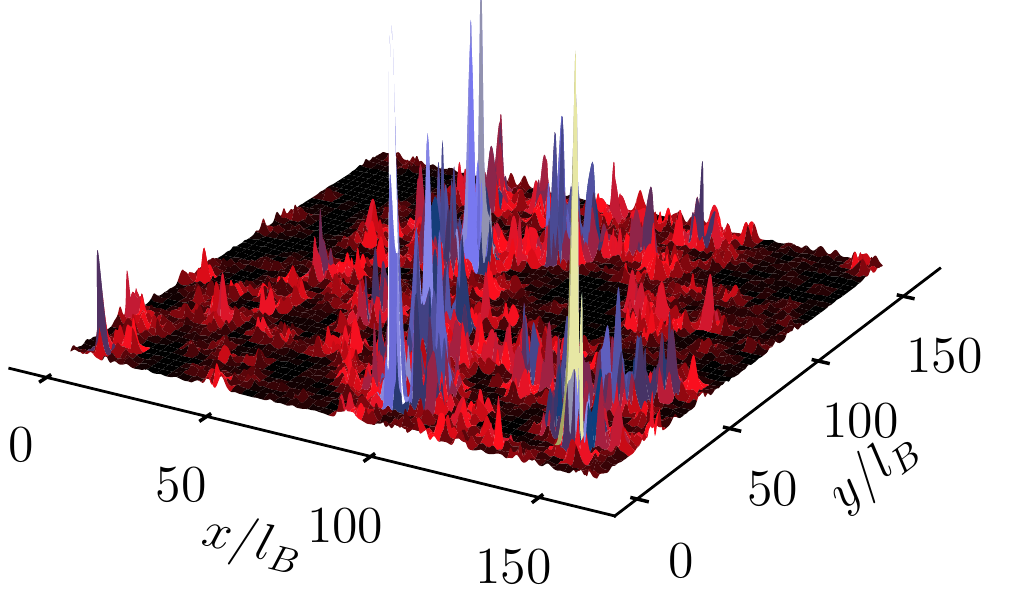}
\end{minipage}
 
 \begin{minipage}{0.04\columnwidth}
(b)
\end{minipage}
\begin{minipage}{0.93\columnwidth}
\includegraphics[width=0.91\columnwidth]{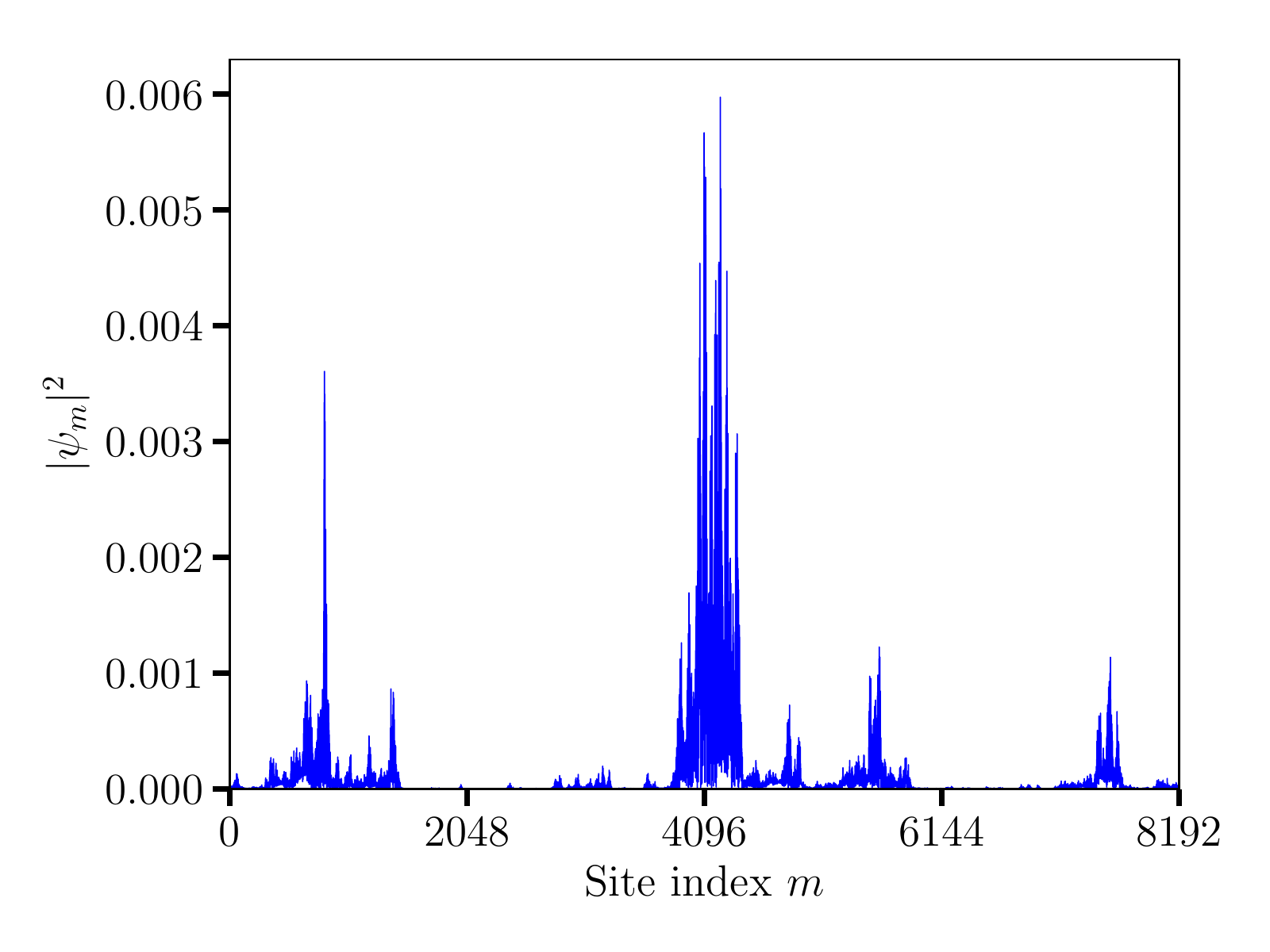} 
\end{minipage}
 
\caption{(a) A typical quantum Hall critical state. The state is generated by exactly diagonalizing the Hamiltonian of the lowest Landau level with Gaussian white noise disorder on a square torus with $N_{\phi} = 4096$ flux quanta. Each side of the torus is $160.4 \  l_B$, where $l_B$ is the magnetic length. The vertical axis is on a linear scale. (b) The wavefunction of the middle panel of Fig.\ \ref{figwfsW1}, plotted on a linear scale for comparison. } \label{figwfQH}
\end{figure}		

In Fig.\ \ref{figwfQH} (a), we show a prototypical critical state calculated by us for the case of non-interacting electrons in two dimensions subjected to a large perpendicular magnetic field (the Landau level limit). and a Gaussian random potential \cite{Huckestein1995}, small compared to the cyclotron energy, so Landau level mixing can be neglected. Such states have large fluctuations in the value of $|\psi|^2$, and appear neither localized nor extended. This will be used to compare with critical-like states in our model, e.g. Fig.\ \ref{figwfsW1}(b), shown on a linear scale in Fig.\ \ref{figwfQH}(b).
	
	For a metallic extended state in a $d$-dimensional system of linear dimension $L$, each of the $|\psi_m|^2$ is of order $L^{-d}$, so $\langle P_q \rangle_{metal} \sim L^{d-dq}$. The ``anomalous exponent'' $\Delta_q$ is defined as $\tau_q = d(q-1) + \Delta_q$, with $\Delta_0 = \Delta_1 = 0$. For a metal $\Delta_q = 0$ for all $q$. On the other hand, for an insulating state localized at site $m_0$, we have $|\psi_m|^2 = \delta_{m,m_0}$. The $q^{th}$ IPR is identically unity for positive $q$ and infinite for negative $q$. In the thermodynamic limit $L \to \infty$, therefore \begin{align*}
\tau_q &= d(q-1) \qquad \qquad \ \ \text{(metal)} \\
\tau_q &= d(q-1) + \Delta_q \qquad \text{(critical)}\\
\tau_q &= \begin{cases} 0, &q > 0\\ -d, &q = 0\\ -\infty, &q <0 \end{cases} \quad \text{(insulator)}.
\end{align*}

	In early studies, the $f(\alpha)$ spectrum was obtained numerically for the quantum Hall transition in 2-D \cite{Pook91, Huckestein92, Klesse95} as well as the Anderson metal-insulator transition in 3-D \cite{Schreiber1991, Grussbach1992}. Later studies \cite{Vasquezetal08, Vasquezetal08b, Obuse2008, Evers2008, RodriguezSlevinPRL2010, Rodriguezetal11, Rodriguezarxiv17} have focussed on calculating $\tau_q$ to high precision at the critical point to obtain critical parameters and verify the nature of the field theory at the critical point.
	
	  We use the quantity $\Delta_{\frac{1}{2}} \equiv \Delta(q = \frac{1}{2})$ \cite{Mirlin2006} to  distinguish extended, localized and ``critical'' states in our model. $\Delta_{\frac{1}{2}}$ is well-defined on both sides of the transition, with a unique value at the critical point that can be found easily by finite size scaling. If $f(\alpha)$ is exactly quadratic, the peak $\alpha_0 = d + 4 \Delta_{\frac{1}{2}}$. Table \ref{tabfavals} gives $\alpha_0$ and $\Delta_{\frac{1}{2}}$ obtained from the literature for a variety of systems.
	
\begin{table}[ht!]
\begin{ruledtabular}
\begin{tabular}{l c c c}
 System & Dim. & $\alpha_0$ & $\Delta_{\frac{1}{2}}$ \\ \hline
Metal & $d$ & $d$ & $0$\\
Insulator & $d$ & $+ \infty$ & $\frac{d}{2}$\\
Critical & $d$ & $d + \Delta'(0)$ & $[0, \frac{d}{2}]$ \\ 
-- QH \cite{Evers2008} & $2$ & $2.2596 \pm 0.0004$ & $0.0645 \pm 0.0001$  \\
-- Anderson\footnote{pure potential scattering} \cite{EversMirlin08, Vasquezetal08b} &  $3$ & $4.027 \pm 0.003$ & $0.265 \pm 0.003$ \\
\end{tabular}
\caption{Summary of multifractal signatures of metallic, insulating and critical states. In this work, we use $\Delta_{\frac{1}{2}}$ (last column) to characterize states of the truncated Anderson Hamiltonian.}\label{tabfavals}
\end{ruledtabular}
\end{table}	
	
	\begin{figure}[ht!]
\centering
 \includegraphics[width=1.0\columnwidth]{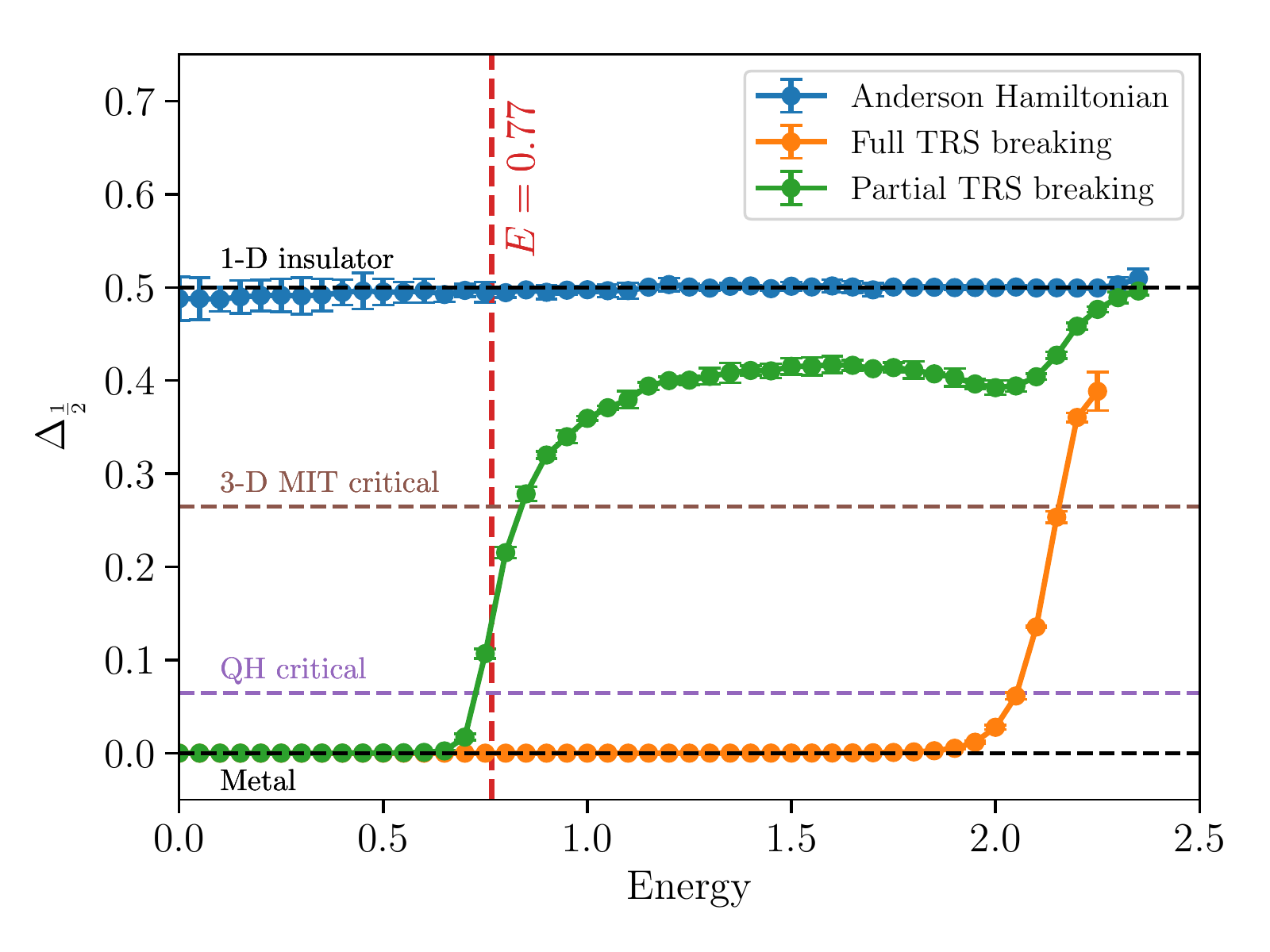}
\caption{Energy resolved value of the anomalous multifractal exponent $\Delta_{\frac{1}{2}}$ for the three Hamiltonians considered in this paper at small disorder ($W = 1$). For reference, the known values of $\Delta_{\frac{1}{2}}$ for an extended metallic wavefunction, localized insulating wavefunction, Quantum Hall critical state (purple) and a critical state at the 3-D Anderson metal-insulator transition (brown) from table \ref{tabfavals} are also shown.} \label{figtauW1}
\end{figure}		
	
In Fig.\ \ref{figtauW1}, we show the energy-resolved values of $\Delta_{\frac{1}{2}}$ for the three Hamiltonians considered at small disorder ($W = 1$). We first compute the ensemble averaged IPR $\langle P_{0.5} \rangle$ for all sizes $N$, and then extract $\tau_{\frac{1}{2}}$ from the slope of a linear regression between $\ln \langle P_{0.5} \rangle$ and $\ln N$ for the largest four sizes in our study. The anomalous exponent is obtained from the relation $\Delta_{\frac{1}{2}} = \tau_{\frac{1}{2}} + 0.5$. We notice that the Anderson Hamiltonian has an anomalous multifractal exponent of $0.5$ for all energies as expected for an insulator in one dimension. The Hamiltonian with $F=1$ has an anomalous exponent of zero in the bulk of the spectrum, consistent with our claim that its eigenstates are extended. For partial Hilbert space truncation ($F = 1/4$), we notice that states at $E < E_c$ have $\Delta_{\frac{1}{2}} = 0$, as in the metallic case. As we cross the cut-off energy $E_c$, the anomalous exponent increases and settles to a value of $0.41 \pm 0.03$, which is higher than that for known critical states in the Quantum Hall transition and Anderson metal-insulator transition in 3-D, but lower than that of a standard Anderson insulator. We believe that these states are essentially localized, with some critical-like character arising from the incompleteness of the Hilbert space.

\section{Eigenstate currents and Eigenvalue spacings}

In this section we investigate two other metrics that are expected to show disparate behaviour in the localized and extended regimes. First, we examine the current carried by the eigenstates in the Anderson and truncated models. Second, we look at the eigenvalues themselves, as opposed to the eigenstates, to see if they have any information that can help distinguish the two phases.

The current $\v{J}$ of a state $\Ket{\psi}$ is given by $\Braket{\psi | \v{J} | \psi}$. It is related to the momentum $\v{p}$ by $\v{J} = \frac{ne}{m}\v{p} $ where the momentum operator $p = -i\hbar \frac{d}{dx}$ in one dimension. Setting $n$ (particle density), $e$ (charge), and $m$ (mass) all equal to unity, and replacing the derivative by a difference for our discrete case, we obtain \begin{align}
	J_\psi &= -\frac{i}{2} \sum\limits_{m=1}^{N} \psi_m^\ast(\psi_{m+1} - \psi_{m-1})\\
	&= \text{Im} \sum\limits_{m=1}^{N} \psi_m^\ast \psi_{m+1} \label{eqJ}.
	\end{align} Anderson localized wavefunctions are time-reversal symmetric, and therefore have real coefficients. It follows from Eq.\ \ref{eqJ} that all Anderson localized states carry zero current. On the other hand, a momentum space eigenstate $\Ket{k_r}$ is also an eigenstate of the current operator, and carries current \begin{align} J_{k_r} = \sin (\frac{2 \pi r}{N}) = \sqrt{1 - \frac{E_{k_r}^2}{4}}. \end{align}
		
\begin{figure}[ht!]
\centering
\includegraphics[width=1.0\columnwidth]{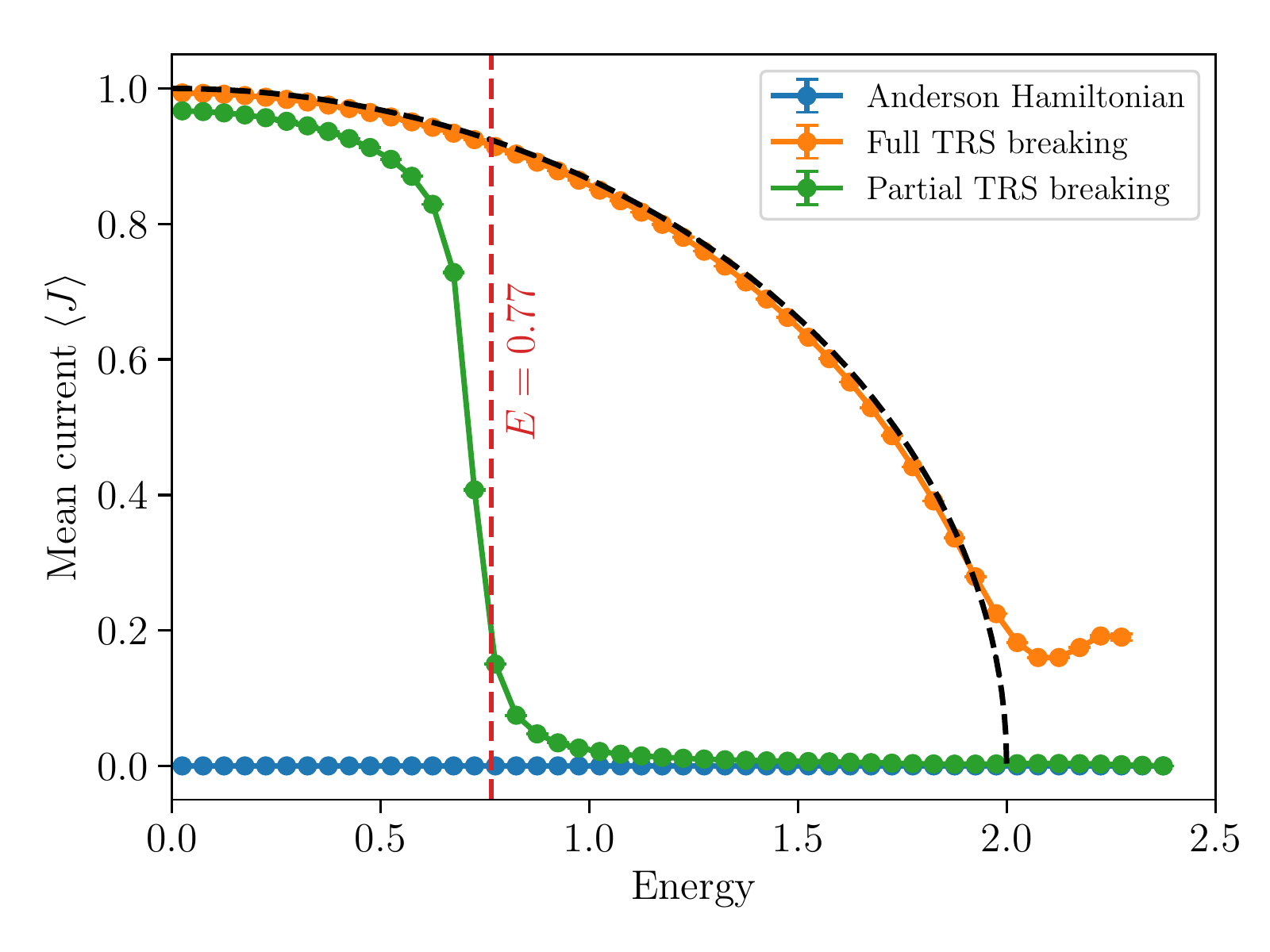}
\caption{Mean current as a function of energy for the three different kinds of Hamiltonians at $N = 8192$ sites in the small disorder regime ($W = 1$). The black dashed curve shows the current for the zero-disorder case $J(E) = \sqrt{1 - \frac{E^2}{4}}$.}\label{figJ}
\end{figure}
	
	The ensemble-averaged current over all states, calculated from Eq.\ \ref{eqJ}, is plotted in an energy-resolved manner in figure \ref{figJ}. Anderson-localized states carry no current, as expected. On the other hand, when TRS is broken completely ($F=1$), all states are expected to carry non-zero current. What is surprising to note, however, is that the current carried by these states is almost exactly the same as that by zero-disorder states at the same energy. 
	
	The truncated Hamiltonian with $F = 1/4$ (partial TRS breaking), with $k \in [-\frac{5\pi}{8}, -\frac{3\pi}{8}]$ projected out seems to interpolate between the two kinds of behaviour. At low energies ($E < E_c$), all eigenstates are strongly chiral, as indicated by the amount of current carried. There is a sharp downturn in the current at $E = E_c$, above which states do not carry current.
	
	So far, we have examined individual eigenstates of the Hamiltonian using various measures of localization. The robustness of TRS-broken eigenstates to disorder has yet another significant consequence. which is revealed when we study the distribution of eigenvalue spacings in the spectrum. In usual disorder models, the probability distribution of eigenvalue spacings is also known to hold information about the nature of eigenstates. Define the $n$\textsuperscript{th} eigenvalue spacing $\Delta E_n \equiv E_{n+1} - E_n$, where $\{E_i\}$ are the eigenvalues sorted in ascending order. One may calculate the probability density $P(s)$ of the scaled eigenvalue spacings $s$, where $s = \frac{\Delta E}{\langle \Delta E \rangle}$, where $\langle \Delta E \rangle$ is the ensemble averaged spacing at that energy. 
	
\begin{table}[ht!]
\begin{ruledtabular}
\begin{tabular}{c c c cc }
 Ensemble & Properties & Symmetry & $\langle r \rangle$ \cite{Atas2013} & $P(s \to 0)$\\ \hline
Poisson & Localized & TRS & $0.3862$ & $e^{-s}$\\
GOE & Delocalized & TRS & $0.5307 \pm 0.0006$ & $s$\\
GUE & Delocalized & No TRS & $0.5996 \pm 0.0006$ & $s^2$\\
\end{tabular}
\caption{Summary of random-matrix eigenvalue statistics}\label{tabRMT}
\end{ruledtabular}
\end{table}	
	
	If the spectrum is localized, then the eigenvalues are uncorrelated, and $P(s)$ shows Poisson statistics with no level repulsion. In the delocalized phase, however, the eigenvalue statistics show characteristics of the standard random-matrix ensembles -- namely the Gaussian Orthogonal Ensemble (GOE) for spinless systems with TRS, and Gaussian Unitary Ensemble (GUE) for spinless systems with no TRS. A key feature of these ensembles is level repulsion, i.e, $P(s = 0) = 0$, and $P(s) \sim s^{\beta} $ for small $s$, where the exponent $\beta$ depends on the universality class of the system \cite{BeenakkerRMP97, HaakeBook}. A convenient single parameter characterizing these distributions is the level spacing ratio \cite{OganesyanHuse07}, denoted by $r$. This quantity has been found to work well in the case of data that is limited to small sizes. In terms of the eigenvalue spacing,
	 \begin{align} r_n \equiv \frac{\min(\Delta E_n, \Delta E_{n-1})}{ \max (\Delta E_n, \Delta E_{n-1} )}. \end{align} 
	 The energy resolved ensemble-averaged mean $r$ value gives clear signatures of the localization information of the underlying phase and has been used effectively in recent numerical studies of disordered many-body systems \cite{PalHuse2010, Cuevas2012, Johri2015, Luitz2015, Geraedts2017}. The properties of eigenvalue statistics are summarized in Table \ref{tabRMT}.
	
\begin{figure}[ht!]
\centering
\includegraphics[width=1.0\columnwidth]{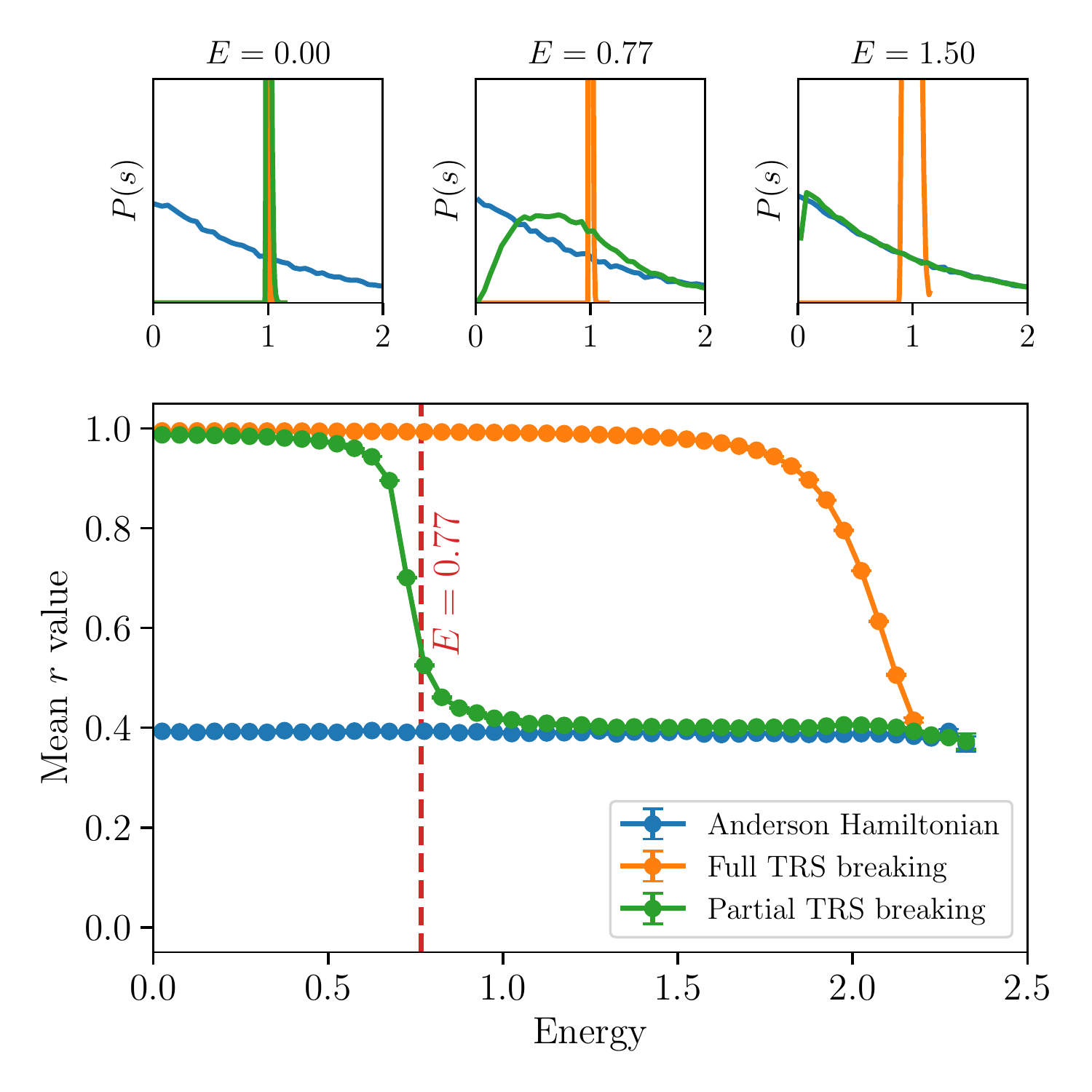}
\caption{Mean eigenvalue spacing ratio $\langle r \rangle$ as a function of energy for the three different kinds of Hamiltonians at $N = 8192$ sites at small disorder ($W = 1$). In the upper panels, we plot the distributions $P(s)$ of scaled eigenvalue spacings.}\label{figevF1}
\end{figure}	
	
	In Fig.\ \ref{figevF1}, we plot the mean $r$ value as a function of energy for the Anderson model and the two truncated Hamiltonians considered in this paper. We first note that the entire spectrum of the Anderson model has an $\langle r \rangle $ value of $0.389 \pm 0.004$ consistent with the Poisson statistics $P(s) \sim e^{-s}$ of localized spectra as shown in the upper panels. The Hamiltonian with $F=1$ shows an $ \langle r \rangle = 0.995 \pm 0.002$ over a large range of its spectrum. One might have na\"ively expected this case to have $\langle r \rangle \sim 0.53 - 0.60$, as in table \ref{tabRMT}, since the states are all delocalized and the Hamiltonian is random. However, not only is $\langle r \rangle \approx 1$, but the distribution of the scaled eigenvalue spacings $s$ is extremely narrow, sharply peaked around $s = 1$. For the tight-binding model with no disorder, where eigenvalues $E_{k_r} = -2 \cos(k_r)$ are regularly spaced, with $P(s) = \delta(s-1)$ and $r = 1$. In fact the $\langle r \rangle$ value is a useful metric only when applied to eigenstates in a given symmetry sector. In the translationally invariant case, since $k$ is good quantum number, we effectively have a block diagonal matrix consisting of $1 \times 1$ blocks, and it makes no sense to talk of an $r$ value for these blocks. The fact that $\langle r \rangle$ is so close to 1 for the case of complete TRS breaking suggests that weak disorder is not relevant when negative $k$ states are projected out. 
	
	The most interesting case is when we project out momentum states $k \in [-\frac{5 \pi}{8}, -\frac{3 \pi}{8}]$ as in the case of partial TRS breaking. Here states at energies $E < E_c$, have $\langle r \rangle \to 1$ as in the case of complete TRS breaking, with $P(s)$ sharply peaked around $s=1$. States in the region $E > E_c$ have $\langle r \rangle  = 0.400 \pm 0.008$, suggesting Poissonian statistics. There is a sharp transition between the two kinds of behaviour at $E = E_c$. This result further corroborates the fact that the case of partial TRS breaking has a spectrum with two very different kinds of states, separated by an energy scale $E_c$. 

\section{The large disorder regime}

In this paper, we have focussed most of analysis of the truncated Anderson model on the small disorder ($W = 1$) case. In this section we briefly address the case of large disorder. The terms small and large disorder are with reference to the bandwidth ($B=4$) of the tight-binding Hamiltonian in 1-D. 

As we increase the disorder in the standard Anderson model, the localization length of all eigenstates drops dramatically. At $W = 4$, when the disorder bandwidth is equal to the tight-binding bandwidth, all localization lengths $\xi_{IPR}$ and $\xi_{M_2}$ are less than 10. At $W = 16 \gg B$, all eigenstates are localized primarily on one or two sites. In this regime, wavefunctions are not localized in momentum space. As a result, the effect of truncating the Hilbert space by projecting out a set of momentum states is very different from that for the small disorder Hamiltonian. We cannot start from Bloch wavefunctions and argue that the mixing between different $\Ket{k}$ modes is well-controlled. In fact, since disorder is the dominant energy scale, the system does not care about the cut-off energy $E_c$ anymore. All eigenstates in both the $F = 1/4$ and $F = 1$ cases have traces of localized character, with large amplitude fluctuations arising from the truncation procedure. 

\begin{figure}[ht!]
\centering
 \includegraphics[width=1.0\columnwidth]{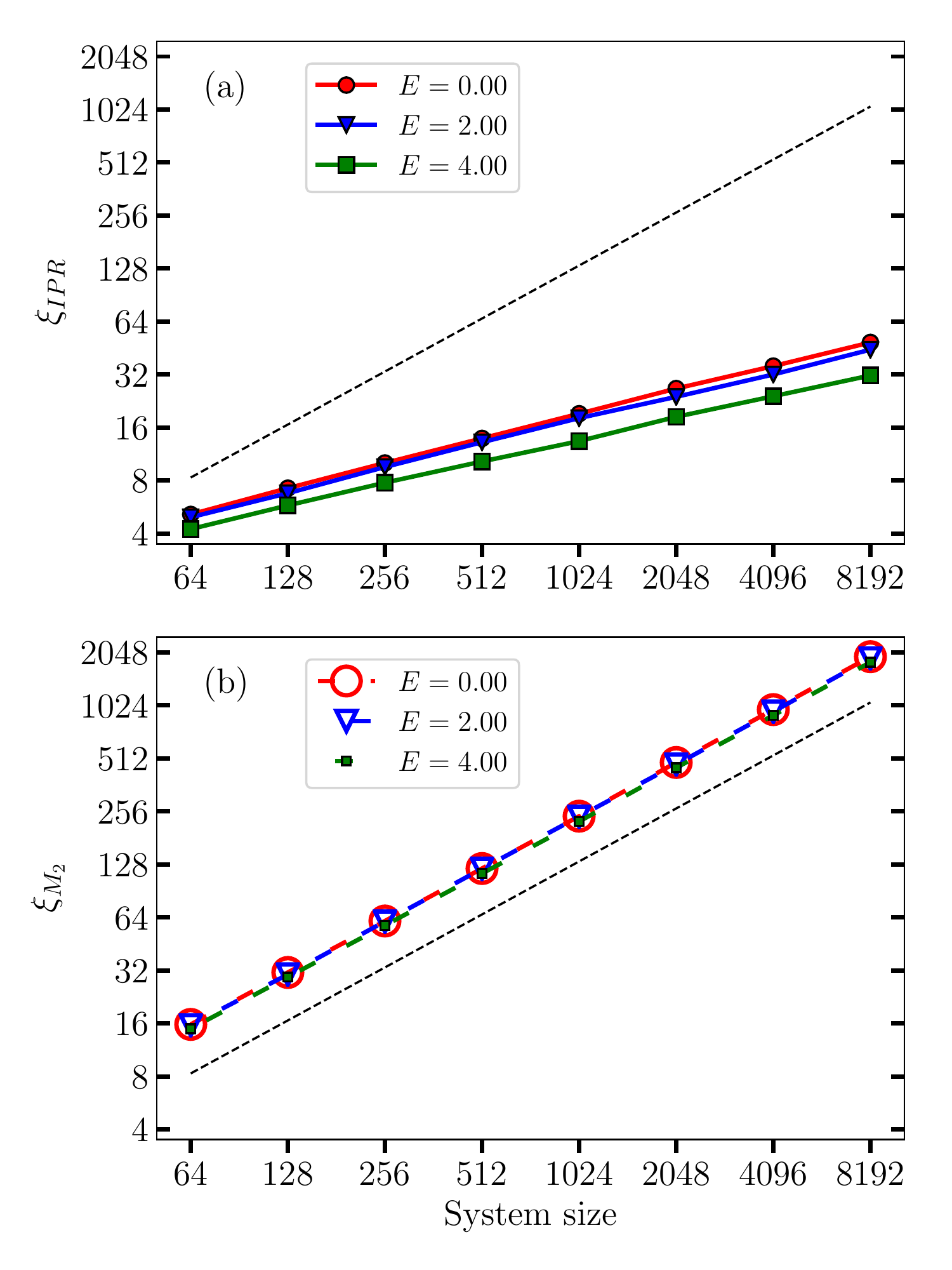} 
\caption{Log-log plot of the IPR localization length $\xi_{IPR}$ (a) and the second moment based localization length $\xi_{M_2}$ (b) as a function of system size $N$ in the Anderson model with partial TRS breaking ($F = 1/4$) at large disorder ($W = 16$). All $k$ states in $[-\frac{5\pi}{8}, -\frac{3 \pi}{8}]$ are projected out. The dotted black line is the same in both panels and has slope 1. The error bars are not visible on this scale.} \label{figxilargeW}
\end{figure}

  In Fig.\ \ref{figxilargeW}, we plot the localization lengths as a function of system size for the case of partial TRS breaking ($F = 1/4$). A larger disorder strength leads to a larger bandwidth, so we choose a different set of representative energies ($E = 0, 2 \text{ and } 4$) than for the small disorder case. We do not see any energy-resolved differences in the scaling of localization lengths $\xi_{IPR}$ and $\xi_{M_2}$ as a function of system size, in contrast to the small disorder behaviour seen in Fig.\ \ref{figxiTRS025}. $\xi_{M_2}$ scales linearly with system size (b), but $\xi_{IPR}$ (a) scales with some sub-linear power law  ($\xi_{IPR} \sim N^{0.44 \pm 0.03}$), characteristic of critical states. The power law does not depend strongly on the energy. A very similar plot may be obtained for the case of complete TRS breaking ($F = 1$). Even in this case, we do \emph{not} see any pure extended states such as those seen in the small disorder case in $E < E_c$.

\begin{figure}[ht!]
\centering
 \includegraphics[width=1.0\columnwidth]{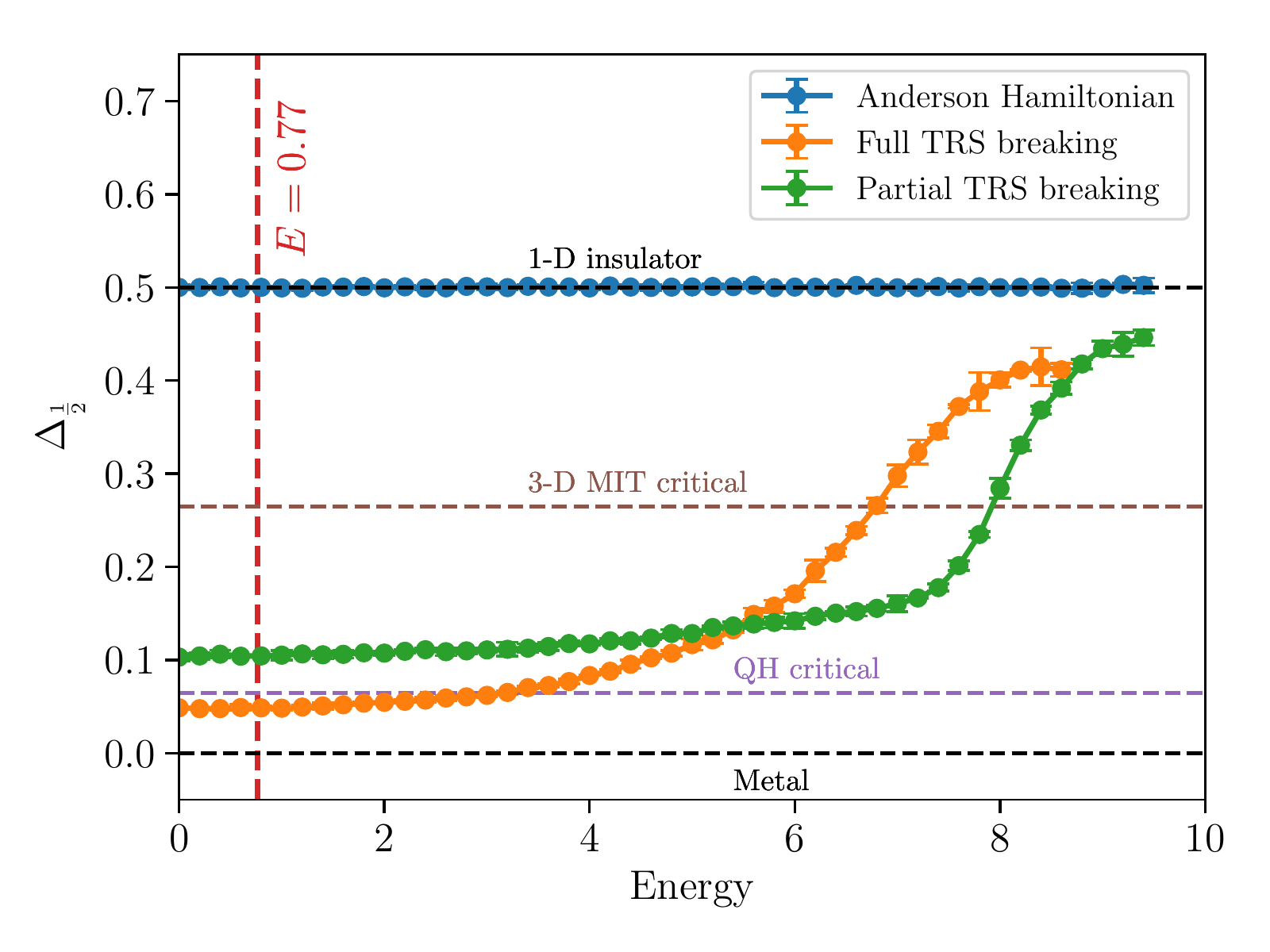}
\caption{Energy resolved value of the anomalous multifractal exponent $\Delta_{\frac{1}{2}}$ for the three Hamiltonians considered in this paper at large disorder ($W = 16$). For reference, the known values of $\Delta_{\frac{1}{2}}$ for an extended metallic wavefunction, localized insulating wavefunction, Quantum Hall critical state (purple) and a critical state at the 3-D Anderson metal-insulator transition (brown) from table \ref{tabfavals} are also shown.} \label{figtauW16}
\end{figure}

The energy resolved mean overlap $\langle v \rangle$ (plotted in figs.\ \ref{figovTRS} and \ref{figov_nok_-0625_-0375} for small disorder) seems to have a power-law scaling with system size $N$ for \emph{all} energies when disorder is large. We recall that in the small disorder case, only states in the neighbourhood of $E = E_c$ have a power law scaling. In Fig. \ref{figtauW16}, we plot the anomalous multifractal exponent $\Delta_{\frac{1}{2}}$ for the three Hamiltonians under discussion at large disorder ($W = 16$). We notice that the standard Anderson Hamiltonian has $\Delta_{\frac{1}{2}} = 0.5$, as expected for an insulator. However, both $F = 1/4$ and $F = 1$ have a critical-like $\Delta_{\frac{1}{2}}$ throughout the spectrum, moving monotonically from more metal-like at the centre of the band, to more insulator-like at the edges. In the case of partial TRS breaking, (green curve in Fig.\ \ref{figtauW16}), the effect of the cut-off energy $E_c$ is completely washed away for $F = 1/4$ as discussed above. In addition, the multifractal spectrum $f(\alpha)$, as a whole, attains a non-trivial shape, for both the case of full and partial TRS breaking, consistent with them being critical-like states. Details of this behaviour are displayed and discussed in Appendix A.

The energy-resolved current too mirrors the story above, with a relatively constant non-zero value across all energies. The projection procedure induces a total non-zero current in the low-energy subspace, which is spread nearly equally among all states. 

\begin{figure}[ht!]
\centering
\includegraphics[width=1.0\columnwidth]{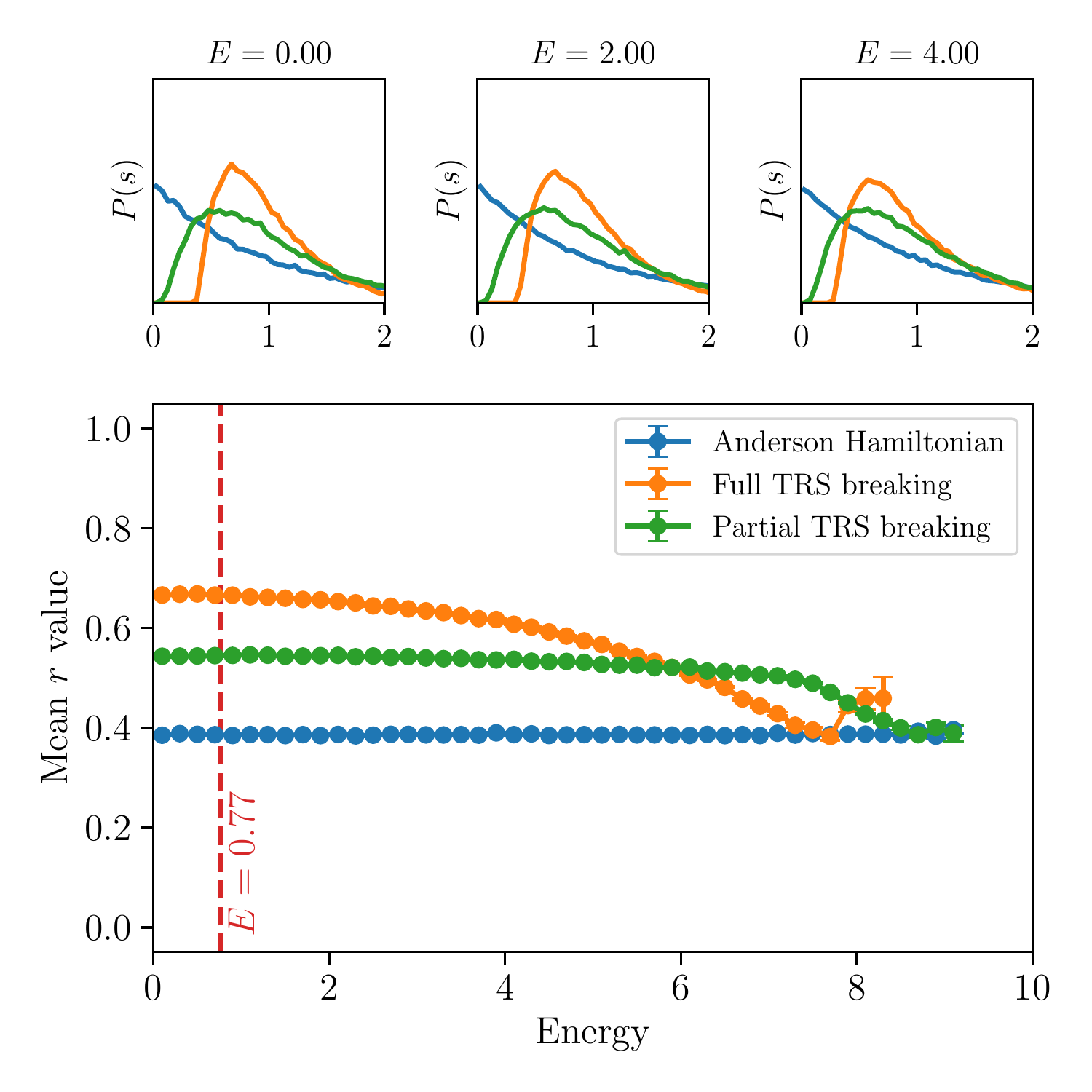}
\caption{Mean eigenvalue spacing ratio $\langle r \rangle$ as a function of energy for the three different kinds of Hamiltonians at $N = 8192$ sites at large disorder ($W = 16$). In the upper panels, we plot the distributions $P(s)$ of scaled eigenvalue spacings.}\label{figevF16}
\end{figure}	

 In Fig.\ \ref{figevF16}, we plot the $\langle r \rangle$ value for the case of large disorder ($W = 16$). For the case of full TRS breaking, $\langle r \rangle$ gradually moves from $\approx 0.65$ at the centre to $\approx 0.4$ at the tails. However eigenvalues still show a tendency to repel as seen by the distribution $P(s)$ of scaled eigenvalue spacings (upper panels). While $P(s)$ is no longer as sharp as a delta function, $P(s) = 0$ for $s < 0.3$, which is very unlike typical behaviour of random matrix ensembles.  For the case of partial TRS breaking ($F = 1/4$), $r \approx 0.6$ over the entire spectrum, suggestive of GUE statistics. This again demonstrates that \emph{large} disorder is required to change eigenvalue statistics from the uniform case.

\section{Discussion and conclusions}

In this paper, we have introduced a new theoretical tool: breaking time-reversal symmetry through selective projection of momentum sectors in the Hilbert space. This method has allowed us to explore the effect of breaking time-reversal symmetry in a controlled manner in the single-particle problem of nearest neighbour hopping on a 1-D lattice in the presence of diagonal disorder (the Anderson localization problem). The interplay of disorder strength and the ``extent'' of time-reversal breaking leads to the possibility of a localization-delocalization transition. 

We first focus on the weak (or small) disorder limit (disorder width $W \ll$ the clean bandwidth $B$), represented in our study by $W/B = 1/4$, where we compute eigenvalues and eigenfunctions (in real space) through numerical exact diagonalization. We use that to study ensemble-averaged eigenvalue spectra and spacing as well as various ensemble-averaged wavefunction characteristics -- inverse participation ratio, second moment, current carrying character, and multifractality, through established measures in the literature, as well as related quantities we find informative.

Without projection (i.e. for the full Hilbert space), we obtain the canonical results for the 1D Anderson model -- states are localized for all energies, as evidenced by the saturation of the localization length as the system size is increased, defined either through the inverse participation ratio or second moment. The longest localization length occurs at zero energy (band center), and decreases monotonically as one moves away from the band center (see, however, \cite{JohriBhattPRL12} for non-monotonic behavior towards the band edge). The wavefunctions in real space assume an exponential form asymptotically. The multifractal function $f(\alpha)$ (see Appendix A for details) for finite sizes is found to become wider and wider with increasing sample size, and the data becomes concentrated at the two ends, consistent with the thermodynamic limit of two points -- at the origin and at infinity.

At the opposite end of the model, we studied the case where TRS is broken completely by projecting out all negative $k$ states, we obtain a spectrum whose wavefunctions have many characteristics of extended states. The localization lengths defined through the inverse participation ratio $\xi_{IPR}$ and second moment $\xi_{M_2}$ scale linearly with system size. They are current carrying and have an anomalous multifractal exponent of zero, typical of extended states. However these states have very rigid eigenvalue statistics, implying that weak disorder is irrelevant in this situation.

Finally, when we project out a \emph{subset} of states $k \in [-\frac{5 \pi}{8}, - \frac{3 \pi}{8}]$, we break TRS partially. In this case, the behaviour changes depending on whether the states have energies in the regime where TRS is broken ($|E| < E_c$) or not ($|E| > E_c$). States at $E < E_c$ are similar to those obtained for the $F = 1$ case. States at $E > E_c$ are Anderson localized with some sinusoidal-like background arising from the incompleteness of the Hilbert space,  i.e.\ as sort of localized with imperfect fidelity. These states can be understood through a perturbative approach, with a large overlap with their fully Anderson localized counterpart. The overlap is similar in spirit to the fidelity defined for the many-body case in \cite{Geraedts17}. They are not current carrying and have Poissonian eigenvalue statistics, like typical localized states. However their multifractal spectrum is similar to critical states (see Appendix A for details, in particular Fig.\ \ref{faW1}). This is also evident from a plot of the distribution of $|\psi|^2$ on a logarithmic scale (Fig.\ \ref{figmodpsiF025}) and comparing it to the corresponding quantity for states at the center of the Landau level (Fig.\ \ref{figmodpsiQH}). This suggests that while the situation is somewhat more complex than a simple extended-to-localized transtion, our simplistic scenario described in the introduction is behind much of the phenomena seen.  

Following a thorough analysis of the weak disorder limit, we considered the opposite case of large or strong disorder (disorder width $W \gg$ clean bandwidth $B$), represented in our case by $W/B = 4$. With $W$ being the largest energy in the problem, results are not strongly dependent on the energy of the eigenstate. Here we see for both full TRS breaking and partial TRS breaking, that the localization length (as measured by the IPR) scales \emph{sublinearly} with the size, similar to a critical state. This is further confirmed by analyzing the multifractal spectrum $f(\alpha)$ for eigenstates. As shown in Fig.\ \ref{faW16} of Appendix A, the $f(\alpha)$ curves peak at a non-trivial value of $\alpha = \alpha_0 = 1.43 \pm 0.01$, and have a roughly parabolic shape. The size dependence is immeasurably weak on the left side ($\alpha < \alpha_0$) and only weakly (logarithmically) dependent with size \footnote{Such logarithmic dependence has been suggested for the multifractality spectrum $f(\alpha)$ at the integer quantum Hall plateau transition by R.\ Bondesan, D.\ Wieczorek and M.\ R.\ Zirnbauer, Nuclear Physics B \textbf{918}, 52 (2017).} on the right side of the curve ($\alpha > \alpha_0$). Barring this small variation, it seems as if the whole spectrum has become critical in this large or strong disorder limit.
   
	Another difference with the weak disorder is evident when one looks at the distribution of eigenvalue splittings. Unlike the weak disorder case, the splittings are no longer rigid (i.e., the distribution is no longer a weakly broadened delta function). On the other hand, neither is it Poissonian, as in the case of the localized phase of the Anderson model (i.e., $F = 0$). For $F = 1/4$, as shown in Fig. \ref{figevF16}, we see a distinct hole at zero splitting, indicative of level repulsion. For $F = 1$, there appears a distinct gap around zero splitting, which is not what one might expect for a critical phase. This may be because a larger disorder is needed to see the gap close for the full TRS breaking case.

	The observation of critical like states (at least in several aspects, such as $f(\alpha)$) suggests looking into the actual form in real space of the truncated Hamiltonian \footnote{We are indebted to Kartiek Agarwal for suggesting this.}. This is done in detail in Appendix B. As can be seen there, truncation of the Hilbert space leads to a power law dependent hopping term $\sim r^{-x}$ coming from both the original hopping part of the Hamiltonian as well as the on-site disorder part. Further, the power law exponent of the effective hopping $x = 1$. This is precisely the value that separates localized and extended regime for power law hopping models \cite{ZhouBhatt} and power-law banded matrices \cite{MirlinPRBM1996, MirlinPRBM2000}. While our model does not have exactly the same form, it has both randomness and $1/r$ power-law hopping, and so may be expected to show critical behavior. Further investigation of the correspondence between our approach and more conventional approaches using power-law hopping appears warranted.

While the technique of Hilbert space truncation as presented in this work is a purely theoretical construct, it would be interesting to investigate if any of the phenomena seen in this numerical study could be seen in experiment, e.g., optical studies of random media  \cite{Segev07a, Segev07b, Segev13, Roati08, Billy08} using filters that provide partial as opposed to total information. Should this be feasible, it would provide an experimental handle on delocalization in a one dimensional system. Further, it would potentially allow such models to be useful in analyzing experimental data where perfect information is rarely attainable.

\section{Acknowledgments}

This research was supported by Department of Energy, Office of Basic Energy Sciences through Grant No. DE-SC0002140. We acknowledge helpful discussions with Matteo Ippoliti, Alberto Rodriguez and especially Kartiek Agarwal. RNB thanks the Aspen Center for Physics for hospitality during the initial stages of writing this manuscript.


\appendix

\section{Multifractal $f(\alpha)$ spectra for the eigenstates of the truncated Anderson Hamiltonians}

In the main text, we showed the multifractal behaviour through the anomalous exponent $\Delta_{\frac{1}{2}}$ as a proxy for the full multifractal spectrum $f(\alpha)$. $\Delta_{\frac{1}{2}}$, unlike $f(\alpha)$, allowed us to encapsulate the metal-critical-insulating behaviour of the entire spectrum of all the three Hamiltonians in one snapshot. Also finite size issues were easier to deal with for $\Delta_{\frac{1}{2}}$ than for $f(\alpha)$  as described below. Here, for completeness, we provide the $f(\alpha)$ curves for some representative cases.

In section \ref{secmfa}, we described the process of computing multifractal exponents $f(\alpha)$ by first calculating the ensemble averaged IPRs $\langle P_q \rangle$. We may use the relation between the IPRs and the multifractal exponents (Eq.\ \ref{eqIPRtauq}) and then perform a  Legendre transformation (Eq.\ \ref{legendrefa}) to obtain $f(\alpha)$. However this procedure of calculating the exponents first and then doing a numerical differentiation tends to be inefficient and increase the error bars, so we use the method of Chhabra and Jensen \cite{ChhabraJensen89} (see also \cite{Huckestein1995}) to calculate $f(\alpha)$ from the eigenfunctions directly as described below. For a wavefunction $\Ket{\psi}$ with real space probability amplitudes $\psi_m = \Braket{x_m | \psi}$, \begin{align}
\mu_m^{(q)} &\equiv \frac{|\psi_m|^{2q}}{\sum\limits_{m=1}^N |\psi_m|^{2q}},\\
f_q &= \frac{\sum\limits_m \mu_m^{(q)} \ln \mu_m^{(q)}}{- \ln N},\\
\alpha_q &= \frac{\sum\limits_m \mu_m^{(q)} \ln |\psi_m|^2}{- \ln N}.
\end{align}

As discussed in other works \cite{Rodriguezetal11, Varga2015}, error bars in $f(\alpha)$ tend to increase rapidly for both small and large $q$. We therefore  limit our calculations in this appendix to the region $|q| < 1$ and plot ensemble averaged values of $\langle f_q \rangle$ as a function of $\langle \alpha_q \rangle$.

\begin{figure}[ht!]
\centering
 \includegraphics[width=1.0\columnwidth]{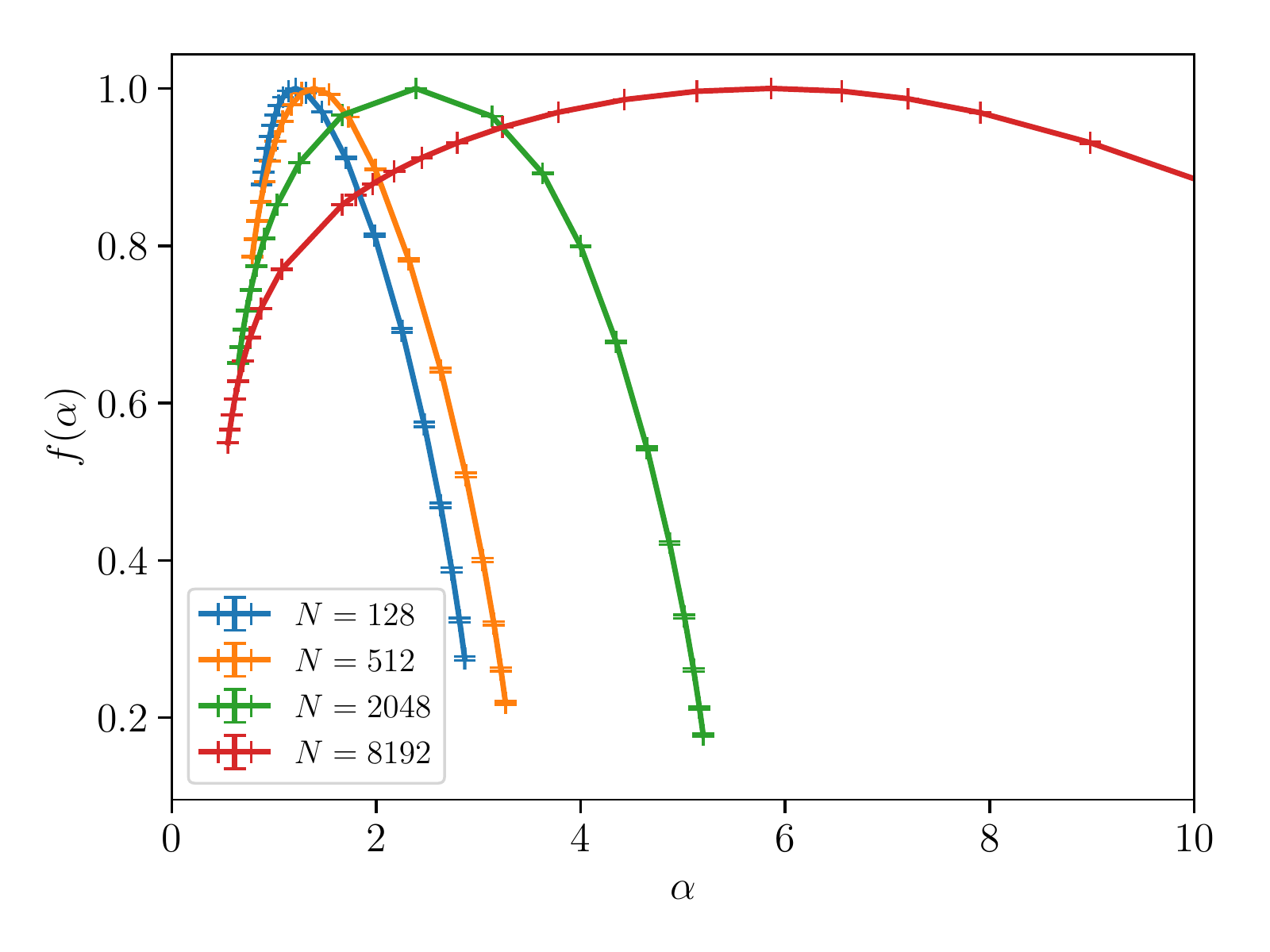}
\caption{Multifractal spectrum $f(\alpha)$ for wavefunctions in the Anderson model at small disorder ($W=1$) at fixed energy $E = 0.77$. The strong finite-size effects and lack of universality are characteristic of exponentially localized wavefunctions as discussed in the text.} \label{faAnd}
\end{figure}

A key feature of the $f(\alpha)$ spectrum is that for a true multifractal, such as a Quantum Hall critical state, it is independent of the system size. A metallic system in 1D has $f(\alpha)$ reduced to a single point $(1,1)$, while on the other hand an insulating wavefunctions have $f(\alpha)$ consisting of two disconnected points -- $(0,0)$ and $(+\infty, 0)$. However, in numerical simulations on finite systems, the two points will be seen to be connected, with the location of the peak $\alpha_0$ moving to $+\infty$ as the system size is increased. To understand this finite size effect, consider a pure exponential wavefunction $\psi_m = \mathcal{C} \exp \left(-\frac{|m|}{\xi} \right)$ on $N$ sites with localization length $\xi$. The normalization constant $\mathcal{C}$ ensures that $\sum\limits_N |\psi_m|^2 = 1$. For this wavefunction, the multifractal spectrum $f(\alpha)$ is a non-trivial curve with its peak at \begin{align}
\alpha_0 &= -\frac{1}{\ln N} \ln \left( \frac{\tanh \frac{1}{\xi}}{2 \sinh \frac{N}{2 \xi}} \right)\\
&= \frac{1}{\ln N} \left[ \left( \frac{N}{2\xi} + O(e^{-N/\xi}) \right) + \left( \ln \xi + O \left( \frac{1}{\xi^2} \right) \right) \right]. \label{faexploc}
\end{align}

\begin{figure}[ht!]
\centering
 \includegraphics[width=1.0\columnwidth]{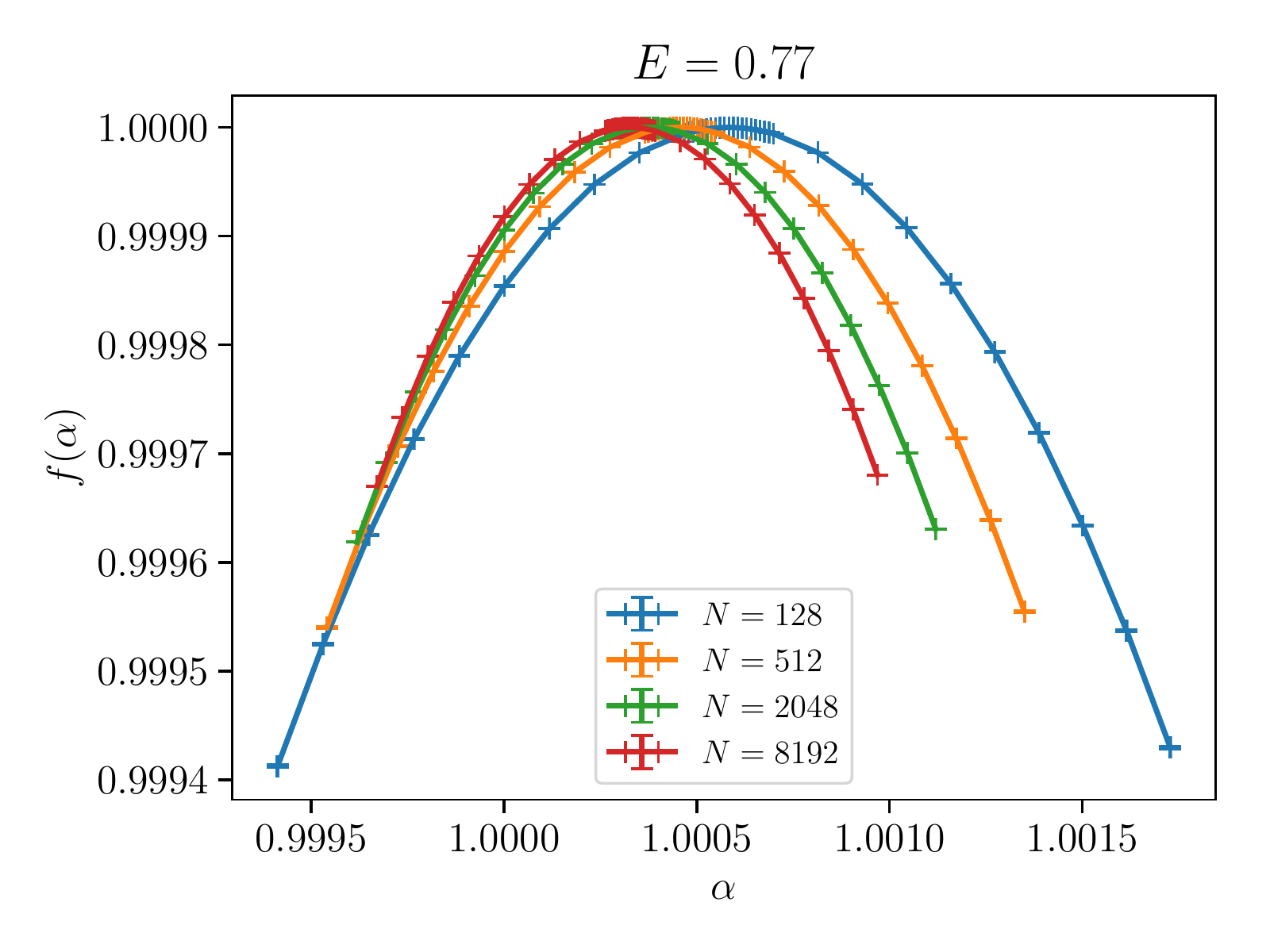}
\caption{Multifractal spectrum $f(\alpha)$ for wavefunctions in the case of complete TRS breaking ($F =1$, with $k \in [-\pi, 0]$ removed) at small disorder ($W=1$). The energy is fixed at $E = 0.77$. Note the scale is different than that for Fig.\ \ref{faAnd}.} \label{faTRS}
\end{figure}

\begin{figure}[ht!]
\centering
 \includegraphics[width=0.96\columnwidth]{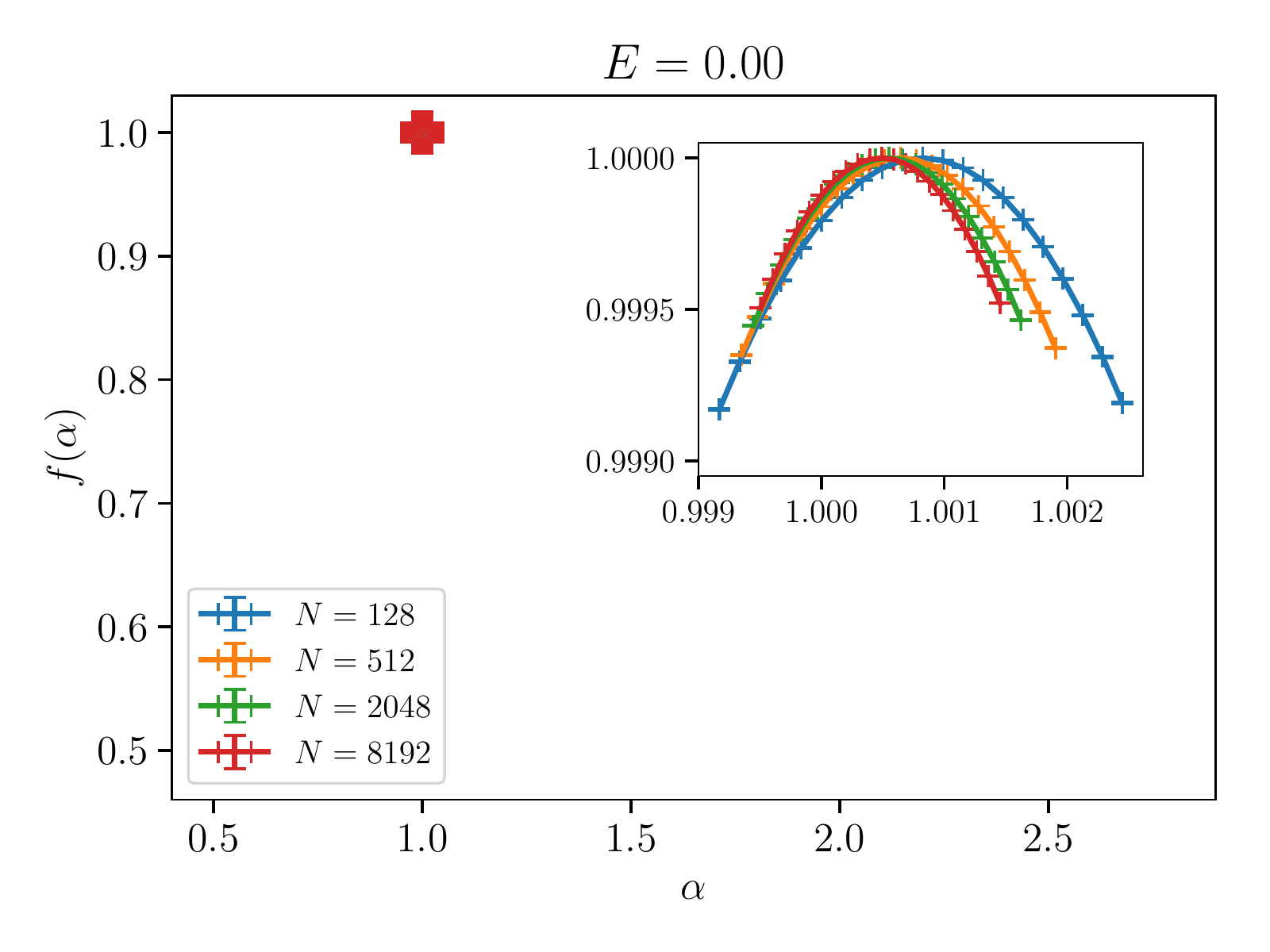}\\
 \includegraphics[width=0.96\columnwidth]{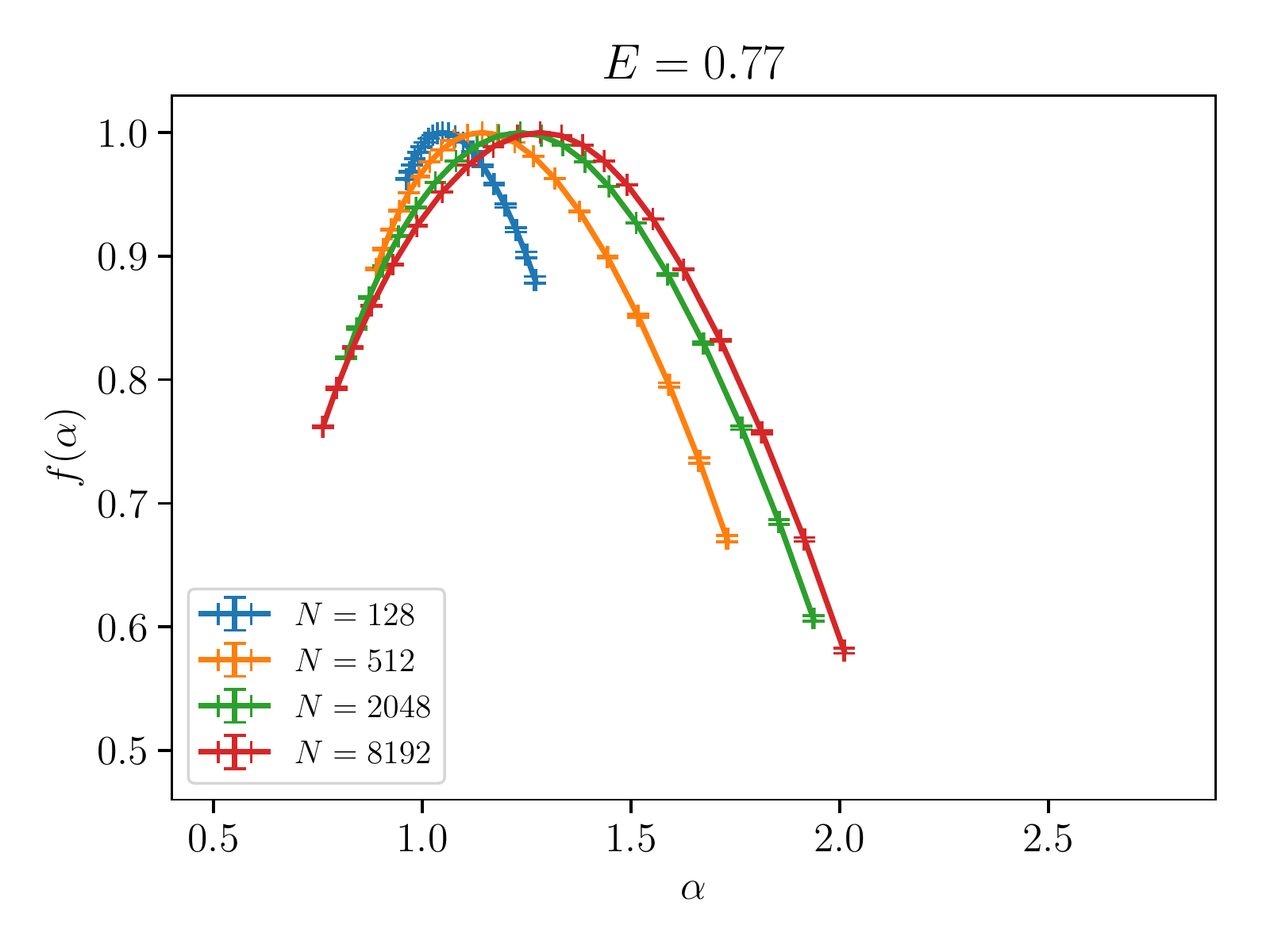}\\
 \includegraphics[width=0.96\columnwidth]{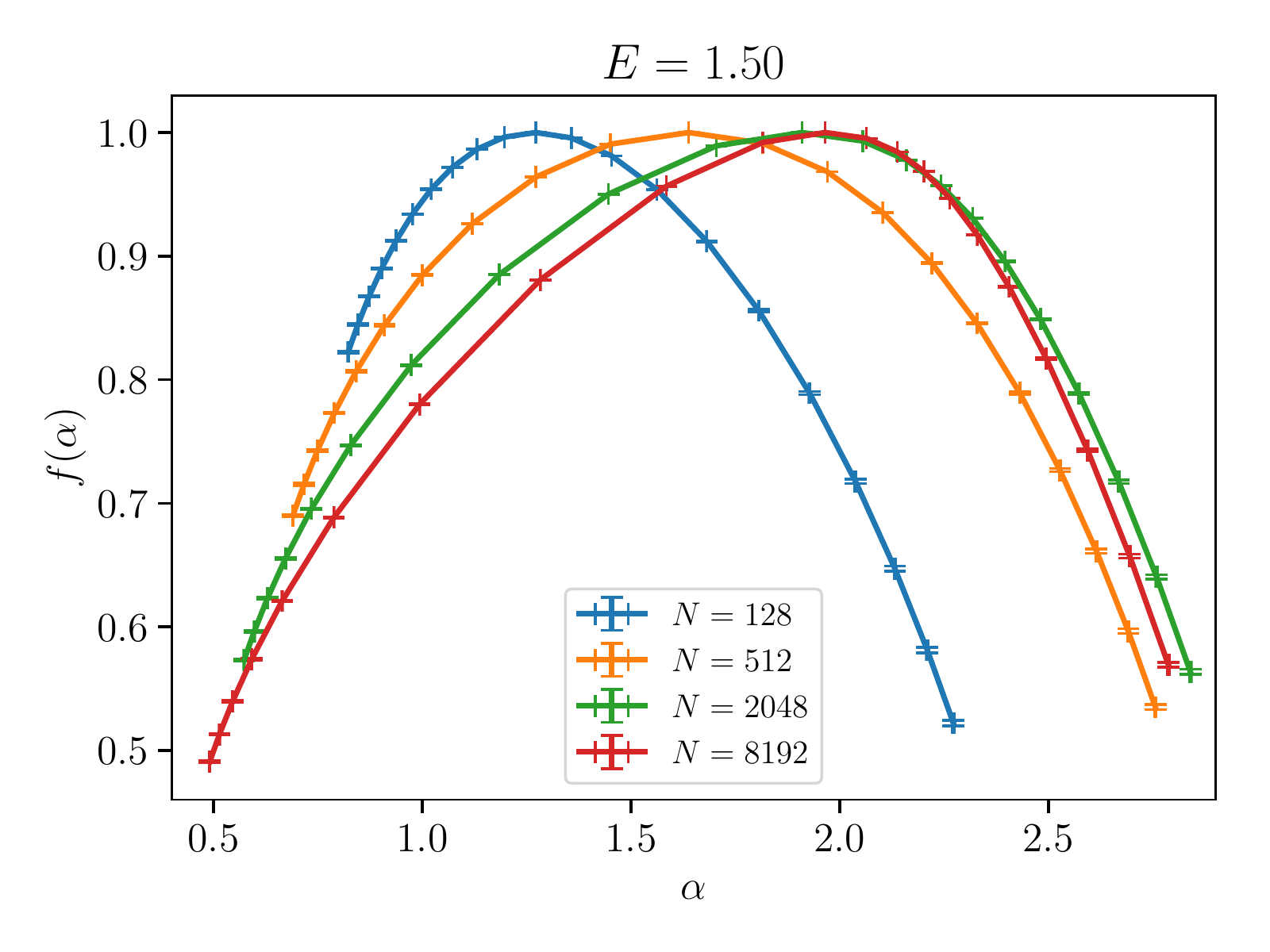}
\caption{Multifractal spectrum $f(\alpha)$ for the case of partially broken TRS ($F = 1/4$, with $k \in [-\frac{5 \pi}{8}, -\frac{3 \pi}{8}]$ removed) at small disorder ($W=1$). We show four different system sizes at each of three representative energies: $E = 0$ (top), $E = E_c = 0.77$ (middle) and $E = 1.5$ (bottom). The scales for each of these figures are the same. On this scale, $f(\alpha)$ for states in the centre of the band reduces to a single point (top). In the inset of the top panel, we show a magnified version of $f(\alpha)$, showing it is tightly concentrated around $(1,1)$.} \label{faW1}
\end{figure}

The $\ln N$ term in Eq.\ (\ref{faexploc}) shows us that even if we are in the regime where we may write $1 \ll \xi \ll N$, with a well-converged IPR localization length ($\xi_{IPR}$) and second moment localization length ($\xi_{M_2}$), we still have to deal with strong finite-size effects in $f(\alpha)$. For example with $\xi \approx 100$ and $N \approx 8000$ as in the centre of Anderson localized system with $W = 1$ (see Table \ref{tabxiAnd}), we have $\alpha_0 = 4.96 \ll +\infty$.

By analyzing the behaviour of the $f(\alpha)$ curve and of $\alpha_0$ as the system size is increased, we may be able to comment on the metallic / critical / insulating like behaviour of the system.  In Fig.\ \ref{faAnd}, we show the multifractal spectrum for Anderson localized wavefunctions for four different sizes. As described in the previous paragraph, we see strong finite-size effects, with the support of $f(\alpha)$ broadening, and the peak $\alpha_0$ of the spectrum shifting to the right, as system size is increased. This is consistent with the analytical result in Eq.\ \ref{faexploc}. In Fig. \ref{faTRS}, we plot the multifractal spectrum of typical metallic wavefunctions in the case of complete TRS breaking ($F = 1$). We note that $f(\alpha)$ is tightly concentrated around $(1,1)$, just as expected from the theory.

In Fig.\ \ref{faW1}, we show the multifractal spectra for the case F = 1/4 (partially broken TRS) for wavefunctions at three representative energies and four different system sizes in the small disorder ($W=1$) regime. The top panel shows the point-like nature of $f(\alpha)$ at $E=0$, consistent with its anomalous exponent $\Delta_{\frac{1}{2}} = 0$ (see Fig.\ \ref{figtauW1}) and in agreement with our characterization of those states as metallic. This figure looks very similar to that of the $F=1$ case (Fig.\ \ref{faTRS}). In the middle and bottom panels, we see the broadening of $f(\alpha)$ from $(1,1)$ and its departure from metallic behaviour for $E = 0.77$ and $E = 1.50$ respectively. We notice that $f(\alpha)$ shows finite size effects; however, there seems to be a tendency to universalize at the largest system sizes in our study. Note that this is clearly qualitatively different from the Anderson-like insulating $f(\alpha)$ seen in Fig.\ \ref{faAnd}. We note that this is consistent with our characterization of these states as critical-like using $\Delta_{\frac{1}{2}}$ in section \ref{secmfa}.

\begin{figure}[ht!]
\centering
\begin{minipage}{0.04\columnwidth}
(a)
\end{minipage}
\begin{minipage}{0.93\columnwidth}
\includegraphics[width=1.0\columnwidth]{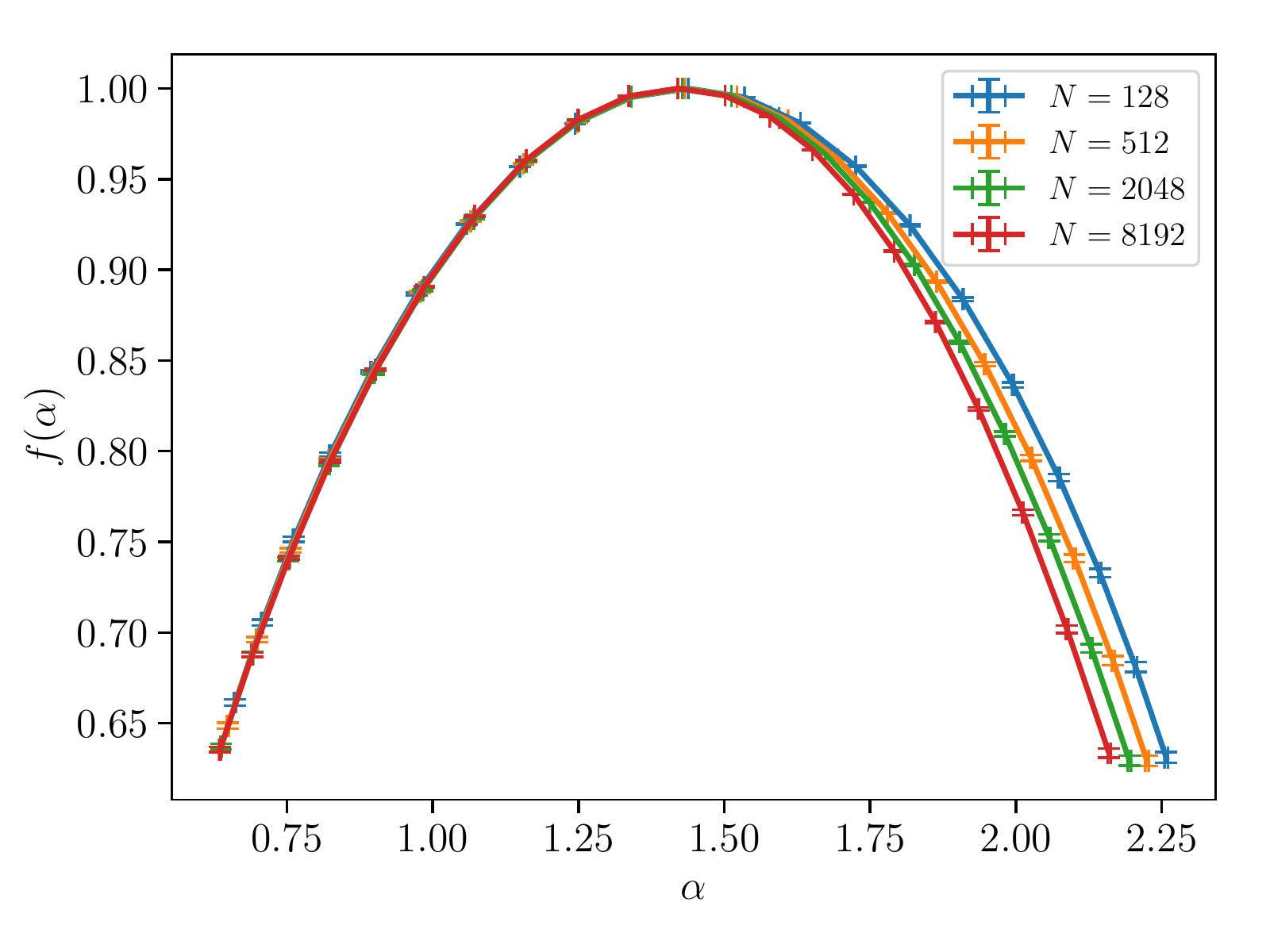}
\end{minipage}
\begin{minipage}{0.04\columnwidth}
(b)
\end{minipage}
\begin{minipage}{0.93\columnwidth}
\includegraphics[width=1.0\columnwidth]{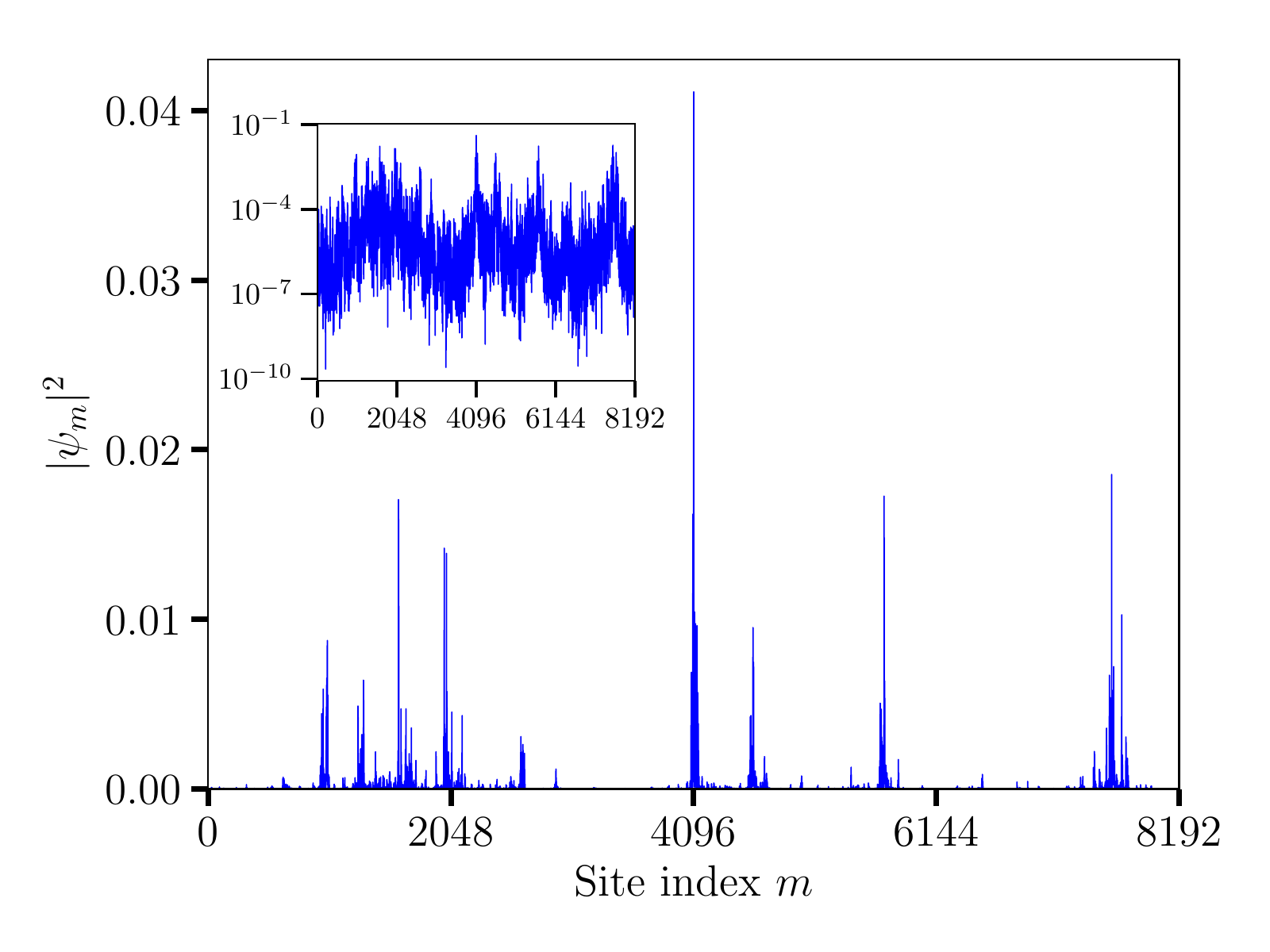}
\end{minipage}

\caption{(a) Multifractal spectrum $f(\alpha)$ for the case of partially broken TRS ($F = 1/4$, with $k \in [-\frac{5 \pi}{8}, -\frac{3 \pi}{8}]$ removed) at large disorder ($W=16$) at fixed energy $E = 0.77$. (b) A typical wavefunction at $E = 0.77$. The inset shows the same wavefunction on a log scale. Similar behaviour is seen at other energies as well as other values of $F$.} \label{faW16}
\end{figure}

We plot in Fig.\ \ref{faW16} (a) the multifractal spectrum for states at $E=0.77$ in the case $F = 1/4$ (partial TRS breaking) for large disorder ($W = 16$). We find that $f(\alpha)$ nearly collapses on itself for all system sizes, with a peak at $\alpha_0 = 1.43 \pm 0.01$. The universality of the multifractal spectrum is suggestive of a critical state, as seen for a typical wavefunction in Fig.\ \ref{faW16} (b), similar to that of the Quantum Hall wavefunction in Fig.\ \ref{figwfQH}. Similar curves can be obtained over a wide range of energies in both the partially broken TRS case as well as the fully broken TRS case (F=1). This is consistent with our interpretation of the anomalous exponent $\Delta_{\frac{1}{2}}$ (see Fig.\ \ref{figtauW16}) for the case of large disorder. All states in the spectrum appear critical-like, with a non-trivial value of $\Delta_{\frac{1}{2}}$. This is in contrast to the small disorder case, where we see a clear demarcation between metallic states in $E < E_c$ and critical-like states in $E > E_c$.

A related quantity we calculate is the probability distribution of wavefunction site densities per decade $\log_{10} |\psi_m|^2$, denoted by $P_N(\log_{10} |\psi_m|^2)$, where the subscript $N$ makes explicit the fact that this distribution is dependent on system size. For a metal, we would expect this quantity to be sharply peaked as most sites have have $|\psi|^2$ in the vicinity of $1/N$. On the other hand, for an insulator with exponentially localized states, this probability distribution would be flat, as there are equal numbers of sites in each decade of $|\psi|^2$.  $P_N(\log_{10} |\psi|^2)$ is related to the probability distribution $P_N(\alpha)$ of $\alpha$ by the relation \begin{align}
\alpha = -\frac{\log_{10} |\psi^2| }{\log_{10} N}. \label{eqalpmodpsi}
\end{align}
It was shown in \cite{Rodriguezetal2009} that the distribution $P_N(\alpha)$ provides an alternative method for calculating the multifractal spectrum $f(\alpha)$ by modelling the size dependence as \begin{align}
P_N(\alpha) = N^{f(\alpha) - 1} P_N(\alpha_0),
\end{align}
where $P_N(\alpha_0) = \max[P_N(\alpha)]$.

\begin{figure}[ht!]
\centering
 \includegraphics[width=1.0\columnwidth]{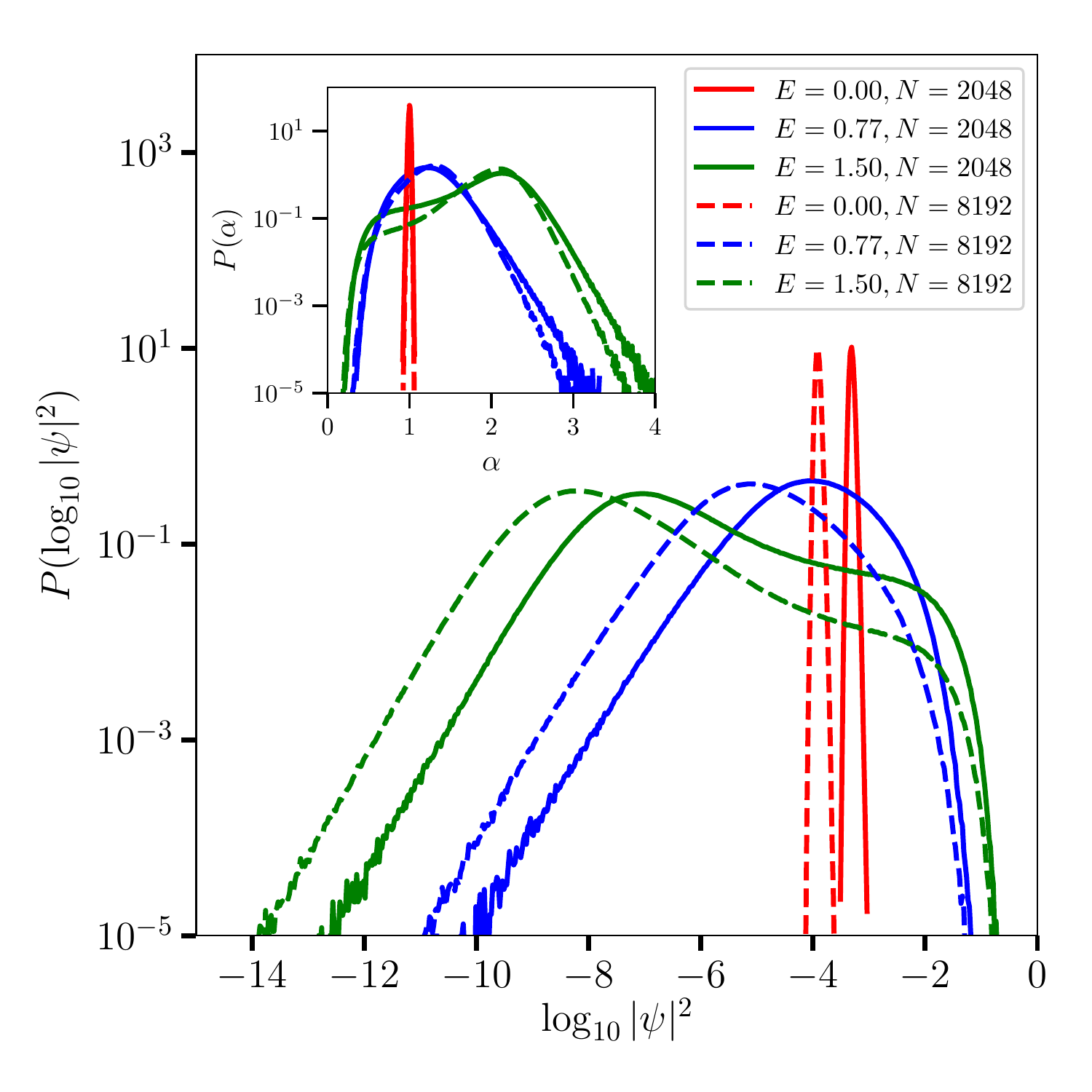}
\caption{Probability distribution by decade of the site probability $P(\log_{10} |\psi|^2)$ for the case of partially broken TRS ($F = 1/4$, with $k \in [-\frac{5 \pi}{8}, -\frac{3 \pi}{8}]$ removed) at small disorder ($W=1$). We show two different system sizes at each of three representative energies: $E = 0$,  $E = E_c = 0.77$ and $E = 1.5$. The inset shows the same probability distribution, plotted as a function of the system-size independent variable $\alpha$.} \label{figmodpsiF025}
\end{figure}

\begin{figure}[ht!]
\centering
 \includegraphics[width=1.0\columnwidth]{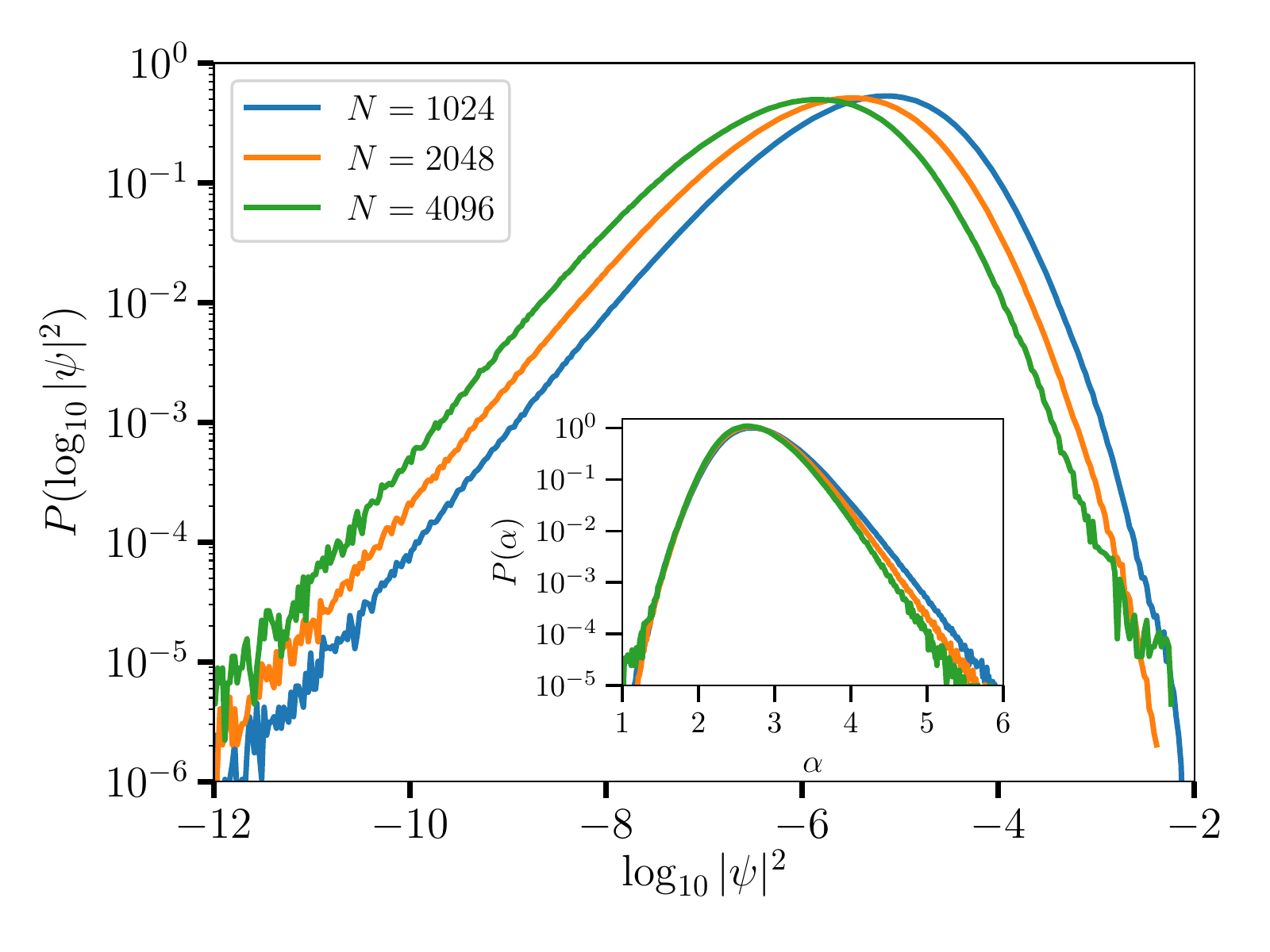}
\caption{Probability distribution by decade of the site probability $P(\log_{10} |\psi|^2)$ for the case of non-interacting electrons at the centre of the band in the lowest Landau level with Gaussian white noise-disorder. We obtain eigenstates closest to $E = 0$ on a square torus with three different values of $N_\phi$ (number of flux quanta). To obtain this plot, we then divide the torus into plaquettes of $\frac{l_B}{3} \times \frac{l_B}{3}$ and bin the values of $|\psi(\v{r})|^2$ integrated over each plaquette over an ensemble of several disorder realizations. The inset shows the scaling collapse of the the three distributions when plotted against the variable $\alpha$ (Eqn.\ \ref{eqalpmodpsi}).} \label{figmodpsiQH}
\end{figure}

In Fig.\ \ref{figmodpsiF025}, we plot $P(\log_{10} |\psi|^2)$ for $N = 2048$ and $N = 8192$ at three characteristic energies. The distributions at $E = 0$ are very narrow, as expected for metal-like states. The distributions for $E = 0.77$ and $E = 1.50$ are broad and have a non-trivial shape. They are neither metal-like nor insulator-like and the variation of $|\psi|^2$ over several orders of magnitude is consistent with our characterization of these states, throughout the paper, as critical-like. We also point out the slope of these curves is equal to $1$, within error bars, for small $\log_{10} |\psi|^2$. This suggests that the functional form of the probability distribution for wavefunction magnitudes is $\mathcal{P}( |\psi| ) \approx c(N) |\psi|$ for $|\psi|^2 \ll N^{-1}$, where $c(N)$ is some normalization constant. 

In Fig.\ \ref{figmodpsiQH}, we plot the same quantity for the critical states obtained at the centre of the lowest Landau level. We see the same straight-line behaviour of $P(\log_{10} |\psi|^2)$ for small $|\psi|^2$ in this log-log plot, suggesting that this is common for critical states.

\section{Link between TRS broken model and long-range hopping}

By expressing the truncated Hamiltonian (Eq.\ \ref{eqHamK}) in position space, it becomes clear that the Hamiltonian is no longer local. However, one can show that the coupling between far-off sites $\Ket{x_m}$ and $\Ket{x_n}$ decays asymptotically as a power law. In the position basis, the elements of the Hamiltonian $H$ may be written as the sum of a hopping term $H^{(hop)}$ and a disorder term $H^{(dis)}$ as
\begin{align}
H_{mn} &= \braket{x_m | \hat{H} | x_n} = H^{(hop)}_{mn} + H^{(dis)}_{mn}\\
\intertext{where}
H^{(hop)}_{mn} &= \sum_{r}^{}{'} \frac{1}{N} \cos(\frac{2 \pi r}{N}) e^{\frac{i 2 \pi (m - n)r}{N}} \\
\text{ and } 
H^{(dis)}_{mn} &= \sum_{r, s}^{}{'} \frac{\sum_m \epsilon_m}{N^2} e^{-\frac{i 2 \pi m(r-s)}{N}} e^{\frac{i 2 \pi (mr - ns)}{N}}.
\end{align}

In the equations above, $\sum'$ denotes the sum over the projected part of the Hilbert space. When the \emph{negative} half of the band is removed ($F = 1$), then the hopping terms may be summed up to give
\begin{align}
H_{mn}^{(hop)} &= \begin{cases}
-\frac{1}{2},  \qquad |m-n| = 1\\
0, \qquad \text{for other odd } m-n\\
-\frac{2}{N} \left[ \frac{1}{1 - e^{i 2 \pi \frac{m-n+1}{N}}} + \frac{1}{1 - e^{i 2 \pi \frac{m-n-1}{N}}} \right],\\ \qquad\text{for even } m-n.
\end{cases}
\end{align}

The disorder terms, in this case, are $H_{mn}^{(dis)} = \sum\limits_r \epsilon_r c_{rm} c^\ast_{rn}$, where
\begin{align}
c_{rm} = \begin{cases}
\frac{1}{2},  &  m = r\\
0, &\text{for other even } m-r\\
-\frac{i}{N \sin \frac{\pi (r-m)}{N}}, &\text{for odd } m-r.
\end{cases}
\end{align}

Due to the symmetric nature of the random disorder $P(\epsilon_m) = P(-\epsilon_m)$, the disorder term is on average zero ($\langle H_{mn}^{(dis)} \rangle = 0$). However its magnitude is on average non-zero, and decays with increasing site distance. In the thermodynamic limit, where $N \to \infty$, we find that both the hopping and the disorder terms fall off similarly for finite $m-n$:

\begin{align}
H_{mn}^{(hop)} &\approx \begin{cases}
-\frac{2i}{\pi} \frac{(m-n)  }{(m-n)^2-1}, \quad\text{for even } m-n\\
0,  \text{ for odd } m-n
\end{cases}
\\
\langle |H_{mn}^{(dis)}|^2 \rangle &\approx \frac{\langle \epsilon^2 \rangle}{2 \pi^2 (m-n)^2}, \qquad m-n \ll N
\end{align}

\begin{figure}[ht!]
\includegraphics[width=\columnwidth]{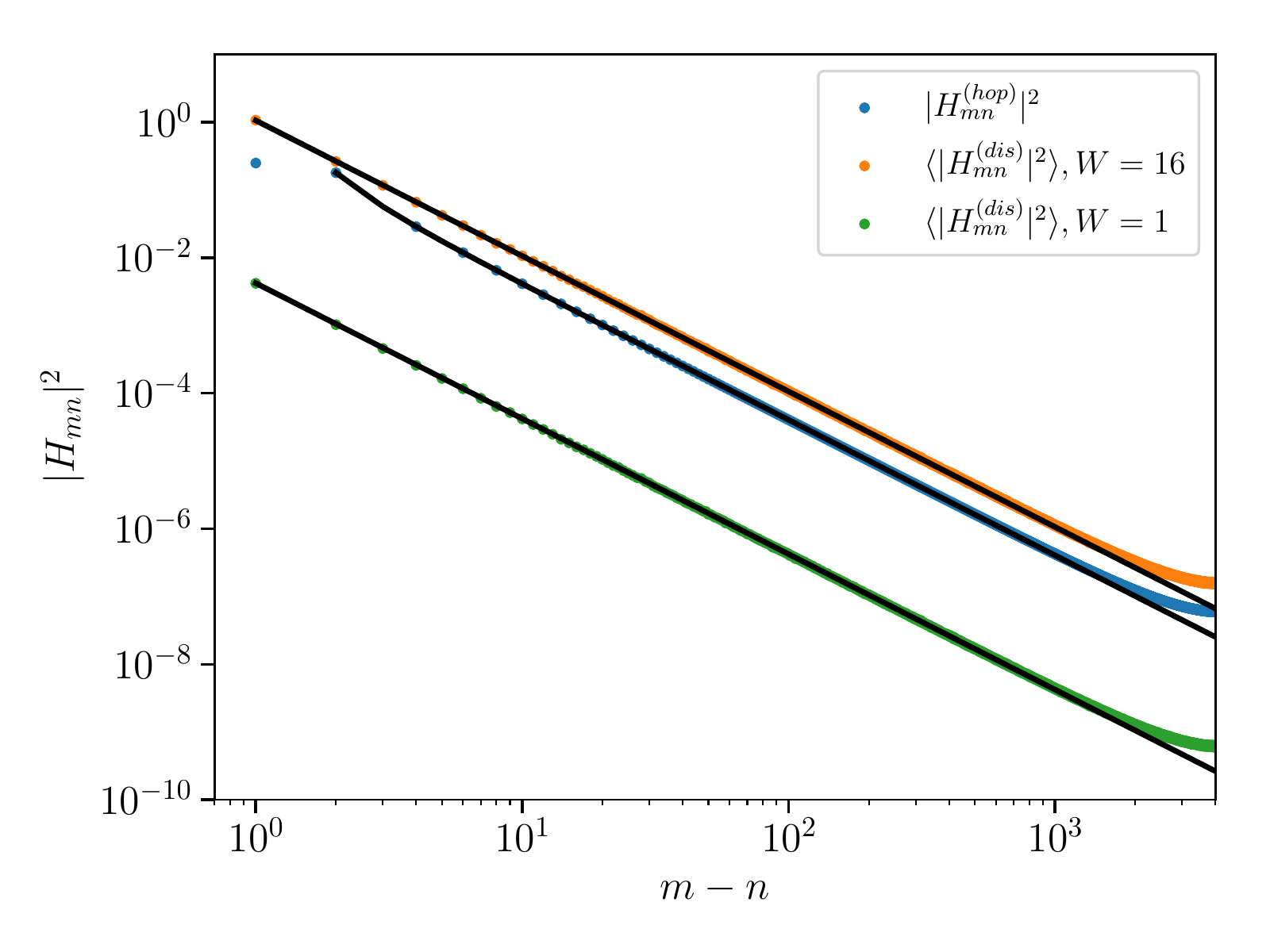}
\caption{Comparison of ensemble averaged absolute values of Hamiltonian matrix elements. At small disorder ($W=1$), the hopping term (blue) dominates the disorder term (green). At large disorder, the disorder term (yellow) dominates. The black lines are the approximations Eq. (B6) and Eq. (B7). Data is cut off at half the chain length due to periodic boundary conditions.}\label{powerlaw2}
\end{figure}

So there is a competition between the random disorder terms and deterministic hopping terms (both of which fall off as $1/r$), which seems to drive a transition from metallic to critical. In Fig.\ \ref{powerlaw2}, we plot the two kinds of terms for both small and large disorder. In the case of small disorder $W = 1$, we have $|H_{mn}^{(hop)}|^2 > \langle |H_{mn}^{(dis)}|^2 \rangle$, but at large disorder ($W = 16$), the random term dominates.

The behaviour described above suggests a connection with the family of critical models in the Periodic Random Banded Matrix (PRBM) ensemble \cite{MirlinPRBM1996, MirlinPRBM2000}. In this model, the Hamiltonian describes a 1-D chain with random power-law hopping. At criticality in this framework, the matrix elements $H_{ij}$ are independent identically distributed Gaussian variables with zero mean and variance given by $\langle |H_{ij}| \rangle = a^2(|i-j|)$, with $a^2(r) = \frac{1}{1 + (r/b)^2}$. For $r\gg b$, these hopping terms fall off as $1/r$, similar to our truncated Anderson model. The parameter $b$ defines a whole family of Anderson critical models, with $b \ll 1$ describing the quasi-insulating limit and $b \gg 1$ describing the quasi-metallic limit.

The case of complete TRS breaking ($F=1$) in our truncated Anderson model is analogous to the PRBM ensemble at criticality and possibly several other random Hamiltonians with broken time reversal symmetry \cite{Fyodorov94}, with the strength of the disorder term in relation to the hopping term driving a transition from metallic-like to critical behaviour. A key difference is that the $\mathcal{O} (N^2)$ random hopping terms in the PRBM ensemble on a system of $N$ sites are all uncorrelated, while in the truncated Anderson model they are linear combinations of $N$ independent disorder terms, and therefore highly correlated.

It would be interesting to further explore the connection between the PRBM ensemble and the truncated Anderson models to shed more light on the role of  randomness and symmetry-breaking in driving a metal-insulator transition in 1 dimension,  in a similar spirit to previous works \cite{Mendez06, Mendez12,*Mendez14}.


\begin{thebibliography}{100}%
\makeatletter
\providecommand \@ifxundefined [1]{%
 \@ifx{#1\undefined}
}%
\providecommand \@ifnum [1]{%
 \ifnum #1\expandafter \@firstoftwo
 \else \expandafter \@secondoftwo
 \fi
}%
\providecommand \@ifx [1]{%
 \ifx #1\expandafter \@firstoftwo
 \else \expandafter \@secondoftwo
 \fi
}%
\providecommand \natexlab [1]{#1}%
\providecommand \enquote  [1]{``#1''}%
\providecommand \bibnamefont  [1]{#1}%
\providecommand \bibfnamefont [1]{#1}%
\providecommand \citenamefont [1]{#1}%
\providecommand \href@noop [0]{\@secondoftwo}%
\providecommand \href [0]{\begingroup \@sanitize@url \@href}%
\providecommand \@href[1]{\@@startlink{#1}\@@href}%
\providecommand \@@href[1]{\endgroup#1\@@endlink}%
\providecommand \@sanitize@url [0]{\catcode `\\12\catcode `\$12\catcode
  `\&12\catcode `\#12\catcode `\^12\catcode `\_12\catcode `\%12\relax}%
\providecommand \@@startlink[1]{}%
\providecommand \@@endlink[0]{}%
\providecommand \url  [0]{\begingroup\@sanitize@url \@url }%
\providecommand \@url [1]{\endgroup\@href {#1}{\urlprefix }}%
\providecommand \urlprefix  [0]{URL }%
\providecommand \Eprint [0]{\href }%
\providecommand \doibase [0]{http://dx.doi.org/}%
\providecommand \selectlanguage [0]{\@gobble}%
\providecommand \bibinfo  [0]{\@secondoftwo}%
\providecommand \bibfield  [0]{\@secondoftwo}%
\providecommand \translation [1]{[#1]}%
\providecommand \BibitemOpen [0]{}%
\providecommand \bibitemStop [0]{}%
\providecommand \bibitemNoStop [0]{.\EOS\space}%
\providecommand \EOS [0]{\spacefactor3000\relax}%
\providecommand \BibitemShut  [1]{\csname bibitem#1\endcsname}%
\let\auto@bib@innerbib\@empty
\bibitem [{\citenamefont {Anderson}(1958)}]{Anderson58}%
  \BibitemOpen
  \bibfield  {author} {\bibinfo {author} {\bibfnamefont {P.~W.}\ \bibnamefont
  {Anderson}},\ }\href {\doibase 10.1103/PhysRev.109.1492} {\bibfield
  {journal} {\bibinfo  {journal} {Phys. Rev.}\ }\textbf {\bibinfo {volume}
  {109}},\ \bibinfo {pages} {1492} (\bibinfo {year} {1958})}\BibitemShut
  {NoStop}%
\bibitem [{\citenamefont {Mott}\ and\ \citenamefont
  {Twose}(1961)}]{MottTwose61}%
  \BibitemOpen
  \bibfield  {author} {\bibinfo {author} {\bibfnamefont {N.}~\bibnamefont
  {Mott}}\ and\ \bibinfo {author} {\bibfnamefont {W.}~\bibnamefont {Twose}},\
  }\href {\doibase 10.1080/00018736100101271} {\bibfield  {journal} {\bibinfo
  {journal} {Advances in Physics}\ }\textbf {\bibinfo {volume} {10}},\ \bibinfo
  {pages} {107} (\bibinfo {year} {1961})}\BibitemShut {NoStop}%
\bibitem [{\citenamefont {Abrahams}\ \emph {et~al.}(1979)\citenamefont
  {Abrahams}, \citenamefont {Anderson}, \citenamefont {Licciardello},\ and\
  \citenamefont {Ramakrishnan}}]{GangofFour79}%
  \BibitemOpen
  \bibfield  {author} {\bibinfo {author} {\bibfnamefont {E.}~\bibnamefont
  {Abrahams}}, \bibinfo {author} {\bibfnamefont {P.~W.}\ \bibnamefont
  {Anderson}}, \bibinfo {author} {\bibfnamefont {D.~C.}\ \bibnamefont
  {Licciardello}}, \ and\ \bibinfo {author} {\bibfnamefont {T.~V.}\
  \bibnamefont {Ramakrishnan}},\ }\href {\doibase 10.1103/PhysRevLett.42.673}
  {\bibfield  {journal} {\bibinfo  {journal} {Phys. Rev. Lett.}\ }\textbf
  {\bibinfo {volume} {42}},\ \bibinfo {pages} {673} (\bibinfo {year}
  {1979})}\BibitemShut {NoStop}%
\bibitem [{\citenamefont {Lee}\ and\ \citenamefont
  {Ramakrishnan}(1985)}]{LeeRamakrishnan85}%
  \BibitemOpen
  \bibfield  {author} {\bibinfo {author} {\bibfnamefont {P.~A.}\ \bibnamefont
  {Lee}}\ and\ \bibinfo {author} {\bibfnamefont {T.~V.}\ \bibnamefont
  {Ramakrishnan}},\ }\href {\doibase 10.1103/RevModPhys.57.287} {\bibfield
  {journal} {\bibinfo  {journal} {Rev. Mod. Phys.}\ }\textbf {\bibinfo {volume}
  {57}},\ \bibinfo {pages} {287} (\bibinfo {year} {1985})}\BibitemShut
  {NoStop}%
\bibitem [{\citenamefont {Altshuler}\ and\ \citenamefont
  {Aronov}(1985)}]{AltshulerAronov}%
  \BibitemOpen
  \bibfield  {author} {\bibinfo {author} {\bibfnamefont {B.}~\bibnamefont
  {Altshuler}}\ and\ \bibinfo {author} {\bibfnamefont {A.}~\bibnamefont
  {Aronov}},\ }\href@noop {} {\emph {\bibinfo {title} {Electron-Electron
  Interaction in Disordered Conductors. edited by AL Efros and M. Pollak}}}\
  (\bibinfo  {publisher} {Amsterdam: Elsevier Scientific Publishing},\ \bibinfo
  {year} {1985})\BibitemShut {NoStop}%
\bibitem [{\citenamefont {MacKinnon}\ and\ \citenamefont
  {Kramer}(1981)}]{MacKinnonKramer1981}%
  \BibitemOpen
  \bibfield  {author} {\bibinfo {author} {\bibfnamefont {A.}~\bibnamefont
  {MacKinnon}}\ and\ \bibinfo {author} {\bibfnamefont {B.}~\bibnamefont
  {Kramer}},\ }\href {\doibase 10.1103/PhysRevLett.47.1546} {\bibfield
  {journal} {\bibinfo  {journal} {Phys. Rev. Lett.}\ }\textbf {\bibinfo
  {volume} {47}},\ \bibinfo {pages} {1546} (\bibinfo {year}
  {1981})}\BibitemShut {NoStop}%
\bibitem [{\citenamefont {Slevin}\ and\ \citenamefont
  {Ohtsuki}(1997)}]{SlevinOhtsuki1997}%
  \BibitemOpen
  \bibfield  {author} {\bibinfo {author} {\bibfnamefont {K.}~\bibnamefont
  {Slevin}}\ and\ \bibinfo {author} {\bibfnamefont {T.}~\bibnamefont
  {Ohtsuki}},\ }\href {\doibase 10.1103/PhysRevLett.78.4083} {\bibfield
  {journal} {\bibinfo  {journal} {Phys. Rev. Lett.}\ }\textbf {\bibinfo
  {volume} {78}},\ \bibinfo {pages} {4083} (\bibinfo {year}
  {1997})}\BibitemShut {NoStop}%
\bibitem [{\citenamefont {Slevin}\ and\ \citenamefont
  {Ohtsuki}(1999)}]{SlevinOhtsuki1999}%
  \BibitemOpen
  \bibfield  {author} {\bibinfo {author} {\bibfnamefont {K.}~\bibnamefont
  {Slevin}}\ and\ \bibinfo {author} {\bibfnamefont {T.}~\bibnamefont
  {Ohtsuki}},\ }\href {\doibase 10.1103/PhysRevLett.82.382} {\bibfield
  {journal} {\bibinfo  {journal} {Phys. Rev. Lett.}\ }\textbf {\bibinfo
  {volume} {82}},\ \bibinfo {pages} {382} (\bibinfo {year} {1999})}\BibitemShut
  {NoStop}%
\bibitem [{\citenamefont {Slevin}\ and\ \citenamefont
  {Ohtsuki}(2001)}]{SlevinOhtsuki2001}%
  \BibitemOpen
  \bibfield  {author} {\bibinfo {author} {\bibfnamefont {K.}~\bibnamefont
  {Slevin}}\ and\ \bibinfo {author} {\bibfnamefont {T.}~\bibnamefont
  {Ohtsuki}},\ }\href {\doibase 10.1103/PhysRevB.63.045108} {\bibfield
  {journal} {\bibinfo  {journal} {Phys. Rev. B}\ }\textbf {\bibinfo {volume}
  {63}},\ \bibinfo {pages} {045108} (\bibinfo {year} {2001})}\BibitemShut
  {NoStop}%
\bibitem [{\citenamefont {Kramer}\ and\ \citenamefont
  {MacKinnon}(1993)}]{MacKinnonKramer1993}%
  \BibitemOpen
  \bibfield  {author} {\bibinfo {author} {\bibfnamefont {B.}~\bibnamefont
  {Kramer}}\ and\ \bibinfo {author} {\bibfnamefont {A.}~\bibnamefont
  {MacKinnon}},\ }\href@noop {} {\bibfield  {journal} {\bibinfo  {journal}
  {Reports on Progress in Physics}\ }\textbf {\bibinfo {volume} {56}},\
  \bibinfo {pages} {1469} (\bibinfo {year} {1993})}\BibitemShut {NoStop}%
\bibitem [{\citenamefont {Evers}\ and\ \citenamefont
  {Mirlin}(2008)}]{EversMirlin08}%
  \BibitemOpen
  \bibfield  {author} {\bibinfo {author} {\bibfnamefont {F.}~\bibnamefont
  {Evers}}\ and\ \bibinfo {author} {\bibfnamefont {A.~D.}\ \bibnamefont
  {Mirlin}},\ }\href {\doibase 10.1103/RevModPhys.80.1355} {\bibfield
  {journal} {\bibinfo  {journal} {Rev. Mod. Phys.}\ }\textbf {\bibinfo {volume}
  {80}},\ \bibinfo {pages} {1355} (\bibinfo {year} {2008})}\BibitemShut
  {NoStop}%
\bibitem [{\citenamefont {Halperin}(1982)}]{Halperin1982}%
  \BibitemOpen
  \bibfield  {author} {\bibinfo {author} {\bibfnamefont {B.~I.}\ \bibnamefont
  {Halperin}},\ }\href@noop {} {\bibfield  {journal} {\bibinfo  {journal}
  {Physical Review B}\ }\textbf {\bibinfo {volume} {25}},\ \bibinfo {pages}
  {2185} (\bibinfo {year} {1982})}\BibitemShut {NoStop}%
\bibitem [{\citenamefont {Huckestein}\ and\ \citenamefont
  {Kramer}(1990)}]{HuckesteinKramer1990}%
  \BibitemOpen
  \bibfield  {author} {\bibinfo {author} {\bibfnamefont {B.}~\bibnamefont
  {Huckestein}}\ and\ \bibinfo {author} {\bibfnamefont {B.}~\bibnamefont
  {Kramer}},\ }\href {\doibase 10.1103/PhysRevLett.64.1437} {\bibfield
  {journal} {\bibinfo  {journal} {Phys. Rev. Lett.}\ }\textbf {\bibinfo
  {volume} {64}},\ \bibinfo {pages} {1437} (\bibinfo {year}
  {1990})}\BibitemShut {NoStop}%
\bibitem [{\citenamefont {Huo}\ and\ \citenamefont
  {Bhatt}(1992)}]{HuoBhatt1992}%
  \BibitemOpen
  \bibfield  {author} {\bibinfo {author} {\bibfnamefont {Y.}~\bibnamefont
  {Huo}}\ and\ \bibinfo {author} {\bibfnamefont {R.~N.}\ \bibnamefont
  {Bhatt}},\ }\href {\doibase 10.1103/PhysRevLett.68.1375} {\bibfield
  {journal} {\bibinfo  {journal} {Phys. Rev. Lett.}\ }\textbf {\bibinfo
  {volume} {68}},\ \bibinfo {pages} {1375} (\bibinfo {year}
  {1992})}\BibitemShut {NoStop}%
\bibitem [{\citenamefont {Chalker}\ and\ \citenamefont
  {Coddington}(1988)}]{ChalkerCoddington}%
  \BibitemOpen
  \bibfield  {author} {\bibinfo {author} {\bibfnamefont {J.~T.}\ \bibnamefont
  {Chalker}}\ and\ \bibinfo {author} {\bibfnamefont {P.~D.}\ \bibnamefont
  {Coddington}},\ }\href {http://stacks.iop.org/0022-3719/21/i=14/a=008}
  {\bibfield  {journal} {\bibinfo  {journal} {Journal of Physics C: Solid State
  Physics}\ }\textbf {\bibinfo {volume} {21}},\ \bibinfo {pages} {2665}
  (\bibinfo {year} {1988})}\BibitemShut {NoStop}%
\bibitem [{\citenamefont {Slevin}\ and\ \citenamefont
  {Ohtsuki}(2009)}]{SlevinOhtsukiQH}%
  \BibitemOpen
  \bibfield  {author} {\bibinfo {author} {\bibfnamefont {K.}~\bibnamefont
  {Slevin}}\ and\ \bibinfo {author} {\bibfnamefont {T.}~\bibnamefont
  {Ohtsuki}},\ }\href {\doibase 10.1103/PhysRevB.80.041304} {\bibfield
  {journal} {\bibinfo  {journal} {Phys. Rev. B}\ }\textbf {\bibinfo {volume}
  {80}},\ \bibinfo {pages} {041304} (\bibinfo {year} {2009})}\BibitemShut
  {NoStop}%
\bibitem [{\citenamefont {Bishop}\ \emph {et~al.}(1982)\citenamefont {Bishop},
  \citenamefont {Dynes},\ and\ \citenamefont {Tsui}}]{Bishop83}%
  \BibitemOpen
  \bibfield  {author} {\bibinfo {author} {\bibfnamefont {D.~J.}\ \bibnamefont
  {Bishop}}, \bibinfo {author} {\bibfnamefont {R.~C.}\ \bibnamefont {Dynes}}, \
  and\ \bibinfo {author} {\bibfnamefont {D.~C.}\ \bibnamefont {Tsui}},\ }\href
  {\doibase 10.1103/PhysRevB.26.773} {\bibfield  {journal} {\bibinfo  {journal}
  {Phys. Rev. B}\ }\textbf {\bibinfo {volume} {26}},\ \bibinfo {pages} {773}
  (\bibinfo {year} {1982})}\BibitemShut {NoStop}%
\bibitem [{\citenamefont {Bergmann}(1983)}]{Bergman83}%
  \BibitemOpen
  \bibfield  {author} {\bibinfo {author} {\bibfnamefont {G.}~\bibnamefont
  {Bergmann}},\ }\href {\doibase 10.1103/PhysRevB.28.2914} {\bibfield
  {journal} {\bibinfo  {journal} {Phys. Rev. B}\ }\textbf {\bibinfo {volume}
  {28}},\ \bibinfo {pages} {2914} (\bibinfo {year} {1983})}\BibitemShut
  {NoStop}%
\bibitem [{\citenamefont {Rosenbaum}\ \emph {et~al.}(1981)\citenamefont
  {Rosenbaum}, \citenamefont {Milligan}, \citenamefont {Thomas}, \citenamefont
  {Lee}, \citenamefont {Ramakrishnan}, \citenamefont {Bhatt}, \citenamefont
  {DeConde}, \citenamefont {Hess},\ and\ \citenamefont {Perry}}]{Rosenbaum81}%
  \BibitemOpen
  \bibfield  {author} {\bibinfo {author} {\bibfnamefont {T.~F.}\ \bibnamefont
  {Rosenbaum}}, \bibinfo {author} {\bibfnamefont {R.~F.}\ \bibnamefont
  {Milligan}}, \bibinfo {author} {\bibfnamefont {G.~A.}\ \bibnamefont
  {Thomas}}, \bibinfo {author} {\bibfnamefont {P.~A.}\ \bibnamefont {Lee}},
  \bibinfo {author} {\bibfnamefont {T.~V.}\ \bibnamefont {Ramakrishnan}},
  \bibinfo {author} {\bibfnamefont {R.~N.}\ \bibnamefont {Bhatt}}, \bibinfo
  {author} {\bibfnamefont {K.}~\bibnamefont {DeConde}}, \bibinfo {author}
  {\bibfnamefont {H.}~\bibnamefont {Hess}}, \ and\ \bibinfo {author}
  {\bibfnamefont {T.}~\bibnamefont {Perry}},\ }\href {\doibase
  10.1103/PhysRevLett.47.1758} {\bibfield  {journal} {\bibinfo  {journal}
  {Phys. Rev. Lett.}\ }\textbf {\bibinfo {volume} {47}},\ \bibinfo {pages}
  {1758} (\bibinfo {year} {1981})}\BibitemShut {NoStop}%
\bibitem [{\citenamefont {Rosenbaum}\ \emph {et~al.}(1983)\citenamefont
  {Rosenbaum}, \citenamefont {Milligan}, \citenamefont {Paalanen},
  \citenamefont {Thomas}, \citenamefont {Bhatt},\ and\ \citenamefont
  {Lin}}]{Rosenbaum83}%
  \BibitemOpen
  \bibfield  {author} {\bibinfo {author} {\bibfnamefont {T.~F.}\ \bibnamefont
  {Rosenbaum}}, \bibinfo {author} {\bibfnamefont {R.~F.}\ \bibnamefont
  {Milligan}}, \bibinfo {author} {\bibfnamefont {M.~A.}\ \bibnamefont
  {Paalanen}}, \bibinfo {author} {\bibfnamefont {G.~A.}\ \bibnamefont
  {Thomas}}, \bibinfo {author} {\bibfnamefont {R.~N.}\ \bibnamefont {Bhatt}}, \
  and\ \bibinfo {author} {\bibfnamefont {W.}~\bibnamefont {Lin}},\ }\href
  {\doibase 10.1103/PhysRevB.27.7509} {\bibfield  {journal} {\bibinfo
  {journal} {Phys. Rev. B}\ }\textbf {\bibinfo {volume} {27}},\ \bibinfo
  {pages} {7509} (\bibinfo {year} {1983})}\BibitemShut {NoStop}%
\bibitem [{\citenamefont {Gor'kov}\ \emph {et~al.}(1979)\citenamefont
  {Gor'kov}, \citenamefont {Larkin},\ and\ \citenamefont
  {Khmelnitskii}}]{Gorkov79}%
  \BibitemOpen
  \bibfield  {author} {\bibinfo {author} {\bibfnamefont {L.}~\bibnamefont
  {Gor'kov}}, \bibinfo {author} {\bibfnamefont {A.}~\bibnamefont {Larkin}}, \
  and\ \bibinfo {author} {\bibfnamefont {D.}~\bibnamefont {Khmelnitskii}},\
  }\href@noop {} {\bibfield  {journal} {\bibinfo  {journal} {JETP Letters}\
  }\textbf {\bibinfo {volume} {30}},\ \bibinfo {pages} {248} (\bibinfo {year}
  {1979})}\BibitemShut {NoStop}%
\bibitem [{\citenamefont {Hikami}\ \emph {et~al.}(1980)\citenamefont {Hikami},
  \citenamefont {Larkin},\ and\ \citenamefont {Nagaoka}}]{Hikami80}%
  \BibitemOpen
  \bibfield  {author} {\bibinfo {author} {\bibfnamefont {S.}~\bibnamefont
  {Hikami}}, \bibinfo {author} {\bibfnamefont {A.~I.}\ \bibnamefont {Larkin}},
  \ and\ \bibinfo {author} {\bibfnamefont {Y.}~\bibnamefont {Nagaoka}},\ }\href
  {\doibase 10.1143/PTP.63.707} {\bibfield  {journal} {\bibinfo  {journal}
  {Progress of Theoretical Physics}\ }\textbf {\bibinfo {volume} {63}},\
  \bibinfo {pages} {707} (\bibinfo {year} {1980})}\BibitemShut {NoStop}%
\bibitem [{\citenamefont {Kawabata}(1980)}]{Kawabata80}%
  \BibitemOpen
  \bibfield  {author} {\bibinfo {author} {\bibfnamefont {A.}~\bibnamefont
  {Kawabata}},\ }\href {\doibase https://doi.org/10.1016/0038-1098(80)90644-4}
  {\bibfield  {journal} {\bibinfo  {journal} {Solid State Communications}\
  }\textbf {\bibinfo {volume} {34}},\ \bibinfo {pages} {431 } (\bibinfo {year}
  {1980})}\BibitemShut {NoStop}%
\bibitem [{\citenamefont {Liu}\ and\ \citenamefont
  {Das~Sarma}(1994)}]{LiuDas1994}%
  \BibitemOpen
  \bibfield  {author} {\bibinfo {author} {\bibfnamefont {D.}~\bibnamefont
  {Liu}}\ and\ \bibinfo {author} {\bibfnamefont {S.}~\bibnamefont
  {Das~Sarma}},\ }\href {\doibase 10.1103/PhysRevB.49.2677} {\bibfield
  {journal} {\bibinfo  {journal} {Phys. Rev. B}\ }\textbf {\bibinfo {volume}
  {49}},\ \bibinfo {pages} {2677} (\bibinfo {year} {1994})}\BibitemShut
  {NoStop}%
\bibitem [{\citenamefont {Xie}\ \emph {et~al.}(1996)\citenamefont {Xie},
  \citenamefont {Liu}, \citenamefont {Sundaram},\ and\ \citenamefont
  {Niu}}]{XieLiu1996}%
  \BibitemOpen
  \bibfield  {author} {\bibinfo {author} {\bibfnamefont {X.~C.}\ \bibnamefont
  {Xie}}, \bibinfo {author} {\bibfnamefont {D.~Z.}\ \bibnamefont {Liu}},
  \bibinfo {author} {\bibfnamefont {B.}~\bibnamefont {Sundaram}}, \ and\
  \bibinfo {author} {\bibfnamefont {Q.}~\bibnamefont {Niu}},\ }\href {\doibase
  10.1103/PhysRevB.54.4966} {\bibfield  {journal} {\bibinfo  {journal} {Phys.
  Rev. B}\ }\textbf {\bibinfo {volume} {54}},\ \bibinfo {pages} {4966}
  (\bibinfo {year} {1996})}\BibitemShut {NoStop}%
\bibitem [{\citenamefont {Liu}\ \emph {et~al.}(1996)\citenamefont {Liu},
  \citenamefont {Xie},\ and\ \citenamefont {Niu}}]{LiuXieNiu1996}%
  \BibitemOpen
  \bibfield  {author} {\bibinfo {author} {\bibfnamefont {D.~Z.}\ \bibnamefont
  {Liu}}, \bibinfo {author} {\bibfnamefont {X.~C.}\ \bibnamefont {Xie}}, \ and\
  \bibinfo {author} {\bibfnamefont {Q.}~\bibnamefont {Niu}},\ }\href {\doibase
  10.1103/PhysRevLett.76.975} {\bibfield  {journal} {\bibinfo  {journal} {Phys.
  Rev. Lett.}\ }\textbf {\bibinfo {volume} {76}},\ \bibinfo {pages} {975}
  (\bibinfo {year} {1996})}\BibitemShut {NoStop}%
\bibitem [{\citenamefont {Yang}\ and\ \citenamefont
  {Bhatt}(1996)}]{YangBhatt1996}%
  \BibitemOpen
  \bibfield  {author} {\bibinfo {author} {\bibfnamefont {K.}~\bibnamefont
  {Yang}}\ and\ \bibinfo {author} {\bibfnamefont {R.~N.}\ \bibnamefont
  {Bhatt}},\ }\href {\doibase 10.1103/PhysRevLett.76.1316} {\bibfield
  {journal} {\bibinfo  {journal} {Phys. Rev. Lett.}\ }\textbf {\bibinfo
  {volume} {76}},\ \bibinfo {pages} {1316} (\bibinfo {year}
  {1996})}\BibitemShut {NoStop}%
\bibitem [{\citenamefont {Wan}\ and\ \citenamefont
  {Bhatt}(2001)}]{WanBhatt2001}%
  \BibitemOpen
  \bibfield  {author} {\bibinfo {author} {\bibfnamefont {X.}~\bibnamefont
  {Wan}}\ and\ \bibinfo {author} {\bibfnamefont {R.~N.}\ \bibnamefont
  {Bhatt}},\ }\href {\doibase 10.1103/PhysRevB.64.201313} {\bibfield  {journal}
  {\bibinfo  {journal} {Phys. Rev. B}\ }\textbf {\bibinfo {volume} {64}},\
  \bibinfo {pages} {201313} (\bibinfo {year} {2001})}\BibitemShut {NoStop}%
\bibitem [{\citenamefont {Hubbard}(1963)}]{Hubbard63}%
  \BibitemOpen
  \bibfield  {author} {\bibinfo {author} {\bibfnamefont {J.}~\bibnamefont
  {Hubbard}},\ }\href {\doibase 10.1098/rspa.1963.0204} {\bibfield  {journal}
  {\bibinfo  {journal} {Proceedings of the Royal Society of London A:
  Mathematical, Physical and Engineering Sciences}\ }\textbf {\bibinfo {volume}
  {276}},\ \bibinfo {pages} {238} (\bibinfo {year} {1963})}\BibitemShut
  {NoStop}%
\bibitem [{\citenamefont {Gutzwiller}(1965)}]{Gutzwiller65}%
  \BibitemOpen
  \bibfield  {author} {\bibinfo {author} {\bibfnamefont {M.~C.}\ \bibnamefont
  {Gutzwiller}},\ }\href {\doibase 10.1103/PhysRev.137.A1726} {\bibfield
  {journal} {\bibinfo  {journal} {Phys. Rev.}\ }\textbf {\bibinfo {volume}
  {137}},\ \bibinfo {pages} {A1726} (\bibinfo {year} {1965})}\BibitemShut
  {NoStop}%
\bibitem [{\citenamefont {Anderson}(1961)}]{Anderson61}%
  \BibitemOpen
  \bibfield  {author} {\bibinfo {author} {\bibfnamefont {P.~W.}\ \bibnamefont
  {Anderson}},\ }\href {\doibase 10.1103/PhysRev.124.41} {\bibfield  {journal}
  {\bibinfo  {journal} {Phys. Rev.}\ }\textbf {\bibinfo {volume} {124}},\
  \bibinfo {pages} {41} (\bibinfo {year} {1961})}\BibitemShut {NoStop}%
\bibitem [{\citenamefont {Rice}\ and\ \citenamefont
  {Ueda}(1985)}]{RiceUeda1985}%
  \BibitemOpen
  \bibfield  {author} {\bibinfo {author} {\bibfnamefont {T.~M.}\ \bibnamefont
  {Rice}}\ and\ \bibinfo {author} {\bibfnamefont {K.}~\bibnamefont {Ueda}},\
  }\href {\doibase 10.1103/PhysRevLett.55.995} {\bibfield  {journal} {\bibinfo
  {journal} {Phys. Rev. Lett.}\ }\textbf {\bibinfo {volume} {55}},\ \bibinfo
  {pages} {995} (\bibinfo {year} {1985})}\BibitemShut {NoStop}%
\bibitem [{\citenamefont {Jarrell}(1995)}]{Jarrell1995}%
  \BibitemOpen
  \bibfield  {author} {\bibinfo {author} {\bibfnamefont {M.}~\bibnamefont
  {Jarrell}},\ }\href {\doibase 10.1103/PhysRevB.51.7429} {\bibfield  {journal}
  {\bibinfo  {journal} {Phys. Rev. B}\ }\textbf {\bibinfo {volume} {51}},\
  \bibinfo {pages} {7429} (\bibinfo {year} {1995})}\BibitemShut {NoStop}%
\bibitem [{\citenamefont {Wilson}(1975)}]{Wilson75}%
  \BibitemOpen
  \bibfield  {author} {\bibinfo {author} {\bibfnamefont {K.~G.}\ \bibnamefont
  {Wilson}},\ }\href {\doibase 10.1103/RevModPhys.47.773} {\bibfield  {journal}
  {\bibinfo  {journal} {Rev. Mod. Phys.}\ }\textbf {\bibinfo {volume} {47}},\
  \bibinfo {pages} {773} (\bibinfo {year} {1975})}\BibitemShut {NoStop}%
\bibitem [{\citenamefont {Kondo}(1964)}]{Kondo64}%
  \BibitemOpen
  \bibfield  {author} {\bibinfo {author} {\bibfnamefont {J.}~\bibnamefont
  {Kondo}},\ }\href {\doibase 10.1143/PTP.32.37} {\bibfield  {journal}
  {\bibinfo  {journal} {Progress of Theoretical Physics}\ }\textbf {\bibinfo
  {volume} {32}},\ \bibinfo {pages} {37} (\bibinfo {year} {1964})}\BibitemShut
  {NoStop}%
\bibitem [{\citenamefont {Anderson}\ \emph {et~al.}(1970)\citenamefont
  {Anderson}, \citenamefont {Yuval},\ and\ \citenamefont
  {Hamann}}]{Anderson70}%
  \BibitemOpen
  \bibfield  {author} {\bibinfo {author} {\bibfnamefont {P.~W.}\ \bibnamefont
  {Anderson}}, \bibinfo {author} {\bibfnamefont {G.}~\bibnamefont {Yuval}}, \
  and\ \bibinfo {author} {\bibfnamefont {D.~R.}\ \bibnamefont {Hamann}},\
  }\href {\doibase 10.1103/PhysRevB.1.4464} {\bibfield  {journal} {\bibinfo
  {journal} {Phys. Rev. B}\ }\textbf {\bibinfo {volume} {1}},\ \bibinfo {pages}
  {4464} (\bibinfo {year} {1970})}\BibitemShut {NoStop}%
\bibitem [{\citenamefont {Krishnamurthy}\ \emph {et~al.}(1980)\citenamefont
  {Krishnamurthy}, \citenamefont {Wilkins},\ and\ \citenamefont
  {Wilson}}]{Krishnamurthy80}%
  \BibitemOpen
  \bibfield  {author} {\bibinfo {author} {\bibfnamefont {H.~R.}\ \bibnamefont
  {Krishnamurthy}}, \bibinfo {author} {\bibfnamefont {J.~W.}\ \bibnamefont
  {Wilkins}}, \ and\ \bibinfo {author} {\bibfnamefont {K.~G.}\ \bibnamefont
  {Wilson}},\ }\href {\doibase 10.1103/PhysRevB.21.1003} {\bibfield  {journal}
  {\bibinfo  {journal} {Phys. Rev. B}\ }\textbf {\bibinfo {volume} {21}},\
  \bibinfo {pages} {1003} (\bibinfo {year} {1980})}\BibitemShut {NoStop}%
\bibitem [{\citenamefont {Landau}(1957)}]{Landau57}%
  \BibitemOpen
  \bibfield  {author} {\bibinfo {author} {\bibfnamefont {L.}~\bibnamefont
  {Landau}},\ }\href@noop {} {\bibfield  {journal} {\bibinfo  {journal} {Soviet
  Physics Jetp-Ussr}\ }\textbf {\bibinfo {volume} {3}},\ \bibinfo {pages} {920}
  (\bibinfo {year} {1957})}\BibitemShut {NoStop}%
\bibitem [{\citenamefont {Baym}\ and\ \citenamefont
  {Pethick}(2008)}]{BaymPethick}%
  \BibitemOpen
  \bibfield  {author} {\bibinfo {author} {\bibfnamefont {G.}~\bibnamefont
  {Baym}}\ and\ \bibinfo {author} {\bibfnamefont {C.}~\bibnamefont {Pethick}},\
  }\href@noop {} {\emph {\bibinfo {title} {Landau Fermi-liquid theory: concepts
  and applications}}}\ (\bibinfo  {publisher} {John Wiley \& Sons},\ \bibinfo
  {year} {2008})\BibitemShut {NoStop}%
\bibitem [{\citenamefont {Shankar}(1994)}]{Shankar94}%
  \BibitemOpen
  \bibfield  {author} {\bibinfo {author} {\bibfnamefont {R.}~\bibnamefont
  {Shankar}},\ }\href {\doibase 10.1103/RevModPhys.66.129} {\bibfield
  {journal} {\bibinfo  {journal} {Rev. Mod. Phys.}\ }\textbf {\bibinfo {volume}
  {66}},\ \bibinfo {pages} {129} (\bibinfo {year} {1994})}\BibitemShut
  {NoStop}%
\bibitem [{\citenamefont {Dunlap}\ \emph {et~al.}(1990)\citenamefont {Dunlap},
  \citenamefont {Wu},\ and\ \citenamefont {Phillips}}]{Phillips90}%
  \BibitemOpen
  \bibfield  {author} {\bibinfo {author} {\bibfnamefont {D.~H.}\ \bibnamefont
  {Dunlap}}, \bibinfo {author} {\bibfnamefont {H.-L.}\ \bibnamefont {Wu}}, \
  and\ \bibinfo {author} {\bibfnamefont {P.~W.}\ \bibnamefont {Phillips}},\
  }\href {\doibase 10.1103/PhysRevLett.65.88} {\bibfield  {journal} {\bibinfo
  {journal} {Phys. Rev. Lett.}\ }\textbf {\bibinfo {volume} {65}},\ \bibinfo
  {pages} {88} (\bibinfo {year} {1990})}\BibitemShut {NoStop}%
\bibitem [{\citenamefont {de~Moura}\ and\ \citenamefont
  {Lyra}(1998)}]{MouraLyra1998}%
  \BibitemOpen
  \bibfield  {author} {\bibinfo {author} {\bibfnamefont {F.~A. B.~F.}\
  \bibnamefont {de~Moura}}\ and\ \bibinfo {author} {\bibfnamefont {M.~L.}\
  \bibnamefont {Lyra}},\ }\href {\doibase 10.1103/PhysRevLett.81.3735}
  {\bibfield  {journal} {\bibinfo  {journal} {Phys. Rev. Lett.}\ }\textbf
  {\bibinfo {volume} {81}},\ \bibinfo {pages} {3735} (\bibinfo {year}
  {1998})}\BibitemShut {NoStop}%
\bibitem [{\citenamefont {Izrailev}\ and\ \citenamefont
  {Krokhin}(1999)}]{Izrailev99}%
  \BibitemOpen
  \bibfield  {author} {\bibinfo {author} {\bibfnamefont {F.~M.}\ \bibnamefont
  {Izrailev}}\ and\ \bibinfo {author} {\bibfnamefont {A.~A.}\ \bibnamefont
  {Krokhin}},\ }\href {\doibase 10.1103/PhysRevLett.82.4062} {\bibfield
  {journal} {\bibinfo  {journal} {Phys. Rev. Lett.}\ }\textbf {\bibinfo
  {volume} {82}},\ \bibinfo {pages} {4062} (\bibinfo {year}
  {1999})}\BibitemShut {NoStop}%
\bibitem [{\citenamefont {de~Moura}\ and\ \citenamefont
  {Lyra}(1999)}]{Moura99}%
  \BibitemOpen
  \bibfield  {author} {\bibinfo {author} {\bibfnamefont {F.~A.}\ \bibnamefont
  {de~Moura}}\ and\ \bibinfo {author} {\bibfnamefont {M.~L.}\ \bibnamefont
  {Lyra}},\ }\href {\doibase https://doi.org/10.1016/S0378-4371(98)00632-3}
  {\bibfield  {journal} {\bibinfo  {journal} {Physica A: Statistical Mechanics
  and its Applications}\ }\textbf {\bibinfo {volume} {266}},\ \bibinfo {pages}
  {465 } (\bibinfo {year} {1999})}\BibitemShut {NoStop}%
\bibitem [{\citenamefont {Aubry}\ and\ \citenamefont
  {Andr{\'e}}(1980)}]{AubryAndre80}%
  \BibitemOpen
  \bibfield  {author} {\bibinfo {author} {\bibfnamefont {S.}~\bibnamefont
  {Aubry}}\ and\ \bibinfo {author} {\bibfnamefont {G.}~\bibnamefont
  {Andr{\'e}}},\ }\href@noop {} {\bibfield  {journal} {\bibinfo  {journal}
  {Ann. Israel Phys. Soc}\ }\textbf {\bibinfo {volume} {3}},\ \bibinfo {pages}
  {18} (\bibinfo {year} {1980})}\BibitemShut {NoStop}%
\bibitem [{\citenamefont {Zhou}\ and\ \citenamefont {Bhatt}(2003)}]{ZhouBhatt}%
  \BibitemOpen
  \bibfield  {author} {\bibinfo {author} {\bibfnamefont {C.}~\bibnamefont
  {Zhou}}\ and\ \bibinfo {author} {\bibfnamefont {R.~N.}\ \bibnamefont
  {Bhatt}},\ }\href {\doibase 10.1103/PhysRevB.68.045101} {\bibfield  {journal}
  {\bibinfo  {journal} {Phys. Rev. B}\ }\textbf {\bibinfo {volume} {68}},\
  \bibinfo {pages} {045101} (\bibinfo {year} {2003})}\BibitemShut {NoStop}%
\bibitem [{\citenamefont {Moure}\ \emph {et~al.}(2018)\citenamefont {Moure},
  \citenamefont {Lee}, \citenamefont {Haas}, \citenamefont {Bhatt},\ and\
  \citenamefont {Kettemann}}]{Mourearxiv}%
  \BibitemOpen
  \bibfield  {author} {\bibinfo {author} {\bibfnamefont {N.}~\bibnamefont
  {Moure}}, \bibinfo {author} {\bibfnamefont {H.-Y.}\ \bibnamefont {Lee}},
  \bibinfo {author} {\bibfnamefont {S.}~\bibnamefont {Haas}}, \bibinfo {author}
  {\bibfnamefont {R.~N.}\ \bibnamefont {Bhatt}}, \ and\ \bibinfo {author}
  {\bibfnamefont {S.}~\bibnamefont {Kettemann}},\ }\href {\doibase
  10.1103/PhysRevB.97.014206} {\bibfield  {journal} {\bibinfo  {journal} {Phys.
  Rev. B}\ }\textbf {\bibinfo {volume} {97}},\ \bibinfo {pages} {014206}
  (\bibinfo {year} {2018})}\BibitemShut {NoStop}%
\bibitem [{\citenamefont {Geraedts}\ \emph {et~al.}(2017)\citenamefont
  {Geraedts}, \citenamefont {Bhatt},\ and\ \citenamefont
  {Nandkishore}}]{Geraedts17}%
  \BibitemOpen
  \bibfield  {author} {\bibinfo {author} {\bibfnamefont {S.~D.}\ \bibnamefont
  {Geraedts}}, \bibinfo {author} {\bibfnamefont {R.~N.}\ \bibnamefont {Bhatt}},
  \ and\ \bibinfo {author} {\bibfnamefont {R.}~\bibnamefont {Nandkishore}},\
  }\href {\doibase 10.1103/PhysRevB.95.064204} {\bibfield  {journal} {\bibinfo
  {journal} {Phys. Rev. B}\ }\textbf {\bibinfo {volume} {95}},\ \bibinfo
  {pages} {064204} (\bibinfo {year} {2017})}\BibitemShut {NoStop}%
\bibitem [{\citenamefont {Pipek}\ and\ \citenamefont
  {Varga}(1992)}]{Varga1992}%
  \BibitemOpen
  \bibfield  {author} {\bibinfo {author} {\bibfnamefont {J.}~\bibnamefont
  {Pipek}}\ and\ \bibinfo {author} {\bibfnamefont {I.}~\bibnamefont {Varga}},\
  }\href {\doibase 10.1103/PhysRevA.46.3148} {\bibfield  {journal} {\bibinfo
  {journal} {Phys. Rev. A}\ }\textbf {\bibinfo {volume} {46}},\ \bibinfo
  {pages} {3148} (\bibinfo {year} {1992})}\BibitemShut {NoStop}%
\bibitem [{Note1()}]{Note1}%
  \BibitemOpen
  \bibinfo {note} {This is just the customary definition $M_2 = \protect
  \mathaccentV {bar}016{x^2} - \protect \mathaccentV {bar}016{x}^2$ where
  $\protect \mathaccentV {bar}016{x} = \int \limits _{-\infty }^{\infty } (x-a)
  p(x) \protect \mathrm {d}x$ measured with respect to arbitrary origin $a$.
  While $\protect \mathaccentV {bar}016{x}$ depends on the choice of origin,
  $M_2$ does not.}\BibitemShut {Stop}%
\bibitem [{\citenamefont {Heyer}(1981)}]{Heyer81}%
  \BibitemOpen
  \bibfield  {author} {\bibinfo {author} {\bibfnamefont {H.}~\bibnamefont
  {Heyer}},\ }\href@noop {} {\bibfield  {journal} {\bibinfo  {journal}
  {Internat. J. Math. Math. Sci.}\ }\textbf {\bibinfo {volume} {4}},\ \bibinfo
  {pages} {1} (\bibinfo {year} {1981})}\BibitemShut {NoStop}%
\bibitem [{\citenamefont {L\'{e}vy}(1939)}]{Levy39}%
  \BibitemOpen
  \bibfield  {author} {\bibinfo {author} {\bibfnamefont {P.}~\bibnamefont
  {L\'{e}vy}},\ }\href@noop {} {\bibfield  {journal} {\bibinfo  {journal}
  {Bull. Soc. Math. France}\ }\textbf {\bibinfo {volume} {67}},\ \bibinfo
  {pages} {1} (\bibinfo {year} {1939})}\BibitemShut {NoStop}%
\bibitem [{Note2()}]{Note2}%
  \BibitemOpen
  \bibinfo {note} {As an example, consider a wavefunction with support only on
  two sites $|\psi _1|^2 = |\psi _3|^2 = 1/2$. Its second moment $M_2 = 1$. The
  same wavefunction, with a shifted origin, is $|\psi _2|^2 = |\psi _N|^2 =
  1/2$. The second moment of this wavefunction is $N/2-1$. This example
  illustrates the need to choose an appropriate origin to minimize $M_2$.
  Physically this choice implies that the boundary sites $n=1$ and $n=N$ are as
  far as possible from the bulk of the wavefunction.}\BibitemShut {Stop}%
\bibitem [{\citenamefont {Resta}\ and\ \citenamefont
  {Sorella}(1999)}]{Raffaele1999}%
  \BibitemOpen
  \bibfield  {author} {\bibinfo {author} {\bibfnamefont {R.}~\bibnamefont
  {Resta}}\ and\ \bibinfo {author} {\bibfnamefont {S.}~\bibnamefont
  {Sorella}},\ }\href {\doibase 10.1103/PhysRevLett.82.370} {\bibfield
  {journal} {\bibinfo  {journal} {Phys. Rev. Lett.}\ }\textbf {\bibinfo
  {volume} {82}},\ \bibinfo {pages} {370} (\bibinfo {year} {1999})}\BibitemShut
  {NoStop}%
\bibitem [{Note3()}]{Note3}%
  \BibitemOpen
  \bibinfo {note} {They do not equal each other, as may be expected for a
  perfectly exponentially localized wavefunction, as in table \ref {tabxi}.
  This is due to the fact that Anderson localized wavefunctions are
  exponentials modulated by a sinusoidal component (an example of an Anderson
  localized wavefunction is given in Fig.\ \ref {figAndW1}). In such a case,
  the wavefunction is not purely monotonic and typically $\xi _{M_2} > \xi
  _{IPR}$.}\BibitemShut {Stop}%
\bibitem [{\citenamefont {Edwards}\ and\ \citenamefont
  {Thouless}(1972)}]{EdwardsThouless1971}%
  \BibitemOpen
  \bibfield  {author} {\bibinfo {author} {\bibfnamefont {J.~T.}\ \bibnamefont
  {Edwards}}\ and\ \bibinfo {author} {\bibfnamefont {D.~J.}\ \bibnamefont
  {Thouless}},\ }\href {http://stacks.iop.org/0022-3719/5/i=8/a=007} {\bibfield
   {journal} {\bibinfo  {journal} {Journal of Physics C: Solid State Physics}\
  }\textbf {\bibinfo {volume} {5}},\ \bibinfo {pages} {807} (\bibinfo {year}
  {1972})}\BibitemShut {NoStop}%
\bibitem [{\citenamefont {Johri}\ and\ \citenamefont
  {Bhatt}(2012{\natexlab{a}})}]{JohriBhattPRL12}%
  \BibitemOpen
  \bibfield  {author} {\bibinfo {author} {\bibfnamefont {S.}~\bibnamefont
  {Johri}}\ and\ \bibinfo {author} {\bibfnamefont {R.~N.}\ \bibnamefont
  {Bhatt}},\ }\href {\doibase 10.1103/PhysRevLett.109.076402} {\bibfield
  {journal} {\bibinfo  {journal} {Phys. Rev. Lett.}\ }\textbf {\bibinfo
  {volume} {109}},\ \bibinfo {pages} {076402} (\bibinfo {year}
  {2012}{\natexlab{a}})}\BibitemShut {NoStop}%
\bibitem [{\citenamefont {Johri}\ and\ \citenamefont
  {Bhatt}(2012{\natexlab{b}})}]{JohriBhattPRB12}%
  \BibitemOpen
  \bibfield  {author} {\bibinfo {author} {\bibfnamefont {S.}~\bibnamefont
  {Johri}}\ and\ \bibinfo {author} {\bibfnamefont {R.~N.}\ \bibnamefont
  {Bhatt}},\ }\href {\doibase 10.1103/PhysRevB.86.125140} {\bibfield  {journal}
  {\bibinfo  {journal} {Phys. Rev. B}\ }\textbf {\bibinfo {volume} {86}},\
  \bibinfo {pages} {125140} (\bibinfo {year} {2012}{\natexlab{b}})}\BibitemShut
  {NoStop}%
\bibitem [{\citenamefont {Dasgupta}\ and\ \citenamefont
  {Gupta}(2003)}]{JohnsonLindenstrauss}%
  \BibitemOpen
  \bibfield  {author} {\bibinfo {author} {\bibfnamefont {S.}~\bibnamefont
  {Dasgupta}}\ and\ \bibinfo {author} {\bibfnamefont {A.}~\bibnamefont
  {Gupta}},\ }\href {\doibase 10.1002/rsa.10073} {\bibfield  {journal}
  {\bibinfo  {journal} {Random Structures and Algorithms}\ }\textbf {\bibinfo
  {volume} {22}},\ \bibinfo {pages} {60} (\bibinfo {year} {2003})}\BibitemShut
  {NoStop}%
\bibitem [{\citenamefont {Lakshminarayan}\ \emph {et~al.}(2008)\citenamefont
  {Lakshminarayan}, \citenamefont {Tomsovic}, \citenamefont {Bohigas},\ and\
  \citenamefont {Majumdar}}]{Lakshminarayan2008}%
  \BibitemOpen
  \bibfield  {author} {\bibinfo {author} {\bibfnamefont {A.}~\bibnamefont
  {Lakshminarayan}}, \bibinfo {author} {\bibfnamefont {S.}~\bibnamefont
  {Tomsovic}}, \bibinfo {author} {\bibfnamefont {O.}~\bibnamefont {Bohigas}}, \
  and\ \bibinfo {author} {\bibfnamefont {S.~N.}\ \bibnamefont {Majumdar}},\
  }\href {\doibase 10.1103/PhysRevLett.100.044103} {\bibfield  {journal}
  {\bibinfo  {journal} {Phys. Rev. Lett.}\ }\textbf {\bibinfo {volume} {100}},\
  \bibinfo {pages} {044103} (\bibinfo {year} {2008})}\BibitemShut {NoStop}%
\bibitem [{\citenamefont {Huckestein}(1995)}]{Huckestein1995}%
  \BibitemOpen
  \bibfield  {author} {\bibinfo {author} {\bibfnamefont {B.}~\bibnamefont
  {Huckestein}},\ }\href {\doibase 10.1103/RevModPhys.67.357} {\bibfield
  {journal} {\bibinfo  {journal} {Rev. Mod. Phys.}\ }\textbf {\bibinfo {volume}
  {67}},\ \bibinfo {pages} {357} (\bibinfo {year} {1995})}\BibitemShut
  {NoStop}%
\bibitem [{\citenamefont {Rodriguez}\ \emph {et~al.}(2011)\citenamefont
  {Rodriguez}, \citenamefont {Vasquez}, \citenamefont {Slevin},\ and\
  \citenamefont {R\"omer}}]{Rodriguezetal11}%
  \BibitemOpen
  \bibfield  {author} {\bibinfo {author} {\bibfnamefont {A.}~\bibnamefont
  {Rodriguez}}, \bibinfo {author} {\bibfnamefont {L.~J.}\ \bibnamefont
  {Vasquez}}, \bibinfo {author} {\bibfnamefont {K.}~\bibnamefont {Slevin}}, \
  and\ \bibinfo {author} {\bibfnamefont {R.~A.}\ \bibnamefont {R\"omer}},\
  }\href {\doibase 10.1103/PhysRevB.84.134209} {\bibfield  {journal} {\bibinfo
  {journal} {Phys. Rev. B}\ }\textbf {\bibinfo {volume} {84}},\ \bibinfo
  {pages} {134209} (\bibinfo {year} {2011})}\BibitemShut {NoStop}%
\bibitem [{\citenamefont {Halsey}\ \emph {et~al.}(1986)\citenamefont {Halsey},
  \citenamefont {Jensen}, \citenamefont {Kadanoff}, \citenamefont {Procaccia},\
  and\ \citenamefont {Shraiman}}]{Halsey86}%
  \BibitemOpen
  \bibfield  {author} {\bibinfo {author} {\bibfnamefont {T.~C.}\ \bibnamefont
  {Halsey}}, \bibinfo {author} {\bibfnamefont {M.~H.}\ \bibnamefont {Jensen}},
  \bibinfo {author} {\bibfnamefont {L.~P.}\ \bibnamefont {Kadanoff}}, \bibinfo
  {author} {\bibfnamefont {I.}~\bibnamefont {Procaccia}}, \ and\ \bibinfo
  {author} {\bibfnamefont {B.~I.}\ \bibnamefont {Shraiman}},\ }\href {\doibase
  10.1103/PhysRevA.33.1141} {\bibfield  {journal} {\bibinfo  {journal} {Phys.
  Rev. A}\ }\textbf {\bibinfo {volume} {33}},\ \bibinfo {pages} {1141}
  (\bibinfo {year} {1986})}\BibitemShut {NoStop}%
\bibitem [{\citenamefont {Pook}\ and\ \citenamefont
  {Jan{\ss}en}(1991)}]{Pook91}%
  \BibitemOpen
  \bibfield  {author} {\bibinfo {author} {\bibfnamefont {W.}~\bibnamefont
  {Pook}}\ and\ \bibinfo {author} {\bibfnamefont {M.}~\bibnamefont
  {Jan{\ss}en}},\ }\href {\doibase 10.1007/BF01324339} {\bibfield  {journal}
  {\bibinfo  {journal} {Zeitschrift f{\"u}r Physik B Condensed Matter}\
  }\textbf {\bibinfo {volume} {82}},\ \bibinfo {pages} {295} (\bibinfo {year}
  {1991})}\BibitemShut {NoStop}%
\bibitem [{\citenamefont {Huckestein}\ \emph {et~al.}(1992)\citenamefont
  {Huckestein}, \citenamefont {Kramer},\ and\ \citenamefont
  {Schweitzer}}]{Huckestein92}%
  \BibitemOpen
  \bibfield  {author} {\bibinfo {author} {\bibfnamefont {B.}~\bibnamefont
  {Huckestein}}, \bibinfo {author} {\bibfnamefont {B.}~\bibnamefont {Kramer}},
  \ and\ \bibinfo {author} {\bibfnamefont {L.}~\bibnamefont {Schweitzer}},\
  }\href {\doibase http://dx.doi.org/10.1016/0039-6028(92)90320-6} {\bibfield
  {journal} {\bibinfo  {journal} {Surface Science}\ }\textbf {\bibinfo {volume}
  {263}},\ \bibinfo {pages} {125 } (\bibinfo {year} {1992})}\BibitemShut
  {NoStop}%
\bibitem [{\citenamefont {Klesse}\ and\ \citenamefont
  {Metzler}(1995)}]{Klesse95}%
  \BibitemOpen
  \bibfield  {author} {\bibinfo {author} {\bibfnamefont {R.}~\bibnamefont
  {Klesse}}\ and\ \bibinfo {author} {\bibfnamefont {M.}~\bibnamefont
  {Metzler}},\ }\href@noop {} {\bibfield  {journal} {\bibinfo  {journal} {EPL
  (Europhysics Letters)}\ }\textbf {\bibinfo {volume} {32}},\ \bibinfo {pages}
  {229} (\bibinfo {year} {1995})}\BibitemShut {NoStop}%
\bibitem [{\citenamefont {Schreiber}\ and\ \citenamefont
  {Grussbach}(1991)}]{Schreiber1991}%
  \BibitemOpen
  \bibfield  {author} {\bibinfo {author} {\bibfnamefont {M.}~\bibnamefont
  {Schreiber}}\ and\ \bibinfo {author} {\bibfnamefont {H.}~\bibnamefont
  {Grussbach}},\ }\href {\doibase 10.1103/PhysRevLett.67.607} {\bibfield
  {journal} {\bibinfo  {journal} {Phys. Rev. Lett.}\ }\textbf {\bibinfo
  {volume} {67}},\ \bibinfo {pages} {607} (\bibinfo {year} {1991})}\BibitemShut
  {NoStop}%
\bibitem [{\citenamefont {Grussbach}\ and\ \citenamefont
  {Schreiber}(1992)}]{Grussbach1992}%
  \BibitemOpen
  \bibfield  {author} {\bibinfo {author} {\bibfnamefont {H.}~\bibnamefont
  {Grussbach}}\ and\ \bibinfo {author} {\bibfnamefont {M.}~\bibnamefont
  {Schreiber}},\ }\href {\doibase https://doi.org/10.1016/0378-4371(92)90556-6}
  {\bibfield  {journal} {\bibinfo  {journal} {Physica A: Statistical Mechanics
  and its Applications}\ }\textbf {\bibinfo {volume} {191}},\ \bibinfo {pages}
  {394 } (\bibinfo {year} {1992})}\BibitemShut {NoStop}%
\bibitem [{\citenamefont {Vasquez}\ \emph {et~al.}(2008)\citenamefont
  {Vasquez}, \citenamefont {Rodriguez},\ and\ \citenamefont
  {R\"omer}}]{Vasquezetal08}%
  \BibitemOpen
  \bibfield  {author} {\bibinfo {author} {\bibfnamefont {L.~J.}\ \bibnamefont
  {Vasquez}}, \bibinfo {author} {\bibfnamefont {A.}~\bibnamefont {Rodriguez}},
  \ and\ \bibinfo {author} {\bibfnamefont {R.~A.}\ \bibnamefont {R\"omer}},\
  }\href {\doibase 10.1103/PhysRevB.78.195106} {\bibfield  {journal} {\bibinfo
  {journal} {Phys. Rev. B}\ }\textbf {\bibinfo {volume} {78}},\ \bibinfo
  {pages} {195106} (\bibinfo {year} {2008})}\BibitemShut {NoStop}%
\bibitem [{\citenamefont {Rodriguez}\ \emph {et~al.}(2008)\citenamefont
  {Rodriguez}, \citenamefont {Vasquez},\ and\ \citenamefont
  {R\"omer}}]{Vasquezetal08b}%
  \BibitemOpen
  \bibfield  {author} {\bibinfo {author} {\bibfnamefont {A.}~\bibnamefont
  {Rodriguez}}, \bibinfo {author} {\bibfnamefont {L.~J.}\ \bibnamefont
  {Vasquez}}, \ and\ \bibinfo {author} {\bibfnamefont {R.~A.}\ \bibnamefont
  {R\"omer}},\ }\href {\doibase 10.1103/PhysRevB.78.195107} {\bibfield
  {journal} {\bibinfo  {journal} {Phys. Rev. B}\ }\textbf {\bibinfo {volume}
  {78}},\ \bibinfo {pages} {195107} (\bibinfo {year} {2008})}\BibitemShut
  {NoStop}%
\bibitem [{\citenamefont {Obuse}\ \emph {et~al.}(2008)\citenamefont {Obuse},
  \citenamefont {Subramaniam}, \citenamefont {Furusaki}, \citenamefont
  {Gruzberg},\ and\ \citenamefont {Ludwig}}]{Obuse2008}%
  \BibitemOpen
  \bibfield  {author} {\bibinfo {author} {\bibfnamefont {H.}~\bibnamefont
  {Obuse}}, \bibinfo {author} {\bibfnamefont {A.~R.}\ \bibnamefont
  {Subramaniam}}, \bibinfo {author} {\bibfnamefont {A.}~\bibnamefont
  {Furusaki}}, \bibinfo {author} {\bibfnamefont {I.~A.}\ \bibnamefont
  {Gruzberg}}, \ and\ \bibinfo {author} {\bibfnamefont {A.~W.~W.}\ \bibnamefont
  {Ludwig}},\ }\href {\doibase 10.1103/PhysRevLett.101.116802} {\bibfield
  {journal} {\bibinfo  {journal} {Phys. Rev. Lett.}\ }\textbf {\bibinfo
  {volume} {101}},\ \bibinfo {pages} {116802} (\bibinfo {year}
  {2008})}\BibitemShut {NoStop}%
\bibitem [{\citenamefont {Evers}\ \emph {et~al.}(2008)\citenamefont {Evers},
  \citenamefont {Mildenberger},\ and\ \citenamefont {Mirlin}}]{Evers2008}%
  \BibitemOpen
  \bibfield  {author} {\bibinfo {author} {\bibfnamefont {F.}~\bibnamefont
  {Evers}}, \bibinfo {author} {\bibfnamefont {A.}~\bibnamefont {Mildenberger}},
  \ and\ \bibinfo {author} {\bibfnamefont {A.~D.}\ \bibnamefont {Mirlin}},\
  }\href {\doibase 10.1103/PhysRevLett.101.116803} {\bibfield  {journal}
  {\bibinfo  {journal} {Phys. Rev. Lett.}\ }\textbf {\bibinfo {volume} {101}},\
  \bibinfo {pages} {116803} (\bibinfo {year} {2008})}\BibitemShut {NoStop}%
\bibitem [{\citenamefont {Rodriguez}\ \emph {et~al.}(2010)\citenamefont
  {Rodriguez}, \citenamefont {Vasquez}, \citenamefont {Slevin},\ and\
  \citenamefont {R\"omer}}]{RodriguezSlevinPRL2010}%
  \BibitemOpen
  \bibfield  {author} {\bibinfo {author} {\bibfnamefont {A.}~\bibnamefont
  {Rodriguez}}, \bibinfo {author} {\bibfnamefont {L.~J.}\ \bibnamefont
  {Vasquez}}, \bibinfo {author} {\bibfnamefont {K.}~\bibnamefont {Slevin}}, \
  and\ \bibinfo {author} {\bibfnamefont {R.~A.}\ \bibnamefont {R\"omer}},\
  }\href {\doibase 10.1103/PhysRevLett.105.046403} {\bibfield  {journal}
  {\bibinfo  {journal} {Phys. Rev. Lett.}\ }\textbf {\bibinfo {volume} {105}},\
  \bibinfo {pages} {046403} (\bibinfo {year} {2010})}\BibitemShut {NoStop}%
\bibitem [{\citenamefont {Lindinger}\ and\ \citenamefont
  {Rodr\'{\i}guez}(2017)}]{Rodriguezarxiv17}%
  \BibitemOpen
  \bibfield  {author} {\bibinfo {author} {\bibfnamefont {J.}~\bibnamefont
  {Lindinger}}\ and\ \bibinfo {author} {\bibfnamefont {A.}~\bibnamefont
  {Rodr\'{\i}guez}},\ }\href {\doibase 10.1103/PhysRevB.96.134202} {\bibfield
  {journal} {\bibinfo  {journal} {Phys. Rev. B}\ }\textbf {\bibinfo {volume}
  {96}},\ \bibinfo {pages} {134202} (\bibinfo {year} {2017})}\BibitemShut
  {NoStop}%
\bibitem [{\citenamefont {Mirlin}\ \emph {et~al.}(2006)\citenamefont {Mirlin},
  \citenamefont {Fyodorov}, \citenamefont {Mildenberger},\ and\ \citenamefont
  {Evers}}]{Mirlin2006}%
  \BibitemOpen
  \bibfield  {author} {\bibinfo {author} {\bibfnamefont {A.~D.}\ \bibnamefont
  {Mirlin}}, \bibinfo {author} {\bibfnamefont {Y.~V.}\ \bibnamefont
  {Fyodorov}}, \bibinfo {author} {\bibfnamefont {A.}~\bibnamefont
  {Mildenberger}}, \ and\ \bibinfo {author} {\bibfnamefont {F.}~\bibnamefont
  {Evers}},\ }\href {\doibase 10.1103/PhysRevLett.97.046803} {\bibfield
  {journal} {\bibinfo  {journal} {Phys. Rev. Lett.}\ }\textbf {\bibinfo
  {volume} {97}},\ \bibinfo {pages} {046803} (\bibinfo {year}
  {2006})}\BibitemShut {NoStop}%
\bibitem [{\citenamefont {Atas}\ \emph {et~al.}(2013)\citenamefont {Atas},
  \citenamefont {Bogomolny}, \citenamefont {Giraud},\ and\ \citenamefont
  {Roux}}]{Atas2013}%
  \BibitemOpen
  \bibfield  {author} {\bibinfo {author} {\bibfnamefont {Y.~Y.}\ \bibnamefont
  {Atas}}, \bibinfo {author} {\bibfnamefont {E.}~\bibnamefont {Bogomolny}},
  \bibinfo {author} {\bibfnamefont {O.}~\bibnamefont {Giraud}}, \ and\ \bibinfo
  {author} {\bibfnamefont {G.}~\bibnamefont {Roux}},\ }\href {\doibase
  10.1103/PhysRevLett.110.084101} {\bibfield  {journal} {\bibinfo  {journal}
  {Phys. Rev. Lett.}\ }\textbf {\bibinfo {volume} {110}},\ \bibinfo {pages}
  {084101} (\bibinfo {year} {2013})}\BibitemShut {NoStop}%
\bibitem [{\citenamefont {Beenakker}(1997)}]{BeenakkerRMP97}%
  \BibitemOpen
  \bibfield  {author} {\bibinfo {author} {\bibfnamefont {C.~W.~J.}\
  \bibnamefont {Beenakker}},\ }\href {\doibase 10.1103/RevModPhys.69.731}
  {\bibfield  {journal} {\bibinfo  {journal} {Rev. Mod. Phys.}\ }\textbf
  {\bibinfo {volume} {69}},\ \bibinfo {pages} {731} (\bibinfo {year}
  {1997})}\BibitemShut {NoStop}%
\bibitem [{\citenamefont {Haake}(2013)}]{HaakeBook}%
  \BibitemOpen
  \bibfield  {author} {\bibinfo {author} {\bibfnamefont {F.}~\bibnamefont
  {Haake}},\ }\href@noop {} {\emph {\bibinfo {title} {Quantum signatures of
  chaos}}},\ Vol.~\bibinfo {volume} {54}\ (\bibinfo  {publisher} {Springer
  Science \& Business Media},\ \bibinfo {year} {2013})\BibitemShut {NoStop}%
\bibitem [{\citenamefont {Oganesyan}\ and\ \citenamefont
  {Huse}(2007)}]{OganesyanHuse07}%
  \BibitemOpen
  \bibfield  {author} {\bibinfo {author} {\bibfnamefont {V.}~\bibnamefont
  {Oganesyan}}\ and\ \bibinfo {author} {\bibfnamefont {D.~A.}\ \bibnamefont
  {Huse}},\ }\href {\doibase 10.1103/PhysRevB.75.155111} {\bibfield  {journal}
  {\bibinfo  {journal} {Phys. Rev. B}\ }\textbf {\bibinfo {volume} {75}},\
  \bibinfo {pages} {155111} (\bibinfo {year} {2007})}\BibitemShut {NoStop}%
\bibitem [{\citenamefont {Pal}\ and\ \citenamefont {Huse}(2010)}]{PalHuse2010}%
  \BibitemOpen
  \bibfield  {author} {\bibinfo {author} {\bibfnamefont {A.}~\bibnamefont
  {Pal}}\ and\ \bibinfo {author} {\bibfnamefont {D.~A.}\ \bibnamefont {Huse}},\
  }\href {\doibase 10.1103/PhysRevB.82.174411} {\bibfield  {journal} {\bibinfo
  {journal} {Phys. Rev. B}\ }\textbf {\bibinfo {volume} {82}},\ \bibinfo
  {pages} {174411} (\bibinfo {year} {2010})}\BibitemShut {NoStop}%
\bibitem [{\citenamefont {Cuevas}\ \emph {et~al.}(2012)\citenamefont {Cuevas},
  \citenamefont {Feigel'Man}, \citenamefont {Ioffe},\ and\ \citenamefont
  {Mezard}}]{Cuevas2012}%
  \BibitemOpen
  \bibfield  {author} {\bibinfo {author} {\bibfnamefont {E.}~\bibnamefont
  {Cuevas}}, \bibinfo {author} {\bibfnamefont {M.}~\bibnamefont {Feigel'Man}},
  \bibinfo {author} {\bibfnamefont {L.}~\bibnamefont {Ioffe}}, \ and\ \bibinfo
  {author} {\bibfnamefont {M.}~\bibnamefont {Mezard}},\ }\href@noop {}
  {\bibfield  {journal} {\bibinfo  {journal} {Nature communications}\ }\textbf
  {\bibinfo {volume} {3}},\ \bibinfo {pages} {1128} (\bibinfo {year}
  {2012})}\BibitemShut {NoStop}%
\bibitem [{\citenamefont {Johri}\ \emph {et~al.}(2015)\citenamefont {Johri},
  \citenamefont {Nandkishore},\ and\ \citenamefont {Bhatt}}]{Johri2015}%
  \BibitemOpen
  \bibfield  {author} {\bibinfo {author} {\bibfnamefont {S.}~\bibnamefont
  {Johri}}, \bibinfo {author} {\bibfnamefont {R.}~\bibnamefont {Nandkishore}},
  \ and\ \bibinfo {author} {\bibfnamefont {R.~N.}\ \bibnamefont {Bhatt}},\
  }\href {\doibase 10.1103/PhysRevLett.114.117401} {\bibfield  {journal}
  {\bibinfo  {journal} {Phys. Rev. Lett.}\ }\textbf {\bibinfo {volume} {114}},\
  \bibinfo {pages} {117401} (\bibinfo {year} {2015})}\BibitemShut {NoStop}%
\bibitem [{\citenamefont {Luitz}\ \emph {et~al.}(2015)\citenamefont {Luitz},
  \citenamefont {Laflorencie},\ and\ \citenamefont {Alet}}]{Luitz2015}%
  \BibitemOpen
  \bibfield  {author} {\bibinfo {author} {\bibfnamefont {D.~J.}\ \bibnamefont
  {Luitz}}, \bibinfo {author} {\bibfnamefont {N.}~\bibnamefont {Laflorencie}},
  \ and\ \bibinfo {author} {\bibfnamefont {F.}~\bibnamefont {Alet}},\ }\href
  {\doibase 10.1103/PhysRevB.91.081103} {\bibfield  {journal} {\bibinfo
  {journal} {Phys. Rev. B}\ }\textbf {\bibinfo {volume} {91}},\ \bibinfo
  {pages} {081103} (\bibinfo {year} {2015})}\BibitemShut {NoStop}%
\bibitem [{\citenamefont {Geraedts}\ and\ \citenamefont
  {Bhatt}(2017)}]{Geraedts2017}%
  \BibitemOpen
  \bibfield  {author} {\bibinfo {author} {\bibfnamefont {S.~D.}\ \bibnamefont
  {Geraedts}}\ and\ \bibinfo {author} {\bibfnamefont {R.~N.}\ \bibnamefont
  {Bhatt}},\ }\href {\doibase 10.1103/PhysRevB.95.054303} {\bibfield  {journal}
  {\bibinfo  {journal} {Phys. Rev. B}\ }\textbf {\bibinfo {volume} {95}},\
  \bibinfo {pages} {054303} (\bibinfo {year} {2017})}\BibitemShut {NoStop}%
\bibitem [{Note4()}]{Note4}%
  \BibitemOpen
  \bibinfo {note} {Such logarithmic dependence has been suggested for the
  multifractality spectrum $f(\alpha )$ at the integer quantum Hall plateau
  transition by R.\ Bondesan, D.\ Wieczorek and M.\ R.\ Zirnbauer, Nuclear
  Physics B \protect \textbf {918}, 52 (2017).}\BibitemShut {Stop}%
\bibitem [{Note5()}]{Note5}%
  \BibitemOpen
  \bibinfo {note} {We are indebted to Kartiek Agarwal for suggesting
  this.}\BibitemShut {Stop}%
\bibitem [{\citenamefont {Mirlin}\ \emph {et~al.}(1996)\citenamefont {Mirlin},
  \citenamefont {Fyodorov}, \citenamefont {Dittes}, \citenamefont {Quezada},\
  and\ \citenamefont {Seligman}}]{MirlinPRBM1996}%
  \BibitemOpen
  \bibfield  {author} {\bibinfo {author} {\bibfnamefont {A.~D.}\ \bibnamefont
  {Mirlin}}, \bibinfo {author} {\bibfnamefont {Y.~V.}\ \bibnamefont
  {Fyodorov}}, \bibinfo {author} {\bibfnamefont {F.-M.}\ \bibnamefont
  {Dittes}}, \bibinfo {author} {\bibfnamefont {J.}~\bibnamefont {Quezada}}, \
  and\ \bibinfo {author} {\bibfnamefont {T.~H.}\ \bibnamefont {Seligman}},\
  }\href {\doibase 10.1103/PhysRevE.54.3221} {\bibfield  {journal} {\bibinfo
  {journal} {Phys. Rev. E}\ }\textbf {\bibinfo {volume} {54}},\ \bibinfo
  {pages} {3221} (\bibinfo {year} {1996})}\BibitemShut {NoStop}%
\bibitem [{\citenamefont {Mirlin}\ and\ \citenamefont
  {Evers}(2000)}]{MirlinPRBM2000}%
  \BibitemOpen
  \bibfield  {author} {\bibinfo {author} {\bibfnamefont {A.~D.}\ \bibnamefont
  {Mirlin}}\ and\ \bibinfo {author} {\bibfnamefont {F.}~\bibnamefont {Evers}},\
  }\href {\doibase 10.1103/PhysRevB.62.7920} {\bibfield  {journal} {\bibinfo
  {journal} {Phys. Rev. B}\ }\textbf {\bibinfo {volume} {62}},\ \bibinfo
  {pages} {7920} (\bibinfo {year} {2000})}\BibitemShut {NoStop}%
\bibitem [{\citenamefont {Schwartz}\ \emph
  {et~al.}(2007{\natexlab{a}})\citenamefont {Schwartz}, \citenamefont {Bartal},
  \citenamefont {Fishman},\ and\ \citenamefont {Segev}}]{Segev07a}%
  \BibitemOpen
  \bibfield  {author} {\bibinfo {author} {\bibfnamefont {T.}~\bibnamefont
  {Schwartz}}, \bibinfo {author} {\bibfnamefont {G.}~\bibnamefont {Bartal}},
  \bibinfo {author} {\bibfnamefont {S.}~\bibnamefont {Fishman}}, \ and\
  \bibinfo {author} {\bibfnamefont {M.}~\bibnamefont {Segev}},\ }\href
  {\doibase 10.1364/OPN.18.12.000035} {\bibfield  {journal} {\bibinfo
  {journal} {Opt. Photon. News}\ }\textbf {\bibinfo {volume} {18}},\ \bibinfo
  {pages} {35} (\bibinfo {year} {2007}{\natexlab{a}})}\BibitemShut {NoStop}%
\bibitem [{\citenamefont {Schwartz}\ \emph
  {et~al.}(2007{\natexlab{b}})\citenamefont {Schwartz}, \citenamefont {Bartal},
  \citenamefont {Fishman},\ and\ \citenamefont {Segev}}]{Segev07b}%
  \BibitemOpen
  \bibfield  {author} {\bibinfo {author} {\bibfnamefont {T.}~\bibnamefont
  {Schwartz}}, \bibinfo {author} {\bibfnamefont {G.}~\bibnamefont {Bartal}},
  \bibinfo {author} {\bibfnamefont {S.}~\bibnamefont {Fishman}}, \ and\
  \bibinfo {author} {\bibfnamefont {M.}~\bibnamefont {Segev}},\ }\href@noop {}
  {\bibfield  {journal} {\bibinfo  {journal} {Nature}\ }\textbf {\bibinfo
  {volume} {446}},\ \bibinfo {pages} {52} (\bibinfo {year}
  {2007}{\natexlab{b}})}\BibitemShut {NoStop}%
\bibitem [{\citenamefont {Segev}\ \emph {et~al.}(2013)\citenamefont {Segev},
  \citenamefont {Silberberg},\ and\ \citenamefont {Christodoulides}}]{Segev13}%
  \BibitemOpen
  \bibfield  {author} {\bibinfo {author} {\bibfnamefont {M.}~\bibnamefont
  {Segev}}, \bibinfo {author} {\bibfnamefont {Y.}~\bibnamefont {Silberberg}}, \
  and\ \bibinfo {author} {\bibfnamefont {D.~N.}\ \bibnamefont
  {Christodoulides}},\ }\href@noop {} {\bibfield  {journal} {\bibinfo
  {journal} {Nature Photonics}\ }\textbf {\bibinfo {volume} {7}},\ \bibinfo
  {pages} {197} (\bibinfo {year} {2013})}\BibitemShut {NoStop}%
\bibitem [{\citenamefont {Roati}\ \emph {et~al.}(2008)\citenamefont {Roati},
  \citenamefont {D’Errico}, \citenamefont {Fallani}, \citenamefont {Fattori},
  \citenamefont {Fort}, \citenamefont {Zaccanti}, \citenamefont {Modugno},
  \citenamefont {Modugno},\ and\ \citenamefont {Inguscio}}]{Roati08}%
  \BibitemOpen
  \bibfield  {author} {\bibinfo {author} {\bibfnamefont {G.}~\bibnamefont
  {Roati}}, \bibinfo {author} {\bibfnamefont {C.}~\bibnamefont {D’Errico}},
  \bibinfo {author} {\bibfnamefont {L.}~\bibnamefont {Fallani}}, \bibinfo
  {author} {\bibfnamefont {M.}~\bibnamefont {Fattori}}, \bibinfo {author}
  {\bibfnamefont {C.}~\bibnamefont {Fort}}, \bibinfo {author} {\bibfnamefont
  {M.}~\bibnamefont {Zaccanti}}, \bibinfo {author} {\bibfnamefont
  {G.}~\bibnamefont {Modugno}}, \bibinfo {author} {\bibfnamefont
  {M.}~\bibnamefont {Modugno}}, \ and\ \bibinfo {author} {\bibfnamefont
  {M.}~\bibnamefont {Inguscio}},\ }\href@noop {} {\bibfield  {journal}
  {\bibinfo  {journal} {Nature}\ }\textbf {\bibinfo {volume} {453}},\ \bibinfo
  {pages} {895} (\bibinfo {year} {2008})}\BibitemShut {NoStop}%
\bibitem [{\citenamefont {Billy}\ \emph {et~al.}(2008)\citenamefont {Billy},
  \citenamefont {Josse}, \citenamefont {Zuo}, \citenamefont {Bernard},
  \citenamefont {Hambrecht}, \citenamefont {Lugan}, \citenamefont
  {Cl{\'e}ment}, \citenamefont {Sanchez-Palencia}, \citenamefont {Bouyer},\
  and\ \citenamefont {Aspect}}]{Billy08}%
  \BibitemOpen
  \bibfield  {author} {\bibinfo {author} {\bibfnamefont {J.}~\bibnamefont
  {Billy}}, \bibinfo {author} {\bibfnamefont {V.}~\bibnamefont {Josse}},
  \bibinfo {author} {\bibfnamefont {Z.}~\bibnamefont {Zuo}}, \bibinfo {author}
  {\bibfnamefont {A.}~\bibnamefont {Bernard}}, \bibinfo {author} {\bibfnamefont
  {B.}~\bibnamefont {Hambrecht}}, \bibinfo {author} {\bibfnamefont
  {P.}~\bibnamefont {Lugan}}, \bibinfo {author} {\bibfnamefont
  {D.}~\bibnamefont {Cl{\'e}ment}}, \bibinfo {author} {\bibfnamefont
  {L.}~\bibnamefont {Sanchez-Palencia}}, \bibinfo {author} {\bibfnamefont
  {P.}~\bibnamefont {Bouyer}}, \ and\ \bibinfo {author} {\bibfnamefont
  {A.}~\bibnamefont {Aspect}},\ }\href@noop {} {\bibfield  {journal} {\bibinfo
  {journal} {Nature}\ }\textbf {\bibinfo {volume} {453}},\ \bibinfo {pages}
  {891} (\bibinfo {year} {2008})}\BibitemShut {NoStop}%
\bibitem [{\citenamefont {Chhabra}\ and\ \citenamefont
  {Jensen}(1989)}]{ChhabraJensen89}%
  \BibitemOpen
  \bibfield  {author} {\bibinfo {author} {\bibfnamefont {A.}~\bibnamefont
  {Chhabra}}\ and\ \bibinfo {author} {\bibfnamefont {R.~V.}\ \bibnamefont
  {Jensen}},\ }\href {\doibase 10.1103/PhysRevLett.62.1327} {\bibfield
  {journal} {\bibinfo  {journal} {Phys. Rev. Lett.}\ }\textbf {\bibinfo
  {volume} {62}},\ \bibinfo {pages} {1327} (\bibinfo {year}
  {1989})}\BibitemShut {NoStop}%
\bibitem [{\citenamefont {Ujfalusi}\ and\ \citenamefont
  {Varga}(2015)}]{Varga2015}%
  \BibitemOpen
  \bibfield  {author} {\bibinfo {author} {\bibfnamefont {L.}~\bibnamefont
  {Ujfalusi}}\ and\ \bibinfo {author} {\bibfnamefont {I.}~\bibnamefont
  {Varga}},\ }\href {\doibase 10.1103/PhysRevB.91.184206} {\bibfield  {journal}
  {\bibinfo  {journal} {Phys. Rev. B}\ }\textbf {\bibinfo {volume} {91}},\
  \bibinfo {pages} {184206} (\bibinfo {year} {2015})}\BibitemShut {NoStop}%
\bibitem [{\citenamefont {Rodriguez}\ \emph {et~al.}(2009)\citenamefont
  {Rodriguez}, \citenamefont {Vasquez},\ and\ \citenamefont
  {R\"omer}}]{Rodriguezetal2009}%
  \BibitemOpen
  \bibfield  {author} {\bibinfo {author} {\bibfnamefont {A.}~\bibnamefont
  {Rodriguez}}, \bibinfo {author} {\bibfnamefont {L.~J.}\ \bibnamefont
  {Vasquez}}, \ and\ \bibinfo {author} {\bibfnamefont {R.~A.}\ \bibnamefont
  {R\"omer}},\ }\href {\doibase 10.1103/PhysRevLett.102.106406} {\bibfield
  {journal} {\bibinfo  {journal} {Phys. Rev. Lett.}\ }\textbf {\bibinfo
  {volume} {102}},\ \bibinfo {pages} {106406} (\bibinfo {year}
  {2009})}\BibitemShut {NoStop}%
\bibitem [{\citenamefont {Fyodorov}\ and\ \citenamefont
  {Mirlin}(1994)}]{Fyodorov94}%
  \BibitemOpen
  \bibfield  {author} {\bibinfo {author} {\bibfnamefont {Y.~V.}\ \bibnamefont
  {Fyodorov}}\ and\ \bibinfo {author} {\bibfnamefont {A.~D.}\ \bibnamefont
  {Mirlin}},\ }\href@noop {} {\bibfield  {journal} {\bibinfo  {journal}
  {International Journal of Modern Physics B}\ }\textbf {\bibinfo {volume}
  {08}},\ \bibinfo {pages} {3795} (\bibinfo {year} {1994})}\BibitemShut
  {NoStop}%
\bibitem [{\citenamefont {M{\'e}ndez-Berm{\'u}dez}\ and\ \citenamefont
  {Varga}(2006)}]{Mendez06}%
  \BibitemOpen
  \bibfield  {author} {\bibinfo {author} {\bibfnamefont {J.}~\bibnamefont
  {M{\'e}ndez-Berm{\'u}dez}}\ and\ \bibinfo {author} {\bibfnamefont
  {I.}~\bibnamefont {Varga}},\ }\href@noop {} {\bibfield  {journal} {\bibinfo
  {journal} {Physical Review B}\ }\textbf {\bibinfo {volume} {74}},\ \bibinfo
  {pages} {125114} (\bibinfo {year} {2006})}\BibitemShut {NoStop}%
\bibitem [{\citenamefont {M{\'e}ndez-Berm{\'u}dez}\ \emph
  {et~al.}(2012)\citenamefont {M{\'e}ndez-Berm{\'u}dez}, \citenamefont
  {Alcazar-L{\'o}pez},\ and\ \citenamefont {Varga}}]{Mendez12}%
  \BibitemOpen
  \bibfield  {author} {\bibinfo {author} {\bibfnamefont {J.}~\bibnamefont
  {M{\'e}ndez-Berm{\'u}dez}}, \bibinfo {author} {\bibfnamefont
  {A.}~\bibnamefont {Alcazar-L{\'o}pez}}, \ and\ \bibinfo {author}
  {\bibfnamefont {I.}~\bibnamefont {Varga}},\ }\href@noop {} {\bibfield
  {journal} {\bibinfo  {journal} {EPL (Europhysics Letters)}\ }\textbf
  {\bibinfo {volume} {98}},\ \bibinfo {pages} {37006} (\bibinfo {year}
  {2012})}\BibitemShut {NoStop}%
\bibitem [{\citenamefont {M{\'e}ndez-Berm{\'u}dez}\ \emph
  {et~al.}(2014)\citenamefont {M{\'e}ndez-Berm{\'u}dez}, \citenamefont
  {Alcazar-L{\'o}pez},\ and\ \citenamefont {Varga}}]{Mendez14}%
  \BibitemOpen
  \bibfield  {author} {\bibinfo {author} {\bibfnamefont {J.}~\bibnamefont
  {M{\'e}ndez-Berm{\'u}dez}}, \bibinfo {author} {\bibfnamefont
  {A.}~\bibnamefont {Alcazar-L{\'o}pez}}, \ and\ \bibinfo {author}
  {\bibfnamefont {I.}~\bibnamefont {Varga}},\ }\href@noop {} {\bibfield
  {journal} {\bibinfo  {journal} {Journal of Statistical Mechanics: Theory and
  Experiment}\ }\textbf {\bibinfo {volume} {2014}},\ \bibinfo {pages} {P11012}
  (\bibinfo {year} {2014})}\BibitemShut {NoStop}%
\end{thebibliography}
\end{document}